\documentclass[twocolumn,floatfix, oldfontcommands]{aastex63}

\usepackage{graphicx}
\usepackage{amssymb}
\usepackage{amsmath}
\usepackage[caption=false]{subfig}
\usepackage{enumerate}
\usepackage{enumitem}
\usepackage{color}
\usepackage{units}
\usepackage{rotating}
\usepackage{wrapfig}
\usepackage{float}
\usepackage{lineno}

\definecolor{update}{rgb}{1,0,0}

\makeatletter
\newcommand\footnoteref[1]{\protected@xdef\@thefnmark{\ref{#1}}\@footnotemark}
\makeatother
\renewcommand{\thempfootnote}{\arabic{mpfootnote}}

\usepackage{chngcntr}
\counterwithout{figure}{section}
\counterwithout{table}{section}

\received{July 2, 2021}
\revised{October 8, 2021}
\accepted{October 13, 2021}
\submitjournal{ApJ}

\shorttitle{LCG Candidates in COSMOS}
\shortauthors{L. J. Prichard et al.}

\graphicspath{{./}{figures/}}
\begin{document}
% \mainmatter    %when using this the numbering changes

\title{Lyman Continuum Galaxy Candidates in COSMOS}

\correspondingauthor{Laura J. Prichard}
\email{lprichard@stsci.edu}

\author[0000-0002-0604-654X]{Laura J. Prichard}
\affiliation{Space Telescope Science Institute, 3700 San Martin Drive, Baltimore, MD 21218, USA}

\author[0000-0002-9946-4731]{Marc Rafelski}
\affiliation{Space Telescope Science Institute, 3700 San Martin Drive, Baltimore, MD 21218, USA}
\affiliation{Department of Physics \& Astronomy, Johns Hopkins University, Baltimore, MD 21218, USA}

\author[0000-0001-5703-2108]{Jeff Cooke}
\affiliation{Centre for Astrophysics and Supercomputing, Swinburne University of Technology, PO Box 218, Hawthorn VIC 3122, Australia}
\affiliation{ARC Centre of Excellence for All Sky Astrophysics in 3 Dimensions (ASTRO 3D), Australia}

\author[0000-0002-0441-8629]{Uros Me\v{s}tri\'{c}}
\affiliation{Centre for Astrophysics and Supercomputing, Swinburne University of Technology, PO Box 218, Hawthorn VIC 3122, Australia}
\affiliation{ARC Centre of Excellence for All Sky Astrophysics in 3 Dimensions (ASTRO 3D), Australia}
\affiliation{INAF—Osservatorio di Astrofisica e Scienza dello Spazio, via Gobetti 93/3, I-40129 Bologna, Italy}

\author{Robert Bassett}
\author[0000-0002-5360-8103]{Emma V. Ryan-Weber}
\affiliation{Centre for Astrophysics and Supercomputing, Swinburne University of Technology, PO Box 218, Hawthorn VIC 3122, Australia}
\affiliation{ARC Centre of Excellence for All Sky Astrophysics in 3 Dimensions (ASTRO 3D), Australia}

\author[0000-0003-3759-8707]{Ben Sunnquist}
\affiliation{Space Telescope Science Institute, 3700 San Martin Drive, Baltimore, MD 21218, USA}

\author[0000-0002-8630-6435]{Anahita Alavi}
\affiliation{IPAC, California Institute of Technology, 1200 E. California Boulevard, Pasadena, CA 91125, USA}

\author[0000-0001-6145-5090]{Nimish Hathi}
\affiliation{Space Telescope Science Institute, 3700 San Martin Drive, Baltimore, MD 21218, USA}

\author[0000-0002-9373-3865]{Xin Wang}
\affiliation{IPAC, California Institute of Technology, 1200 E. California Boulevard, Pasadena, CA 91125, USA}

\author[0000-0002-4917-7873]{Mitchell Revalski}
\author{Varun Bajaj}
\affiliation{Space Telescope Science Institute, 3700 San Martin Drive, Baltimore, MD 21218, USA}

\author[0000-0002-7893-1054]{John M. O'Meara}
\affiliation{W. M. Keck Observatory, 65-1120 Mamalahoa Highway, Kamuela, HI 96743, USA}

\author[0000-0001-5185-9876]{Lee Spitler}
\affiliation{Department of Physics and Astronomy, Macquarie University, Sydney, NSW 2109, Australia}
\affiliation{Research Centre in Astronomy, Astrophysics \& Astrophotonics, Macquarie University, Sydney, NSW 2109, Australia}

%------------------
%ABSTRACT (250 words)
%------------------
\begin{abstract}
Star-forming galaxies are the sources likely to have reionized the universe. As we cannot observe them directly due to the opacity of the intergalactic medium at $z\gtrsim5$, we study $z\sim3\text{--}5$ galaxies as proxies to place observational constraints on cosmic reionization. Using new deep \textit{Hubble Space Telescope} rest-frame UV F336W and F435W imaging (30-orbit, $\sim40$~arcmin$^2$, $\sim29\text{--}30$~mag depth at 5$\sigma$), we attempt to identify a sample of Lyman continuum galaxies (LCGs). These are individual sources that emit ionizing flux below the Lyman break ($<912~\text{\AA}$). This population would allow us to constrain cosmic reionization parameters such as the number density and escape fraction ($f_{\rm esc}$) of ionizing sources. We compile a comprehensive parent sample that does not rely on the Lyman-break technique for redshifts. We present three new spectroscopic candidates at $z\sim3.7\text{--}4.4$, and 32 new photometric candidates. The high-resolution multi-band HST imaging and new Keck/Low Resolution Imaging Spectrometer (LRIS) redshifts make these promising spectroscopic LCG candidates. Using both a traditional and probabilistic approach, we find the most likely $f_{\rm esc}$ values for the three spectroscopic LCG candidates are $>100\%$, and therefore not physical. We are unable to confirm the true nature of these sources with the best available imaging and direct blue Keck/LRIS spectroscopy. More spectra, especially from the new class of 30 m telescopes, will be required to build a statistical sample of LCGs to place firm observational constraints on cosmic reionization. 
\end{abstract}


%------------------
%INTRO
%------------------
\section{Introduction} \label{sec:intro}

During the Epoch of Reionization (EoR), the universe was transformed from almost completely neutral to nearly entirely ionized \citep[e.g.,][]{Loeb2001}. This transition was thought to occur at $z\gtrsim6$ \citep[e.g.,][]{Fan2006, Pentericci2011, Mason2018}, but recent results show that it could have continued inhomogeneously in patches down to $z\gtrsim5$ (e.g., observations: \citealt{Becker2015, Bosman2018, Eilers2018, Yang2020, Kusakabe2020}; theory: \citealt[][]{Kulkarni2019, Keating2020, Nasir2020}). The sources responsible for cosmic reionization are not yet known but it is thought that star-forming galaxies are the main contributors. This is because the number density of quasars has been found not to significantly contribute to the ultraviolet (UV) background beyond $z\sim3$ \citep[e.g.,][]{Hopkins2007, Jiang2008, Fontanot2012}.

Although star-forming galaxies remain the most likely source of ionizing flux during the EoR, the type responsible for the bulk of the UV background radiation remains elusive. Many studies of the EoR have focused on abundant, young, low-mass, star-forming galaxies as the main contributors \cite[e.g.,][]{Ouchi2009, Wise2009, Yajima2011, Finkelstein2012, Bouwens2015, Paardekooper2015, Robertson2015}. However, recent modeling found that most of the ionizing photon budget ($\gtrsim80\%$) could come from massive, UV-bright galaxies with high escape fractions of ionizing radiation \citep[$f_{\rm esc}$;][]{Naidu2020}. This theory is bolstered by observations of ultra-luminous Lyman-$\alpha$ (Ly$\alpha$) emitters during the EoR \citep[$z\sim6.6$; e.g.,][]{Hu2016, Matthee2018, Songaila2018, Meyer2021, Taylor2020}. These powerful UV-flux emitters are thought to be located at the sites of massive reionization bubbles, which is supported by results from simulations \citep[e.g.,][]{Gronke2020}.

Given the opacity of the intergalactic medium (IGM) at $z\gtrsim$ 5 to UV radiation, direct observations of the flux that ionized the universe are not possible \citep[e.g.,][]{Inoue2008, Inoue2014}. However, we require more observational constraints on $f_{\rm esc}$ to understand reionization. While we cannot directly observe the photons that reionized the universe, we can look to $z\sim3\text{--}5$ galaxies that serve as close proxies to those during the EoR. Identifying high-redshift galaxies that emit Lyman continuum (LyC) flux, ionizing UV radiation below the Lyman break ($<912~\text{\AA}$), has long been a focus of observational cosmology. We can calibrate observable properties of galaxies during the EoR, such as optical nebular emission lines, to the LyC flux emission. We can therefore determine this relation for $z\sim3\text{--}5$ galaxies and extrapolate it to $z > 6$ \citep[e.g.,][]{Bassett2019}.

Low-redshift observations of LyC emitting galaxies can be used to better understand $f_{\rm esc}$ \citep[e.g.,][]{Cardamone2009, Izotov2016a, Izotov2016b, Alavi2020}. However, these low-redshift galaxies may not serve as the best proxies compared to the higher-redshift ($z\sim3\text{--}5$) analogs. For example, LyC emitters in the nearby universe exhibit redder UV slopes \citep[e.g.,][]{Borthakur2014, Izotov2016b}. Also, between $z\sim4$ and $z\sim7$ is $\sim0.8$ Gyr, but between $z\sim0.5$ and $z\sim7$ is $\sim8$ Gyr.

Early searches for high-redshift LyC emitting galaxies yielded promising candidates that were later ruled out due to low-redshift contamination \citep[e.g.,][]{Steidel2001, Iwata2009, Siana2010, Vanzella2010b, Nestor2011, Nestor2013, Mostardi2013}. High-resolution \textit{Hubble Space Telescope} (HST) imaging is essential for identifying contamination by overlapping foreground sources \citep[e.g.,][]{Vanzella2010a, Siana2015, Mostardi2015, Vanzella2016, Pahl2021}. Even with HST data, LyC escape can remain undetected from likely high-redshift candidates, but can be used for upper limits on $f_{\rm esc}$ \citep[e.g.,][]{Amorin2014, Vasei2016, Bian2020, Mestric2021}. 

To date there are relatively few confirmed direct spectroscopic detections of genuine leaking LyC flux from high-redshift galaxies: \textit{Ion2} \citep{Vanzella2015, deBarros2016, Vanzella2016},  Q1549-C25 \citep{Shapley2016}, \textit{Ion3} \citep{Vanzella2018}, \textit{Ion1} \citep{Vanzella2012, Ji2020}, and 13 clean detections from the Keck Lyman Continuum Spectroscopic Survey \cite[KLCS;][]{Steidel2018, Pahl2021}. A few more promising candidates have been identified with UV imaging to detect the LyC flux, HST imaging to identify low-redshift contaminants, and spectroscopic redshifts at redder wavelengths \citep[e.g.,][]{Bian2017, Rivera-Thorsen2017, Rivera-Thorsen2019, Fletcher2019, Nakajima2020, Mestric2020, Bassett2022}.

One possible explanation for the lack of detections relative to the extensive LyC search campaigns is the methods used to select possible candidates. Identifying high-redshift LyC-emitter candidates requires secure redshifts and a method of detecting leaking LyC flux. An efficient way of measuring the redshifts of distant UV-bright star-forming galaxies is the Lyman-break technique \citep{Steidel1996b}. This method uses large area multi-broadband imaging to efficiently identify high-redshift galaxies, rather than expensive deep spectroscopy of individual sources. An important question to ask is if LyC flux is added below the Lyman break, is the galaxy always identifiable as a Lyman-break galaxy (LBG)? 

This was the hypothesis that \cite{Cooke2014} explored by artificially adding LyC flux of various strengths to LBG spectra and measuring the resulting colors. \cite{Mestric2020} also tested this theory by searching in new CFHT Large Area $U$-band Deep Survey \citep[CLAUDS;][]{Sawicki2019} images for LyC flux for galaxies with publicly available spectroscopic redshifts. This search resulted in two new candidates. As found by \cite{Cooke2014, Mestric2020}, these ``Lyman continuum galaxies'' (LCGs) are typically outside of the LBG color-color selection boxes. A fraction of LBGs with LyC flux still reside in the traditional LBG selection region \citep[see][]{Cooke2014}. However, their location is inconsistent with the model predictions for their redshift. All galaxies that have LyC flux are included in the LCG definition. However, understanding those outside of the LBG selection region, that may be missed by other searches, is a primary goal of this work.

Even spectroscopic redshifts in the literature can have selection biases that favor LBGs. Targets for spectroscopic surveys are usually pre-selected to have a representative distribution using either color selection or photometric redshifts. Galaxies are likely to have the securest photometric redshifts if they exhibit a clean Lyman break. Therefore, spectroscopic redshift selected samples can also be biased against strong LyC detection. 

In this work, we present new deep HST imaging of the Cosmic Evolution Survey field \citep[COSMOS;][]{Scoville2007} to investigate  LCGs. We use accurate photometric redshifts from the FourStar Galaxy Evolution Survey \citep[ZFOURGE;][]{Straatman2016}, and 18 other public photometric and spectroscopic redshift catalogs, to identify candidates. ZFOURGE uses 30-band photometry in the COSMOS field from $0.5\text{--}8~\mu \rm m$, including medium infrared (IR) bands that sample the Balmer break at $1 \lesssim z \lesssim 4$. This enables accurate photometric redshifts without relying on the Lyman break that may be less prominent due to escaping LyC. We attempted to spectroscopically follow up our LCG candidates using Keck/Low Resolution Imaging Spectrometer (LRIS) and other telescopes. We present our results from the photometric and spectroscopic candidate searches here. 

We describe the HST observations, custom reduction, and photometry in Section \ref{sec:hst}. We outline the redshift catalogs and selection methods used to identify LCG candidates in Section \ref{sec:samp}. We detail our various spectroscopic follow-up campaigns and reduction in Section \ref{sec:spec}. The sample refinement, properties, and analysis are described in Section \ref{sec:an}. We discuss these results in the context of the literature in Section \ref{sec:disc}. We summarize the paper and give our main conclusions in Section \ref{sec:conc}. Throughout this paper we use AB magnitudes \citep[m$_\mathrm{AB}$;][]{Oke1983} and assume the latest Planck cosmological parameters \citep[$H_0 = 67.4 ~\rm km s^{-1} \rm Mpc^{-1}$, $\Omega_{m}=0.315$;][]{Planck2020}.

%------------------
%HST DATA
%------------------
\begin{figure} 
\includegraphics[width=0.5\textwidth, trim=15 20 15 15, clip]{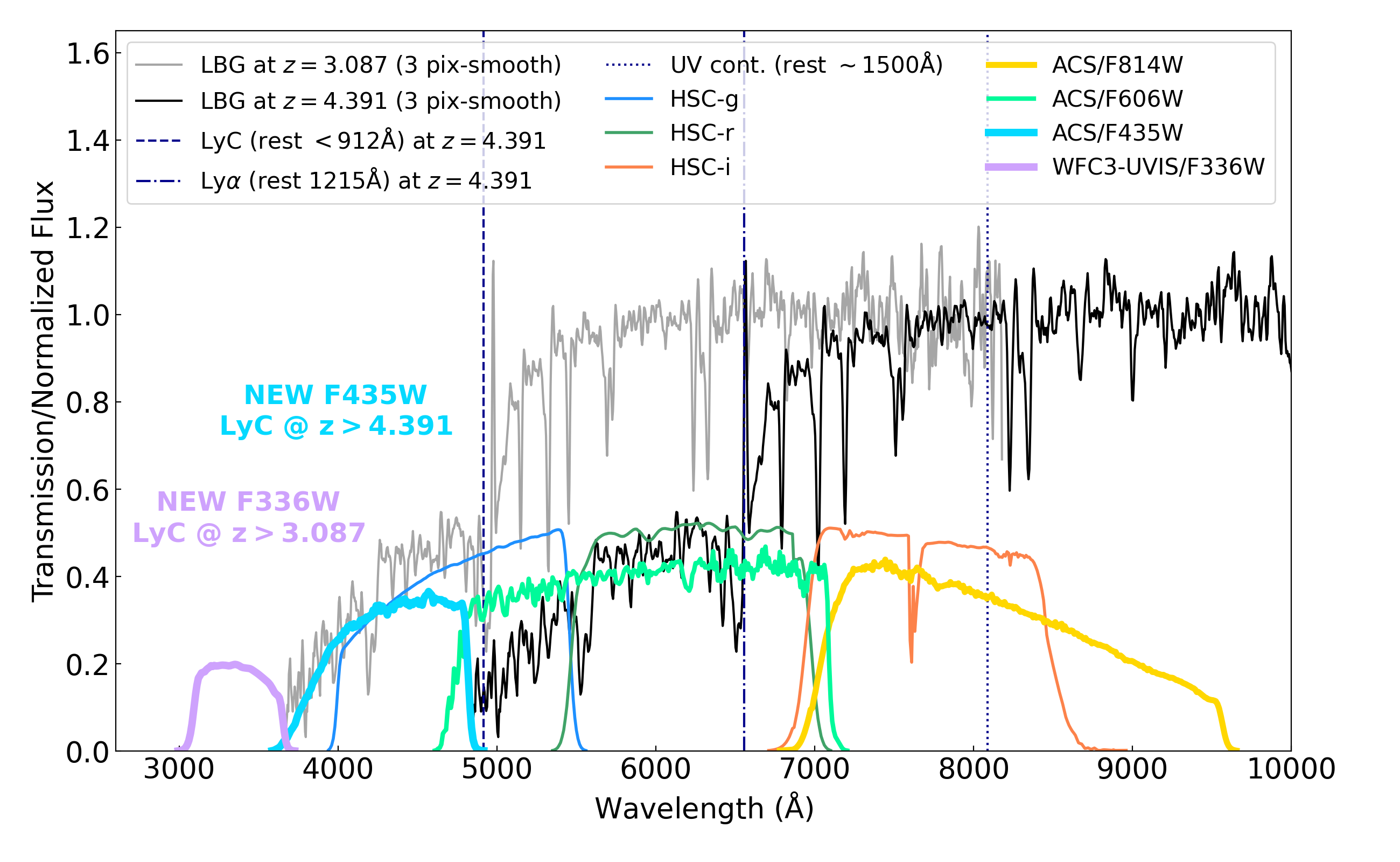}\caption{Photometric bands used in this work shown over a 3-pixel boxcar-smoothed LBG composite spectrum (stack of $\sim200$ spectra with similar Ly$\alpha$ strength) from \cite{Shapley2003}. The LBG spectrum is redshifted to $z=3.087$ (F336W $z_{\rm lim}$; gray) and $z=4.391$ (F435W $z_{\rm lim}$; black). Four HST bands are shown (left to right): new WFC3/F336W (purple), new ACS/F435W (light blue), CANDELS-ACS/F606W (light green), COSMOS-ACS/F814W (gold), and three HSC bands (left to right): \textit{g} (dark blue), \textit{r} (dark green), \textit{i} (orange). Key spectral features are highlighted on the black ($z=4.391$) spectrum with blue vertical lines: Lyman break at rest $\sim912 ~\text{\AA}$ (dashed), Ly$\alpha$ at rest $\sim1216 ~\text{\AA}$ (dot-dashed), non-ionizing UV continuum rest $\sim1500 ~\text{\AA}$ (dotted).\small} 
\label{fig:lbg_bands}
\end{figure}

\begin{figure*}
\includegraphics[width=\textwidth, trim=100 85 110 115, clip]{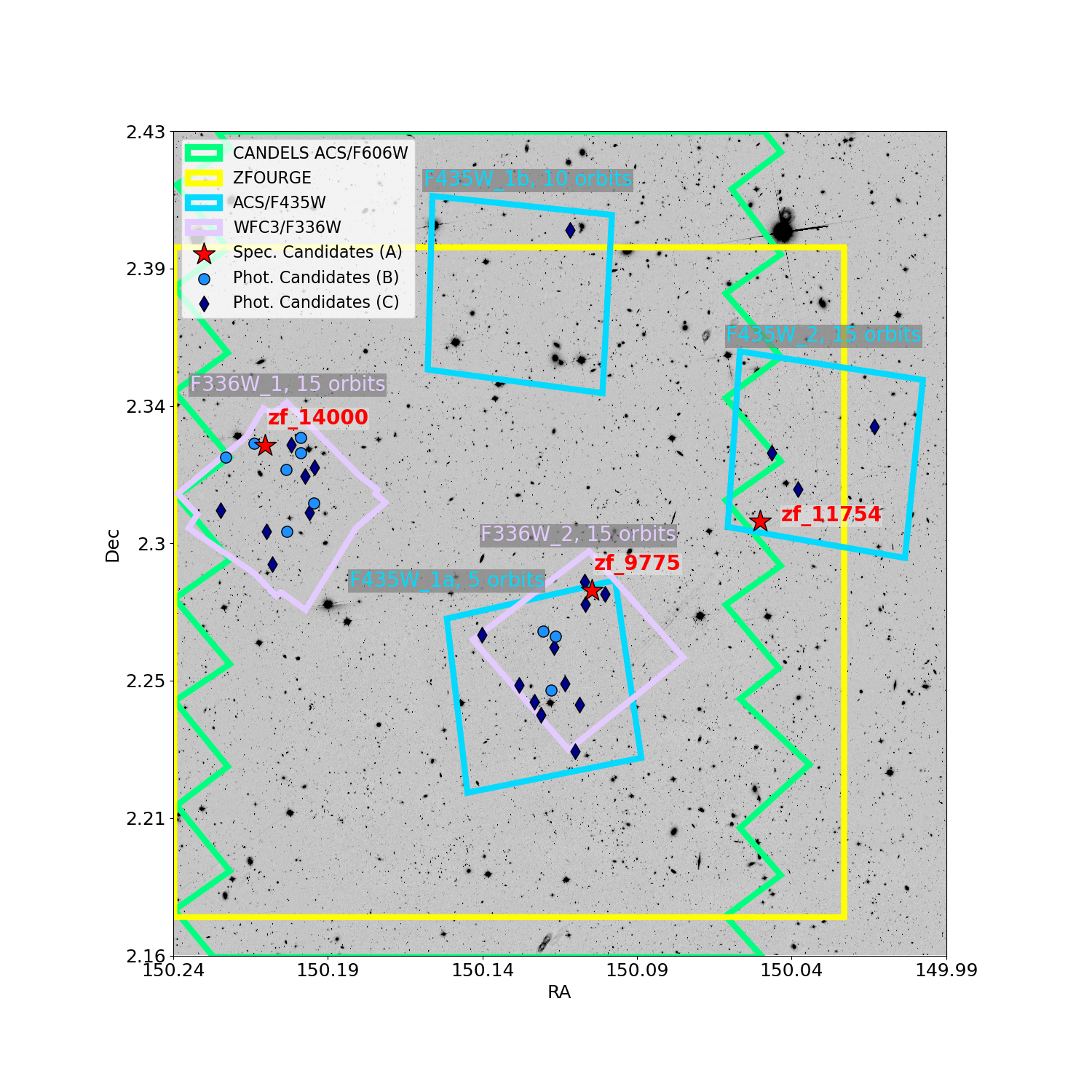}
\caption{COSMOS-ACS/F814W image cutout \citep{Koekemoer2007} with survey footprints relevant to this work and our new LCG candidates overlaid. Two deep WFC3/F336W (purple) and three parallel ACS/F435W (light blue) pointings with their names and orbit depths shown. The CANDELS-ACS/F606W \citep[light green;][]{Grogin2011, Koekemoer2011} and ZFOURGE survey \citep[yellow;][]{Straatman2016} footprints are also shown. The LCG candidates are split into three categories: spectroscopic candidates/Group A (red stars and ZFOURGE IDs), and photometric candidates, both Group B (blue circles) and C (dark blue diamonds, lowest priority candidates). See Section \ref{sec:props} for more details on the LCG candidate group categorization. \small}
\label{fig:space}
\end{figure*}

\section{\textit{Hubble Space Telescope} data} \label{sec:hst}

\subsection{Observations}
\label{sec:hstobs}

High-resolution HST data are required for this work to identify any potential low-redshift contaminants along the line of sight. The new HST data presented here were taken in Cycle 25 (GO 15100; PI Cooke; 30 orbits) in three blocks between 2018 February 21 and 2018 April 15. The data were taken using two filters in parallel: F336W on Wide Field Camera 3 (WFC3)/UVIS and F435W on Advanced Camera for Surveys (ACS)/WFC.

For galaxies at $z=3$--$5$, LyC flux ($<912 ~\text{\AA}$) is cleanly sampled in the WFC3/F336W band at $z>3.087$ and in the ACS/F435W band at $z>4.391$. These limits assume a $<0.3\%$ red leak ($\gtrsim$8 mag fainter than the UV continuum) as used in \cite{Smith2018} and we refer to these limits as $z_{\rm lim}$. Figure \ref{fig:lbg_bands} shows an LBG composite ($\sim200$ spectra binned by Ly$\alpha$ strength) from \cite{Shapley2003}. The LBG composite is redshifted to $z_{\rm lim}=3.087$ (gray) and $z_{\rm lim}=4.391$ (black). Transmission curves for photometric bands used in this work are overlaid. The HST bands: WFC3/F336W, ACS/F435W, ACS/F606W, ACS/F814W, and three ground-based Hyper-Suprime Cam (HSC) bands: \textit{g}, \textit{r}, \textit{i} are shown. We also highlight key features on the LBG template at $z = 4.391$ (black) with vertical dark blue lines: the Lyman break at rest $\sim912 ~\text{\AA}$ (dashed), Ly$\alpha$ at rest $\sim1216 ~\text{\AA}$ (dot-dashed), and the non-ionizing UV continuum at rest $\sim1500 ~\text{\AA}$ (dotted).

The 30 orbits of HST data were originally intended to be taken in two F336W pointings with two parallel F435W pointings. Given observation scheduling constraints, the first F336W pointing (F336W\_1) was observed in two overlapping visits with different position angles (PAs) for a total of 15 orbits. This means that the parallel F435W pointings are split up into F435W\_1a (5 orbits) and F453W\_1b (10 orbits). The second F336W pointing (F336W\_2) and F435W parallel (F435\_2) both include the full 15 orbits. The total area of all the pointings is $\sim40$ arcmin$^2$ at various depths. Figure \ref{fig:space} shows all five HST pointing footprints (two F336W and three F435W), with names and respective orbit depths, overlaid on the COSMOS-ACS/F814W background image cutout \citep{Koekemoer2007}. We also show the CANDELS-ACS/F606W \citep{Grogin2011, Koekemoer2011} footprint (light green) overlaid.

We use full-orbit exposures to reduce the total background contribution caused by post-flashing each exposure to 12 e$^{-}$ \citep[as recommended in][]{Anderson2012}. The longer exposures also reduce read noise and maximize the counts per exposure for faint sources. This enables improved charge transfer efficiency (CTE) correction which is better suited for higher count levels. We use a five-point dither pattern per visit to sub-sample the point-spread function (PSF). We employ a larger dither pattern between each visit to remove the chip gap and low-level, large-scale pattern noise \citep{Rafelski2015}. The total exposure time of both F336W pointings shown in Figure \ref{fig:space} is $\sim44000$ s. The exposure time of the deepest F435W pointing (F435W\_2) is $\sim40000$ s, and this exposure time is split by a 1:2 ratio between F435W\_1a and F435W\_1b respectively. See Table \ref{tab:psf} for a summary.

\subsection{HST data reduction}
\label{sec:hstred}

Given the faint nature of potential LCG candidates in the rest-far UV (FUV), having the best quality HST images is essential. We therefore make several improvements to the standard Space Telescope Science Institute (STScI) WFC3/UVIS reduction pipeline, to clean and calibrate our images beyond what was previously available from the Mikulski Archive for Space Telescopes (MAST). We also developed and apply new additional calibrations to the CTE-corrected single-exposure calibrated images (called FLCs). Some of these improvements have been added to the standard WFC3/UVIS MAST pipeline (detailed in Section \ref{sec:darks}), while others are additional calibration improvements (outlined in Section \ref{sec:hstcal}). We provide further information and plots (Figure \ref{fig:cte}) to demonstrate our improvements in Appendix \ref{sec:appdarks}.

\subsubsection{HST/WFC3 darks pipeline improvements}
\label{sec:darks}

The standard darks pipeline for WFC3/UVIS on HST left some residual systematic and instrumental effects. Previously, \cite{Rafelski2015} identified the cause and corrected for some of the worst of these residual effects in the WFC3/UVIS darks. In this work, we adapt the STScI WFC3/UVIS darks \texttt{Python} pipeline \citep[copied April 19, 2019;][]{Bourque2016}. The improvements include those outlined in \cite{Rafelski2015} that were previously implemented in custom IDL routines, and we add further new improved calibrations. Here we briefly outline the major changes we make to the WFC3/UVIS darks pipeline which have since been adopted by the WFC3 team as standard for their delivered data products (released May 2021). The WFC3/UVIS MAST data also include the new time-dependent photometry flux calibrations that are essential for accurate photometry \citep{Bohlin2020, Bajaj2020, Calamida2021}.

\begin{itemize}[leftmargin=*]
\itemsep0em 
    \item A new charge transfer efficiency (CTE) correction \citep[\textsc{calwf3 v3.6.0};][]{Anderson2020, Anderson2021} is used. The new code uses a pixel-based algorithm that optimizes parameters for the data to mitigate read noise. This reduces background noise and improves the correction of bright sources such as cosmic rays which reduces residual gradients across the images.
    \item The detectors are periodically heated in an anneal to reduce hot pixels. The period between each heating is referred to as an ``anneal cycle'', within which the detector has a different ``fingerprint''. Darks are now processed within concurrent anneal cycles (i.e., not using the darks from the previous anneal cycle for pixel replacement). This results in cleaner corrections to the data, and reduced blotchy patterns and noise (see also appendix in \citealt{Rafelski2015}).
\end{itemize}

\subsubsection{Additional calibrations}
\label{sec:hstcal}

The improvements described so far have been folded into the official WFC3/UVIS pipeline coupled with the new time-dependent photometry. We therefore download the WFC3/F336W data presented in this paper from MAST (on May 28, 2021) as a starting point. We then apply additional corrections described below to these data to further improve their calibration. We make no changes to the ACS reduction pipeline, so the F435W FLCs are those from MAST. A Jupyter notebook of the additional calibrations and some steps required to easily apply them to WFC3 and ACS (if relevant) FLCs is available on GitHub\footnote{\url{https://github.com/lprichard/HST_FLC_corrections} including codes from \url{https://github.com/bsunnquist/uvis-skydarks}}. We briefly summarize these corrections below and give more details in Appendix \ref{sec:appdarks}.
\begin{itemize}[leftmargin=*]
\itemsep0em 
    \item We developed a novel hot pixel flagging routine to apply a variable flagging threshold as a function of distance from the ``readout'' amplifier. The method used in the MAST data uses a constant threshold. Due to the imperfect CTE correction, this results in $\sim30\%$ of hot pixels being missed further from the readout.
    \item We flag the negative divots adjacent to readout cosmic rays (ROCRs) in the data quality (DQ) arrays of each FLC. These ROCRs land on the array while it is being read out and are overcorrected by the CTE code.
    \item We equalize the overall signal levels in the four different amplifier regions (quadrants) of each FLC. This produces smoother images without the background discontinuities.
\end{itemize}

\subsection{Image drizzling}
\label{sec:driz}

\subsubsection{Ancillary data}
\label{sec:hstanc}

We collate publicly available ancillary HST imaging to aid with source detection and for escape fraction estimates. At $z\gtrsim3.1$ ($z_{\rm lim}$ for F336W), the non-ionizing UV continuum around rest $\sim1500 ~\text{\AA}$ is partially covered by F606W. For $z\gtrsim4.4$ ($z_{\rm lim}$ for F435W), the UV continuum lands in F814W. See Figure \ref{fig:lbg_bands} for the positions of the bands on spectra shifted to both $z_{\rm lim}$ values. We therefore use the CANDELS-ACS/F606W \citep{Grogin2011, Koekemoer2011}\footnote{\url{https://archive.stsci.edu/prepds/candels/}} and COSMOS-ACS/F814W \citep{Koekemoer2007}\footnote{\url{https://www.stsci.edu/~koekemoe/cosmos/current/}} images in our search and analysis of LCGs. See Figure \ref{fig:space} for the F606W footprint (light green) and the F814W band (background image). The COSMOS-F814W image spans the entire COSMOS field ($\sim1.4\text{ deg}\times1.4\text{ deg}$), and thus we have F814W spanning our new HST images.

We combine the F814W tiles spanning the new HST pointings and ZFOURGE footprint with the \textsc{Montage v6.0} software\footnote{\url{http://montage.ipac.caltech.edu/}}. We then reconstruct the header World Coordinate System (WCS) information of the resulting mosaic with the \textsc{tweakreg} routine from the \textsc{DrizzlePac} package \citep{Gonzaga2012, Hoffmann2021}. This step makes the WCS compatible for use with \textsc{astrodrizzle} that is used for combining the new HST images. The F606W and F814W mosaics are cropped down to the area of interest covering the new HST pointings, shown in Figure \ref{fig:space}. Both the F606W and F814W images have $0.03^{\prime\prime}$ pixel scales and are aligned to the COSMOS survey coordinates. The CANDELS-F606W images have an exposure time of 3300s and depth of 28.5 mag at $5\sigma$ \citep{Grogin2011, Koekemoer2011}. The COSMOS-F814W images have an exposure time of 2028s with a $27.2$ mag point-source depth at $5\sigma$ \citep{Koekemoer2007}. See Table \ref{tab:psf} for a summary of image depths.

\begin{deluxetable*}{|c|c|c|c|c|c|c|}
\tabletypesize{\footnotesize}
\tablewidth{0pt}
\tablecaption{HST image properties and contamination probabilities in the LyC filters (F336W or F435W).}
\label{tab:psf}
\tablehead{Filter & No. Orbits & Exposure & $5\sigma$-Depth$^{1}$ & PSF FWHM & Zeropoints & LyC Contamination$^{2}$ \\ & &  Time (s) & (m$_\mathrm{AB}$) & ($0.03^{\prime\prime}$ pixels) & (m$_\mathrm{AB}$) &  Probability}
\startdata
COSMOS-ACS/F814W$^{3}$ & 1 & 2028 & 27.20 & 3.9 & 25.952 & -\\
CANDELS-ACS/F606W$^{4}$ & 2 & 3300 & 28.50 & 3.8 & 26.491 & -\\
New ACS/F435W\_1a & 5 & 13326 & 28.52 & 3.3 & 25.650 & 0.025\\
New ACS/F435W\_1b & 10 & 26652 & 29.11 & 3.4 & 25.650 & 0.018\\
New ACS/F435W\_2 & 15 & 39978 & 29.78 & 3.4 & 25.650 & 0.013\\
New WFC3/F336W\_1 & 15 & 44091 & 29.77 & 3.0 & 24.689 & 0.050\\
New WFC3/F336W\_2 & 15 & 44091 & 30.11 & 3.0 & 24.689 & 0.014\\
\enddata
\tablenotetext{}{1. RMS depth measured within a 0.2$^{\prime\prime}$ radius aperture. 2. Probability of a 3$\sigma$ contamination within a $0.5^{\prime\prime}$ aperture in the LyC band. 3. \citep{Koekemoer2007} 4. \citep{Grogin2011, Koekemoer2011}.}
\end{deluxetable*}

\subsubsection{Creating mosaics}
\label{sec:hstmos}

The new F336W and F435W FLCs are drizzled into two separate images using tools from \textsc{DrizzlePac} \citep[v3.1.6 for F435W, v3.2.1 for F336W;][]{Gonzaga2012, Hoffmann2021}. The \textsc{updatewcs} (with \texttt{use\_db=False}), \textsc{tweakreg}, and \textsc{tweakback} packages are used for image alignment. We developed a tool that we use for cleaning image edges to improve alignment with \textsc{tweakreg} (\textsc{cleanedges}\footnote{\url{https://github.com/lprichard/cleanedges}}). The new F336W and F435W FLCs are drizzled onto the COSMOS coordinates using the F814W band as a reference image for the same footprint. We create both an exposure time map (`EXP') and inverse variance weight map (`IVM') for each of the new filter drizzles. The final F336W and F435W mosaics have $0.03^{\prime\prime}$ pixel scales and are aligned with pixel orientation north. See Table \ref{tab:driz} for a summary of the final drizzle parameters. 

We measure the depth of the images by taking the sigma-clipped median root-mean-squared (RMS) of 1000 apertures placed on sky (avoiding sources) in the HST pointings with our \textsc{hst\_sky\_rms} code\footnote{\url{https://github.com/lprichard/hst_sky_rms}}. We then measure the flux required for a $5\sigma$  detection (relative to our median sky RMS) in a $0.2^{\prime\prime}$ radius aperture and convert this to a magnitude for each filter with time-dependent zeropoints from STScI\footnote{WFC3: \url{https://www.stsci.edu/hst/instrumentation/wfc3/data-analysis/photometric-calibration/uvis-photometric-calibration}, ACS: \url{https://www.stsci.edu/hst/instrumentation/acs/data-analysis/zeropoints}}. The resulting RMS image depths of F336W\_1 (15 orbits, misaligned PAs) is m$_\mathrm{AB} = 29.77$ at 5$\sigma$ and for F336W\_2 (15 orbits) is m$_\mathrm{AB} = 30.11$ at 5$\sigma$. The RMS magnitude depths for the three F435W pointings with varying orbit depths are m$_\mathrm{AB} = 28.52$ at 5$\sigma$ for F435W\_1a (5 orbits), m$_\mathrm{AB} = 29.11$ at 5$\sigma$ for F435W\_1b (10 orbits), and m$_\mathrm{AB} = 29.78$ at 5$\sigma$ for F435W\_2 (15 orbits). See Table \ref{tab:psf} for a summary of image properties. All our F336W and F435W data products are available as a High Level Science Product at MAST via \dataset[10.17909/t9-pe8e-s980]{\doi{10.17909/t9-pe8e-s980}}\footnote{\url{https://archive.stsci.edu/hlsp/lcgcosmos/}}.

\begin{deluxetable}{|l|l|}
\tabletypesize{\footnotesize}
\tablewidth{0pt}
\tablecaption{Relevant \textsc{astrodrizzle} parameters for making the new F336W and F435W images.}
\label{tab:driz}
\tablehead{Parameter & Value}
\startdata
\texttt{driz\_cr\_corr} & True \\
\texttt{driz\_combine} & True \\
\texttt{clean} & True \\
\texttt{final\_wcs} & True \\
\texttt{final\_scale} & 0.03 \\
\texttt{final\_pixfrac} & 0.6 \\
\texttt{skymethod} & globalmin+match \\
\texttt{skysub} & True \\
\texttt{combine\_type} & imedian \\
\enddata
\end{deluxetable}

\subsection{Photometry}
\label{sec:phot}

\subsubsection{Point spread functions}
\label{sec:psf}

Obtaining accurate point-spread functions (PSFs) in the UV is notoriously challenging due to the non-existent or faint UV flux from stars that are commonly used for accurate PSF profiles. Fortunately, new tools developed at STScI allow us to accurately model the PSFs in the UV. We use \textsc{psf\_tools}, a module in the \textsc{wfc3\_photometry} package\footnote{\url{https://github.com/spacetelescope/wfc3_photometry}} that provides an interface for performing PSF photometry on WFC3 and ACS images (adapted for this work) using the PSF models developed by \cite{Anderson2000, Anderson2016}. We briefly summarize the method for obtaining accurate PSFs for all the HST images and PSF-matching them for analysis.

We use the \textsc{make\_model\_star\_image} tool from the \textsc{psf\_tools.psfutils} module in the \textsc{wfc3\_photometry} package to plant empirical PSF models into copies of our individual FLCs and drizzle them together. The resulting image includes a grid of $\sim1000$ empirical PSFs that can be stacked like real stars. We stack the PSFs in the images using new PSF stacking tools\footnote{\url{https://github.com/mrevalski/hst_wfc3_psf_modeling}}. We extract each PSF and interpolate them onto a sub-pixel grid. The PSFs are fitted with Moffat profiles, checked for quality, aligned, and the median is taken. We interpolate the sub-pixel sample average PSF profiles back to the native pixel scale of the images and determine a full-width-half-max (FWHM) value (see Table \ref{tab:psf}). 

For the F336W and F435W images we stack the empirical PSFs. For F606W and F814W we stack 77 real stars. We create a convolution kernel using a Hanning Window to translate the PSF of each band to the F814W PSF with the largest FWHM. We convolve all images using their respective kernels to obtain PSF-matched images. For the new F336W and F435W images, we apply different PSF corrections for each pointing given their different depths and PAs.

\subsubsection{Isophotal flux measurements}
\label{sec:srcdet}

To identify sources in the PSF-matched images and extract their flux, we use tools from the \textsc{photutils} \texttt{Python} package. The benefit of isophotal photometry, i.e., extracting flux for each object within segmentation footprints, is that it cleanly extracts flux for the faint blue sources that we are targeting. We first create RMS maps from the inverse variance weight maps (IVM) output by \textsc{astrodrizzle} (RMS = 1/$\sqrt(\rm IVM)$). We then create a noise-equalized (NEQ) image of all our PSF-matched HST bands. To do this, we divide the PSF-matched images by their RMS maps to create signal-to-noise ($S/N$) maps, mask invalid values, and sum them together. We also create negatives of all the images ($\times-1$), including the NEQ image, that we use to aid with source detection. 

We make two different segmentation maps from the NEQ images using the \textsc{photutils.detect\_threshold} and \textsc{photutils.detect\_sources} routines. We create a clean map with few spurious objects for source detection and flux extraction (i.e., isophotes), using a 2D Gaussian kernel with standard deviation of 3 pixels, threshold of 1.8, and 15 connected pixels for detection (\texttt{npixels}). We test values for our clean segmentation map by finding the threshold required to find few sources in our negative images while maximizing flux extraction of the real sources. We also make a low threshold map (0.8, \texttt{npixels}=10) with many spurious sources for rigorously masking objects when measuring a local sky value (sky masks).

We perform isophotal photometry with \textsc{photutils} by measuring flux within the source footprints determined from the NEQ image and giving this as a mask to the \textsc{aperture\_photometry} routine. We first determine a local sky value in an annulus around each source. We heavily mask sources with our rigorous sky mask and take a $3\sigma$-clipped median of the ``sky'' pixels. We multiply this sky value by the isophotal area of the source and subtract it from the source flux. 

To calculate the errors on the isophotal fluxes we use the \textsc{photutils.calc\_total\_error} function. To account for correlated noise that results from drizzling larger native pixels onto a finer grid, we determine a correlated-noise correction factor \citep[as described in][]{Fruchter2002, Hoffmann2021}. The \textsc{photutils.calc\_total\_error} function then takes the image data, the RMS map multiplied by the correlated noise factor, and the exposure-time map to determine an error map that includes shot noise. The \textsc{photutils.aperture\_photometry} routine takes the image data with the local background subtracted, the isophotal extraction footprint, and the total error array to generate accurate photometry and errors.

\subsubsection{LyC contamination calculations}
\label{sec:cont}

We measure the probability of LyC flux contamination for the new F336W and F435W bands using a similar method to determining the image depths (included in our \textsc{hst\_sky\_rms} code\footnote{\url{https://github.com/lprichard/hst_sky_rms}}). We make a 3$\sigma$ segmentation map for each pointing and randomly place 10,000 apertures onto it. We use an aperture width of $0.4^{\prime\prime}$ which is the average size of our LCG candidates (see Section \ref{sec:props}). If there are 15 source pixels within any randomly placed aperture (our \texttt{npixels} value for source detection), it is considered contaminated. Taking a ratio of the clean and contaminated 10,000 randomly placed apertures gives us a contamination rate for each pointing. 

We find a 0.025, 0.018, and 0.013 probability of any F435W\_1a-, F435W\_1b-, F435W\_2--selected source (respectively) being contaminated. We also find a 0.050 and 0.014 contamination rate estimate for F336W\_1 and F336W\_2 respectively (see Table \ref{tab:psf} for a summary). Although both F336W pointings have 15-orbit depths, F336W\_1 has misaligned PAs (see Figure \ref{fig:space}) so the image edges are shallower and more noisy. Toward the image center at full depth, the image properties will be close to that of F336W\_2. These are conservative estimates as the source flux, size, and redshift are not considered, only randomly overlapping sources in the segmentation map.

\subsubsection{Ground-based photometry}
\label{sec:hsc}

As seen in Figure \ref{fig:space}, not all our HST images are covered by CANDELS-F606W that we use in our analysis of LCGs. For F435W\_2, we require ground-based data that span the field and a similar wavelength range as F606W for making color-color plots, for illustrative purposes rather than sample selection. We opt to use HSC on the 8.2~m Subaru telescope on Maunakea. HSC Deep+UltraDeep coverage of COSMOS \citep[Public Data Release 2;][]{Aihara2019}\footnote{\url{https://hsc-release.mtk.nao.ac.jp/doc/}} has depths of 28.4 mag in $g$, 28.0 mag in $r$, and 27.7 mag in $i$ (bands shown in Figure \ref{fig:lbg_bands}). We take the seeing FWHM values from \cite{Aihara2019} of 0.61$^{\prime\prime}$ for HSC-$r$ and 0.81$^{\prime\prime}$ for HSC-$i$ and create a Gaussian PSF for each band. We then create a convolution kernel to PSF-match the F435W images to the HSC-$i$ band. 

Due to the large size of the ground-based PSFs relative to the HST bands, when convolving the F435W band to the HSC-$i$ PSF, the flux from the faint sources that we are interested in is lost altogether. We therefore derive a correction to convert the F435W isophotal fluxes to $1^{\prime\prime}$-aperture- and HSC-PSF-matched fluxes. We achieve this by measuring $1^{\prime\prime}$ aperture fluxes on our F435W image convolved to the HSC-$i$ PSF for all sources. We then plot the HST isophotal fluxes against all the valid PSF- and aperture-matched F435W fluxes and derive the following relation for converting between the two.
\begin{equation}
F_{\rm F435W,1^{\prime\prime}ap} = 0.6 F_{\rm F435W, iso} + 0.063
\label{eq:hsc}
\end{equation}
Here $F_{\rm F435W,1^{\prime\prime}ap}$ is the F435W flux measured in $1^{\prime\prime}$ apertures from the HSC-$i$ PSF-matched image. $F_{\rm F435W, iso}$ is the measured F435W isophotal flux. We apply this correction to the F435W isophotal fluxes to convert them to $1^{\prime\prime}$-aperture HSC-$i$ PSF-convolved fluxes so they can be directly compared to the ground-based HSC bands.

%------------------
%SAMPLE
%------------------
\section{Sample Selection} 
\label{sec:samp}

\subsection{Redshift catalogs} 
\label{sec:zcats}

To select LCGs, we require a sample of high-redshift galaxies not selected with the Lyman-break technique. We use the ZFOURGE survey \citep{Straatman2016} as our primary source of redshifts as it comprises 30 bands of imaging, including medium IR FourStar bands \citep{Persson2013} from the 6.5 m Magellan Baade telescope at Las Campanas Observatory. These medium IR bands accurately sample the region around the Balmer break for well-constrained photometric redshifts \citep[$\delta z \sim2\%$ quoted accuracy;][]{Straatman2016} of $1<z<4$ galaxies \citep{vanDokkum2009}. ZFOURGE spans a roughly $11^{\prime}\times11^{\prime}$ ($13^{\prime}\times13^{\prime}$ with dithering) region of COSMOS (yellow footprint in Figure \ref{fig:space}). As shown in Figure \ref{fig:space}, the ZFOURGE footprint does not span the full area of our new HST F435W images (blue footprints), with around $\sim\nicefrac{1}{3}$ of F435W\_1b and F435W\_2 outside the ZFOURGE area. 

The ZFOURGE survey is useful in the search for LCGs as these medium bands sample the Balmer break. This means that we are less dependent on the Lyman break, which may be less prominent in these galaxies, for determining photometric redshifts. Strong emission lines can bias photometric redshift measurements. Narrow bands that are sensitive to emission were omitted in the ZFOURGE photometric redshift calculations, the medium bands are less sensitive to emission and were included, but may still bias the results. We therefore draw from multiple redshift sources where possible for selecting LCG candidates.

ZFOURGE sources are $K_{s}$-band selected in images with 5$\sigma$ depths of 26.2--26.5 mag at a typical seeing of $\sim0.4^{\prime\prime}$. As ZFOURGE galaxies are $K_{s}$-band selected, the candidates must be sufficiently bright in this red band to be detected and we will miss some faint bluer sources. We therefore collate other available redshift information in the COSMOS field to construct a more complete parent sample. These include photometric redshifts from the COSMOS2015 catalog \citep{Laigle2016} and the CANDELS catalog of COSMOS \citep{Nayyeri2017}.

We compiled a catalog of available spectroscopic redshifts in the field from 3D-HST \citep{Brammer2012, Momcheva2016}, CANDELSz7 \citep[][]{Pentericci2018},
DEIMOS 10K \citep{Hasinger2018}, DEIMOS-C3R2 \citep{Masters2017}, IMACS \citep{Trump2009}, FMOS \citep{Kartaltepe2015}, FORS2 \citep{George2011, Comparat2015}, GEEC2 Gemini-S GMOS \citep{Balogh2014}, hCOSMOS \citep[MMT/Hectospec;][]{Damjanov2018}, LEGA-C \citep[VLT/VIMOS;][]{vanderWel2016}, MOSDEF \citep[MOSFIRE;][]{Kriek2015}, PRIMUS \citep[Magellan/IMACS;][]{Coil2011, Cool2013}, VUDS \citep[VLT/VIMOS;][]{LeFevre2015}, zCOSMOS and zBRIGHT \citep[VLT/VIMOS;][]{Lilly2009}. We note that many targets in spectroscopic surveys are pre-selected using color selection and photometric redshifts. Many of these are biased towards objects with the strongest Lyman breaks. Therefore, many of the publicly available spectroscopic catalogs may be biased against bright leaking LyC.

We first spatially crop all catalogs down to our region of interest (shown in Figure \ref{fig:space}). We then cross-match the redshift catalogs within $1^{\prime\prime}$ apertures ($0.5^{\prime\prime}$ search radii). We start by matching everything in COSMOS to ZFOURGE, then add any candidates from COSMOS that do not overlap. We then cross-match the other catalogs, starting with CANDELS, and then the spectroscopic redshift catalog. The CANDELS catalog contains photometric redshifts produced by six different authors and methods with the same data. We therefore calculate an average CANDELS photometric redshift \citep[as shown to be most accurate in e.g.,][]{Dahlen2013, Barro2019} and standard deviation of the available redshifts and use both in our redshift constraints for selecting our samples. 

Within our F336W and F435W footprints there are 8311 unique sources at all redshifts (our ``\textit{All-$z$ Sample}''). This reference sample splits to 3065 objects in the F336W footprints and 6430 sources in the F435W footprints (with overlap). We create a comprehensive ``\textit{Parent Sample}'' from the \textit{All-$z$ Sample} by selecting all objects from the catalogs with a redshift above the LyC $z_{\rm lim}$ for the F336W or F435W filters. We remain as agnostic as possible at the candidate selection phase and include any galaxy in the \textit{Parent Sample} that has at least one redshift to place it above $z_{\rm lim}$ for F336W or F435W to probe LyC. The CANDELS redshift constraint we use for selecting our \textit{Parent Sample} was our calculated average minus one standard deviation of the CANDELS photometric redshifts. There are 574 unique objects in the \textit{Parent Sample}: 380 objects have at least one redshift $z>3.087$ and are in the F336W footprints and 242 objects have at least one redshift $z>4.391$ and are in the F435W footprints (with overlap).

\subsection{Visual inspection} 
\label{sec:vis}

We visually inspect all objects in the \textit{Parent Sample} with at least three different investigators examining each. We inspect the sources in all four HST bands described here, their NEQ and source-detection segmentation maps to evaluate if there is any visible LyC flux. If any of the investigators notes a detection of possible flux in the filter probing the rest-frame LyC (F336W and/or F435W), they are re-evaluated multiple times. We remove any obvious erroneous redshifts (large bright local galaxies, star spikes, and bright saturated stars) from the sample. We end up with a long list of 77 unique objects that have varying degrees of confidence both in their LyC detection (strength, offset, confusion over the source for close objects) and redshift. We rank these candidates based on their number of visual confirmations, LyC confidence, and redshift confidence. All 77 sources are used as potential candidates for spectroscopic follow up to remain as agnostic as possible and as a relatively high density on the sky is required for multi-object slit spectroscopy (MOS).

%------------------
%SPEC DATA
%------------------
\begin{deluxetable*}{|l|l|l|l|l|l|l|}
\tabletypesize{\footnotesize}
\tablewidth{0pt}
\tablecaption{Summary of spectroscopic follow-up campaigns related to or part of this program.}
\label{tab:spec}
\tablehead{Telescope/Instrument & PID & PI & Project & Awarded & Observed & Conditions \\ & & & Type & (nights) & (nights) & }
\startdata
Keck I/LRIS & 2014A\_W003LA & Cooke & Related & 1.0 & None & Weathered out (WO) \\
Keck I/LRIS & 2015A\_W012LA & Cooke & Related & 1.0 & 1.0 & Good$^{1}$ \\
Keck I/MOSFIRE & 2015A\_W193M & Glazebrook & Related & 0.25 & 0.25 & Good$^{2}$ \\
Keck I/LRIS & 2016A\_W034LA & Cooke & Related & 2 & 0.4 & Poor \\
Keck I/LRIS & 2018A\_W156 & Cooke & Related & 2 & 1 & Poor/WO \\
Keck I/LRIS & 2018B\_N188 & Rafelski & This program & 1.5 & None & WO \\
Keck I/MOSFIRE & 2018B\_W151 & Bassett & Related & 0.5 & 0.5 & Good$^{3}$ \\
LBT/MODS1 & TS\_2019B\_011 & Prichard & This program & 1.5 & None & WO \\
Keck I/LRIS & 2019B\_N010 & Rafelski & This program & 1.5 & 1.5 & Moderate ($\sim80\%$ usable) \\
Keck I/LRIS & 2020B\_N168 & Prichard & This program & 1.5 & None & WO \\
\enddata
\tablenotetext{}{1. Eight early-selected LCG candidates \citep{Bassett2019, Mestric2021}. 2. Seven early-selected LCG candidates ($H$- and $K$-band spectra). 3. \cite{Bassett2022}.}
\end{deluxetable*}

\section{Spectroscopic data} \label{sec:spec}

To spectroscopically confirm the redshift and true nature of our LCG candidates, we require deep, blue spectroscopy. One of the few telescope and instrument combinations that satisfy our constraints is the Low-Resolution Imaging Spectrometer \citep[LRIS;][]{Oke1995, McCarthy1998, Rockosi2010} on Keck I. LRIS is a MOS instrument that allows the efficient observation of targets in the same field. The observing site on Maunakea provides enough elevation to reduce the atmospheric extinction. Thicker atmospheres at lower sites can impede the transmission of blue flux, which would require significantly more observing time to achieve our primary science objectives.

Our team has been attempting to spectroscopically confirm LCGs in COSMOS selected independently of the Lyman-break technique since 2014, with four dedicated runs prior to the HST proposal. This work provided the motivation and foundation for the HST program presented here (PID 15100). We have since been awarded time for four dedicated separate observing runs, including three NASA/Keck and one Large Binocular Telescope (LBT), that are part of the HST program. We also applied for time on two additional related projects in COSMOS with some overlapping targets and science goals to this work. Unfortunately, nearly all this time has been weathered out. In total, we have been awarded 12 dedicated nights (over 20 calendar nights) that are part of, or related to, the HST program presented here. In this paper, we present $\sim1.5$ nights of usable Keck/LRIS data taken in moderate observing conditions (see Table \ref{tab:spec} for a summary).

\subsection{Observations} \label{sec:specobs}

We present Keck/LRIS data of our LCG candidates taken on 2016 February 9 and 10 (2016A\_W034LA, PI Cooke) and 2020 January 20--22 (2019B\_N010, PI Rafelski). We require good seeing ($<1^{\prime\prime}$), clear skies and dark nights for our observations. To spectroscopically detect the LyC, we require the objects to have optical magnitudes of $\sim24$ mag. We then need around a half night ($\sim6$ hours) in our optimal conditions to achieve spectroscopic detection of LyC flux. This assumes there is no contamination from neighboring sources or LRIS detector issues. The conditions for the 2016 run were relatively poor, spread over two nights, but we were able to obtain some usable exposures for zf\_9775. In 2020, around $\sim80\%$ of the time was usable but with a relatively poor average seeing of $\sim1.2^{\prime\prime}$ (relative to what is possible at Keck). This poorer seeing disperses light over the edges of the LRIS slit and therefore increases the exposure time required to get sufficient signal-to-noise.

\begin{figure*}
\begin{center}
    \includegraphics[width=\textwidth, trim=85 45 90 85, clip]{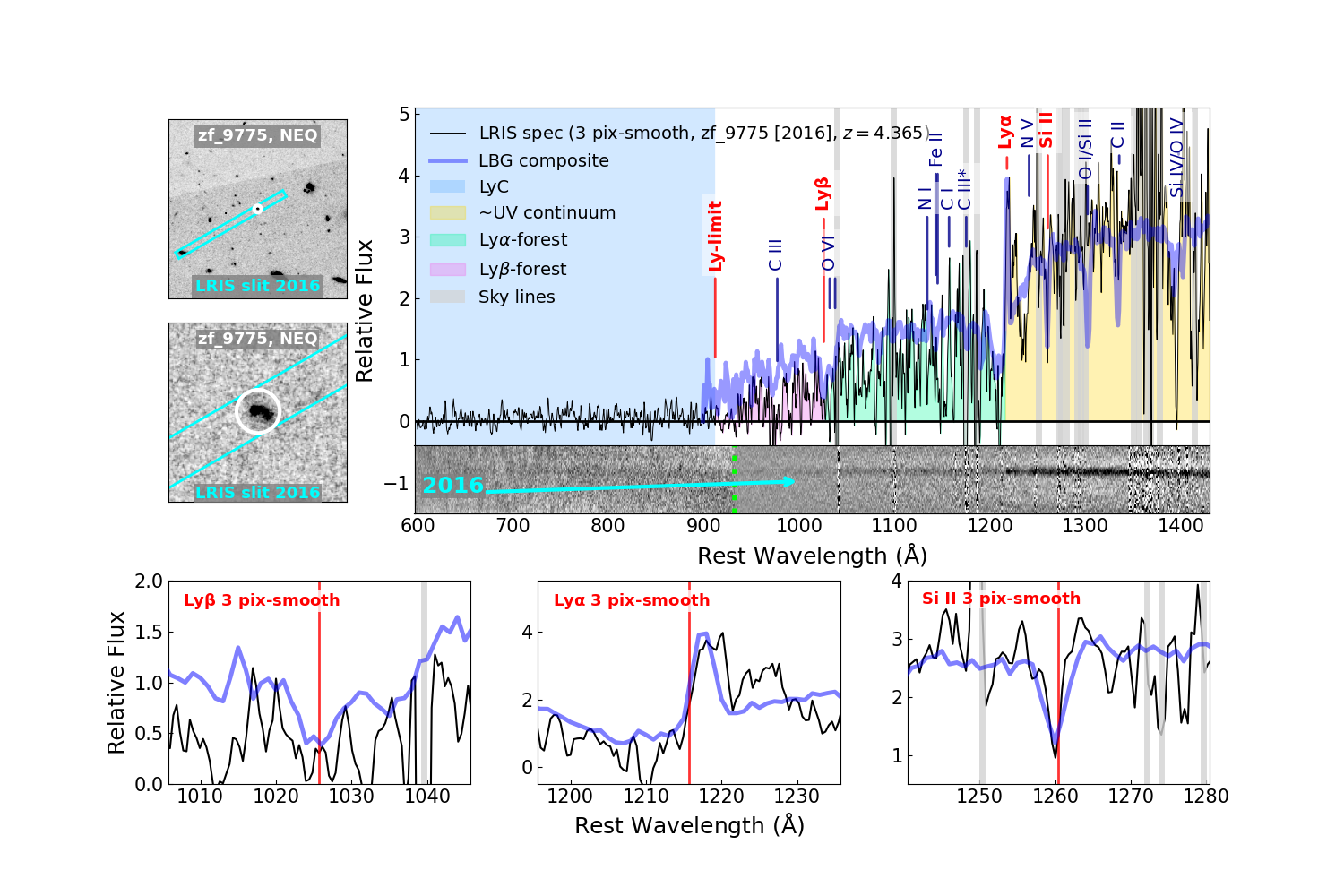}
    \caption{Keck/LRIS spectra (black) for the three Group A spectroscopic LCG candidates (IDs in figure with year of observation, continues over three pages). The spectra are 3-pixel boxcar smoothed and we show zoom-ins of some spectral features (panels below each spectrum). We show the 2D spectra below for reference and point to the trace of the source. The blue-red arm divide is highlighted (green dotted line). Overlaid is a composite of $\sim200$ LBG spectra \citep[blue;][]{Shapley2003} stacked by Ly$\alpha$-line strength. Spectroscopic features are indicated by the vertical lines; the red lines indicate the Lyman limit and features shown in the zoom-ins. Shading under the line shows: UV continuum (yellow), Ly$\alpha$ forest (green), Ly$\beta$ forest (violet), and LyC (blue panel). Noise-equalized image (NEQ) cutouts of each target with $1.2^{\prime\prime}$ apertures (white) and their LRIS slits overlaid are shown (left). The zoom-out stamp (top, 30$^{\prime\prime}$ across) shows the full slit, the zoom-in stamp (bottom, 5$^{\prime\prime}$ across) shows the source orientation in the slit. \small}
\end{center}
\end{figure*}

\begin{figure*}
\ContinuedFloat
\begin{center}
    \includegraphics[width=\textwidth, trim=100 45 90 85, clip]{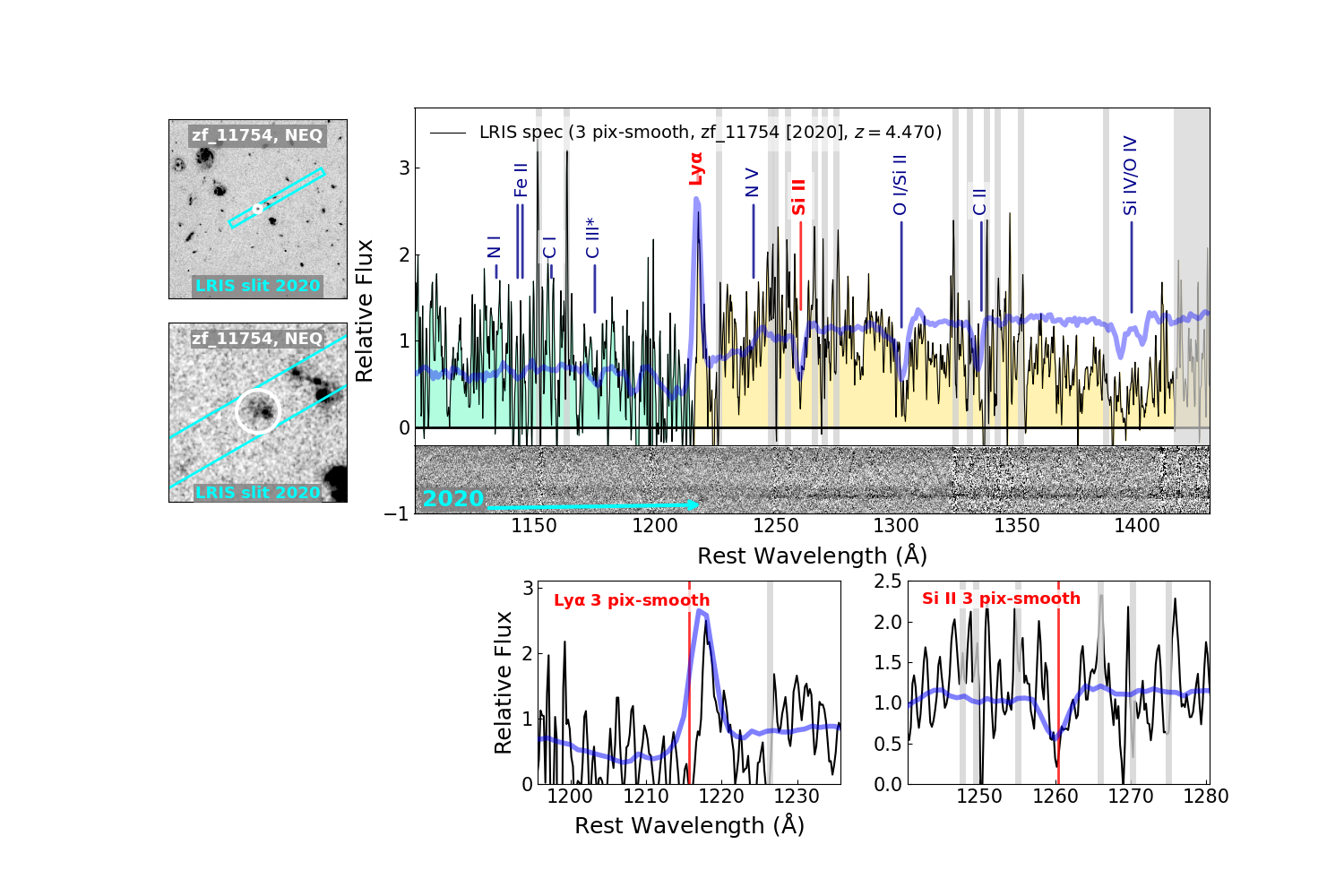}
  \caption{Continued. Only the red side of the spectrum is shown for zf\_11754 due to contamination from a neighboring source in the blue.\small}
  \label{fig:lris}
\end{center}
\end{figure*}

\begin{figure*} 
\ContinuedFloat
\begin{center}
   \includegraphics[width=\textwidth, trim=85 45 90 85, clip]{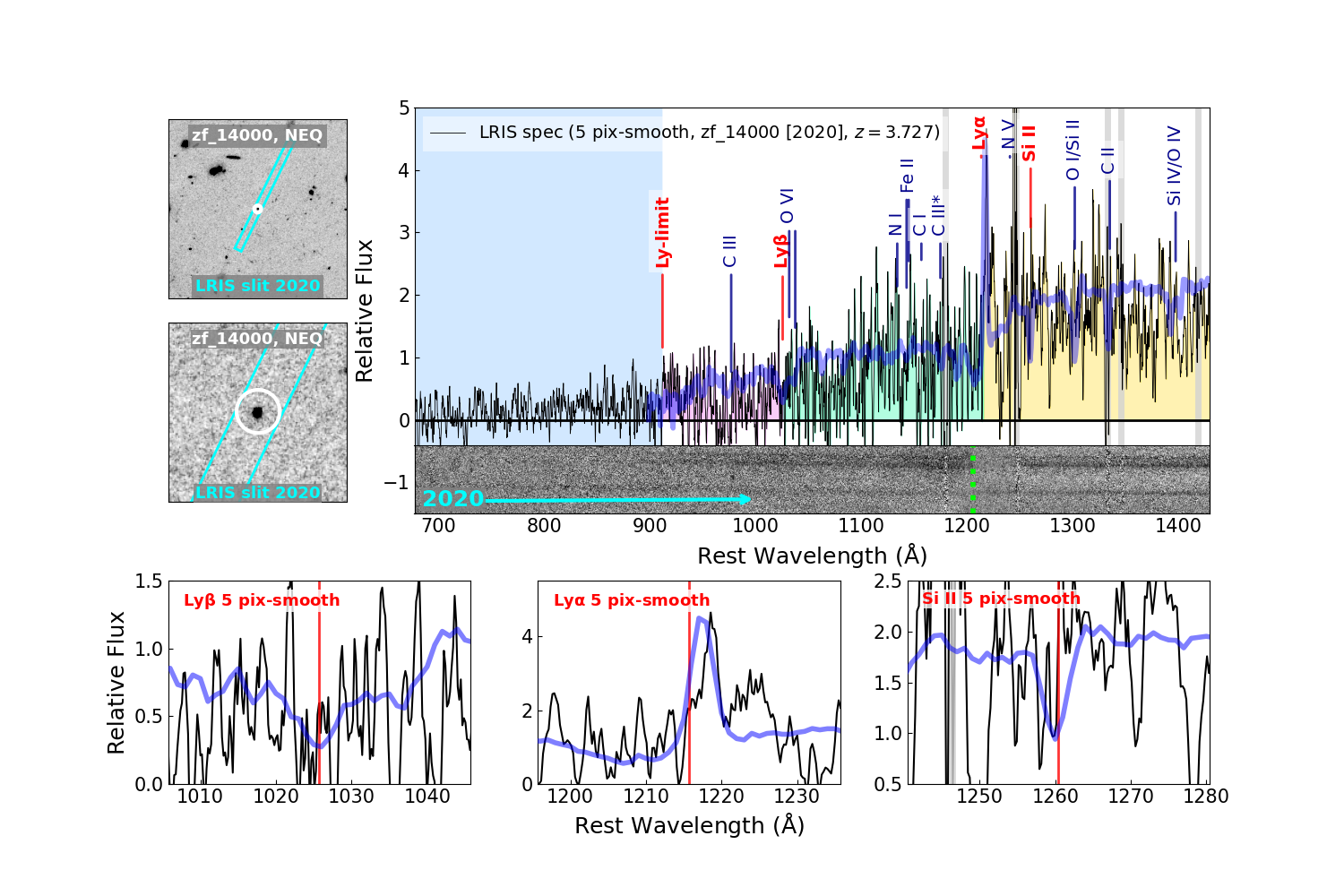}
   \caption{Continued.\small}
   \label{fig:lris}
\end{center}
\end{figure*}

The LRIS instrument uses pre-cut metal masks with slits to disperse the light of our targets across blue and red spectroscopic arms. We use the 400/3400 grism on the blue side and the 400/8500 grating for the red side with the 560 dichroic. We create LRIS masks with the \textsc{autoslit} software. The slit widths are $\sim1.2^{\prime\prime}$ and longer length slits are preferred to improve background subtraction. The masks are designed to encompass as many of our high-priority candidates as possible. We fill any remaining space with lower-priority LCG candidates and fillers that include extreme emission-line galaxies, Ly$\alpha$ emitters and LBGs. We then select at least four stars, ideally five to six, placed as near as possible to the edge of the LRIS masks for acquisition on the sky and mask alignment. 

The highest priority objects are placed as close to the north of the detector (with PA=0) as possible due to amplifier issues with the red side. At the time of observations, both chips could not reliably be read out, leaving no coverage at redder wavelengths for half the mask. The best objects are placed in the central third of the detector when possible (in $x$ and $y$) to accommodate full wavelength coverage and to minimize instrument distortion at the edges. With an optimal distribution on the sky, about $\sim25$ candidates fit on a mask, including $\sim5$ high-priority targets. However, when factoring in all the observational constraints, fewer than that are often in the prime detector and dispersion positions. We aim to observe each mask for a total of $\sim 6$ hours in 1200 s exposures to reduce data loss in variable weather conditions and contamination by cosmic rays. Typically, we observed just a few hours per mask if at all.

The Keck/LRIS spectra are reduced using the Image Reduction and Analysis Facility software \citep[IRAF;][]{Tody1986} following the standard reduction procedure for multi-slit spectroscopy. We also use twilight flats to aid with the blue flux calibrations. A 2$\times$2 binning was used for the 2016 observations and a 1$\times$1 binning was used for the 2020 data.

\subsection{Determining redshifts} \label{sec:specz}

The Keck/LRIS observations yielded redshifts for three LCG candidates: zf\_9775, zf\_11754, and zf\_14000. Only 14 (seven $>3\sigma$ in F336W/F435W) galaxies from our long candidate list were successfully observed due to the weather conditions limiting the number of masks we could observe and their spatial distribution. We present spectra for three of the $>3\sigma$ candidates, but were unable to get redshifts for the other four observed. A full analysis of the spectroscopic properties of the candidates (e.g., Ly$\alpha$ profiles) is planned to be presented in a future paper. Sources with confident spectroscopic LRIS redshifts and non-detections of LyC are the focus of another paper \citep{Mestric2021}. 

Figure \ref{fig:lris} (continues over three pages) shows 3-pixel boxcar-smoothed LRIS spectra (black) and a composite of $\sim200$ LBGs with similar Ly$\alpha$ strength to the galaxy overlaid \citep[blue;][]{Shapley2003}. The postage stamps on the left show the LRIS mask slits (top, $30^{\prime\prime}$ across) and candidate orientation (bottom, $5^{\prime\prime}$ across). Below each spectrum we show zoom-ins on some spectroscopic features. We also highlight other spectroscopic features with shading: the UV continuum red-ward of Ly$\alpha$ ($\gtrsim1216$ \AA; yellow), the Ly$\alpha$ forest (green), the Ly$\beta$ forest (violet), and the LyC ($<912$ \AA; blue panel). 

To determine redshifts for the candidates, we compare each spectrum to an LBG composite from \cite{Shapley2003} and look at the fit to the profile of the UV continuum. We also investigate all the highlighted spectroscopic information available, including the flux decrement as a result of the Ly$\alpha$ forest, the presence of a Ly$\alpha$ emission/absorption feature, and the fit to $\sim6$--12 ISM absorption features (depending on the wavelength coverage of the spectrum). We summarize the properties of the three spectra shown in Figure \ref{fig:lris} and redshifts we determine from each below.

\begin{figure*}
\begin{center}
\includegraphics[width=\textwidth, trim=100 60 20 40, clip]{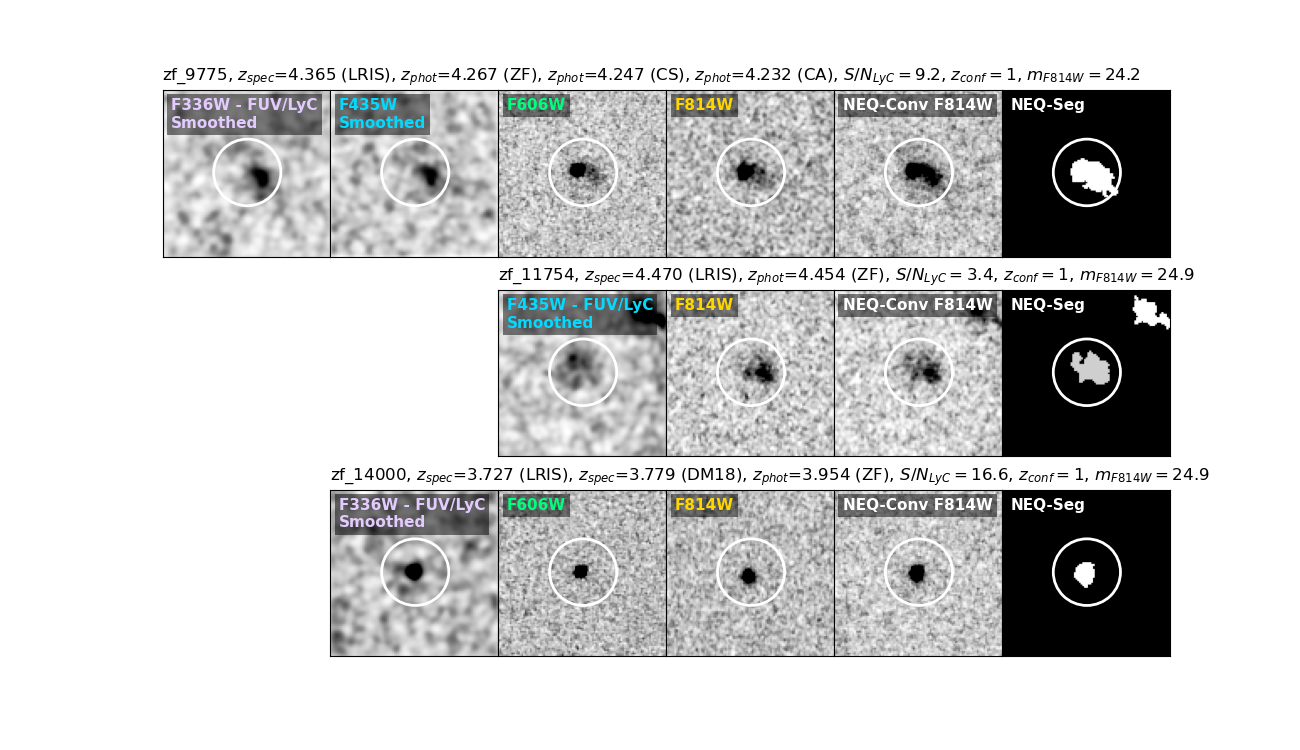}
\caption{HST postage stamps ($3^{\prime\prime}$ across, $1.2^{\prime\prime}$ white circles) of the three Group A ``spectroscopic" LCG candidates. We photometrically select and spectroscopically confirm the candidates with Keck/LRIS. Each row shows a different candidate in the available bands: new F336W, new F435W, CANDELS-F606W, COSMOS-F814W. We show a noise-equalized (NEQ) image that combines all available bands for each target. We create a segmentation map from the NEQ image (NEQ-Seg, right) to use for source detection and isophotal fluxes. The IDs, $>z_{\rm lim}$ redshifts, LyC $S/N$ (band indicated), redshift confidence ($z_{\rm conf}$), and magnitude in F814W (m$_{\rm F814W}$) are shown. The redshifts are either from LRIS (see Section \ref{sec:specz}), ZFOURGE (ZF), COSMOS (CS), CANDELS (CA), or DEIMOS 10K \citep[DM18, see Section \ref{sec:specz} and Appendix \ref{sec:appspec};][]{Hasinger2018} from the public spectroscopic catalog. See Table \ref{tab:lcgs} for a full summary of all available redshift information and Table \ref{tab:lcgsphot} for the photometry of all LCG candidates. See section \ref{sec:props} for details of their categorization and  parameters.\small}
\label{fig:top3}
\end{center}
\end{figure*}

\begin{itemize}[leftmargin=*]
\itemsep0em
    \item \textbf{zf\_9775:} We show the blue and red halves of the 2016 LRIS spectrum. The profile of the LRIS spectrum matches the LBG composite, with distinctive breaks and well-aligned $\sim10$ common strong UV ISM absorption features, all assessed to determine the redshift (e.g., see zoom-ins below). This spectrum has a confident redshift of $z=4.365$ (with $\delta z \sim 0.003$ for all spectra). However, there is some ambiguity over the nature of this detection. The spectrum is dominated by the optically bright component (left), and we are unable to resolve the component emitting the blue flux (right; see components in Figure \ref{fig:top3}). Therefore, zf\_9775 remains a ``spectroscopic candidate'' rather than a confirmed LyC detection.
    \item \textbf{zf\_11754:} We show just the red side of the spectrum as the blue side is contaminated by the close bright source in the upper right of the postage stamp. The red half of the spectrum shows a distinct Ly$\alpha$ line (panel below) and break on either side from the UV continuum (yellow) to Ly$\alpha$ forest (green). We are confident in the redshift from the red side of the spectrum of $z=4.470$ (using $\sim6$ common ISM lines). However, due to the blue side contamination and the fact that it is so close to the other source, there remains some ambiguity as to the nature of the LyC detection seen in the HST F435W images. This contamination could be increasing the blue flux through the Ly$\alpha$ forest (green) and even red-ward of Ly$\alpha$. The red side of the spectrum also shows some deviation from the composite which could be real (i.e., a very young population of stars and clean line of sight), or a product of extraction with such a close neighboring source. zf\_11754 therefore remains a spectroscopic candidate rather than a confirmed detection.
    \item \textbf{zf\_14000:} There is a tentative broad feature that could be Ly$\alpha$ and this exists at a slight break in the spectrum between what would be the UV continuum (yellow) and Ly$\alpha$ forest (green). This smaller decrement, if indeed a physical Ly$\alpha$ forest, could be due to the source being observed along a cleaner line of sight. This could be the case if considering the strong LyC detection which would require a high IGM transmission sight line. There is contamination seen in the 2D spectrum (upper trace) due to slight slit drift throughout the observations, from alignment star catalog inconsistencies or length of observations. However, the upper trace does not affect the extraction of zf\_14000 due to their sufficient separation. The tentative redshift we determine for this source is $z=3.727$, and it remains a spectroscopic candidate rather than confirmation. We rule out the possibility that this object is a star as its blue temperature is consistent with it being a K star, but the spectrum does not support that. Its spectral energy distribution (SED) is also not consistent with a late M-type star. Most catalogs list this source as non-stellar, but they do have a variety of redshifts ($z=0.301\text{--}3.953$, see Table \ref{tab:lcgs}). For it to be a K4--M0 star, it needs to be from 30--128 kpc away to be as faint as it is. Only the reddest M dwarf would be realistic (at $\sim2\text{--}10$ kpc) and they are too red for the SED.
\end{itemize}

Three of the LCG candidates have low-quality ancillary spectra. Two of the spectra place the candidates above the redshift limit of their respective bands for LyC to be detected if present ($z_{\rm lim}$), and one places it below $z_{\rm lim}$. Two spectra (Figure \ref{fig:deimos}) are from the DEIMOS 10K survey and are assigned their lowest-quality redshift flag \citep[DM18 hereafter;][]{Hasinger2018}. The spectra are for zf\_14000 (DM18 ID: L677152, $z_{\rm DM18}=3.779$) and zf\_11423 (DM18 ID: L658665, $z_{\rm DM18}=3.655$). There is tentative feature that could add weight to the LRIS redshift for zf\_14000, but with low confidence. The spectrum for zf\_11423 shows two very tentative features, and conflicts with three photometric redshifts for this source, so we cannot be confident of its DM18 redshift. There is a 3D-HST G141 grism spectrum \citep{Brammer2012, Momcheva2016} for zf\_9775 (3D-HST ID: 9997, $z_{\rm 3DHST}=0.446$) in Figure \ref{fig:3dhst}. There are no strong features to provide evidence for either the 3D-HST or LRIS redshift. The 3D-HST value also conflicts with the closely matching independent redshifts from ZFOURGE, COSMOS, average CANDELS, and our confident LRIS redshift ($z\sim4.2\text{--}4.4$). We therefore deem the 3D-HST redshift unreliable. See Appendix \ref{sec:appspec} for more details.

%------------------
%Analysis
%------------------

\begin{figure*}
\begin{center}
  \includegraphics[height=.9\textheight, trim=110 225 60 195, clip]{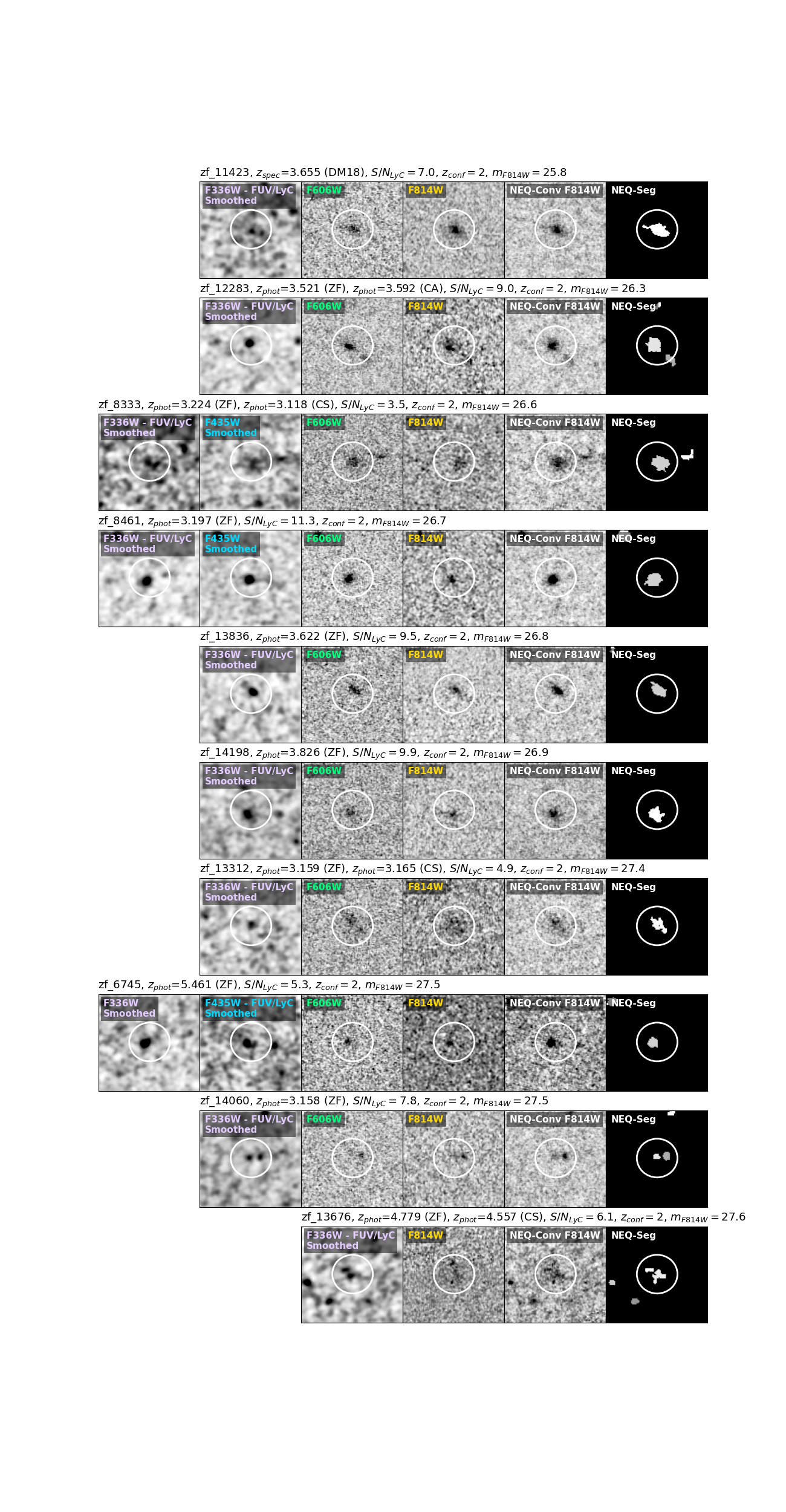}
  \caption{Photometric LCG candidates in Group B with the same labeling convention as Figure \ref{fig:top3}. These are photometrically detected candidates with unconfirmed redshifts. However, based on the available information from 30-band photometry and medium-band ZFOURGE infrared data measuring the Balmer break ($z_{\rm phot}$ ZF for most candidates), we are relatively confident in their redshifts ($z_{\rm conf}=2$). The candidates are ordered by m$_{\rm F814W}$ from brightest to faintest. See section \ref{sec:props} for more details on candidate categorizations.\small} 
  \label{fig:3sigB}
\end{center}
\end{figure*}

\begin{figure*}
\begin{center}
    \includegraphics[height=.9\textheight, trim=110 265 95 240, clip]{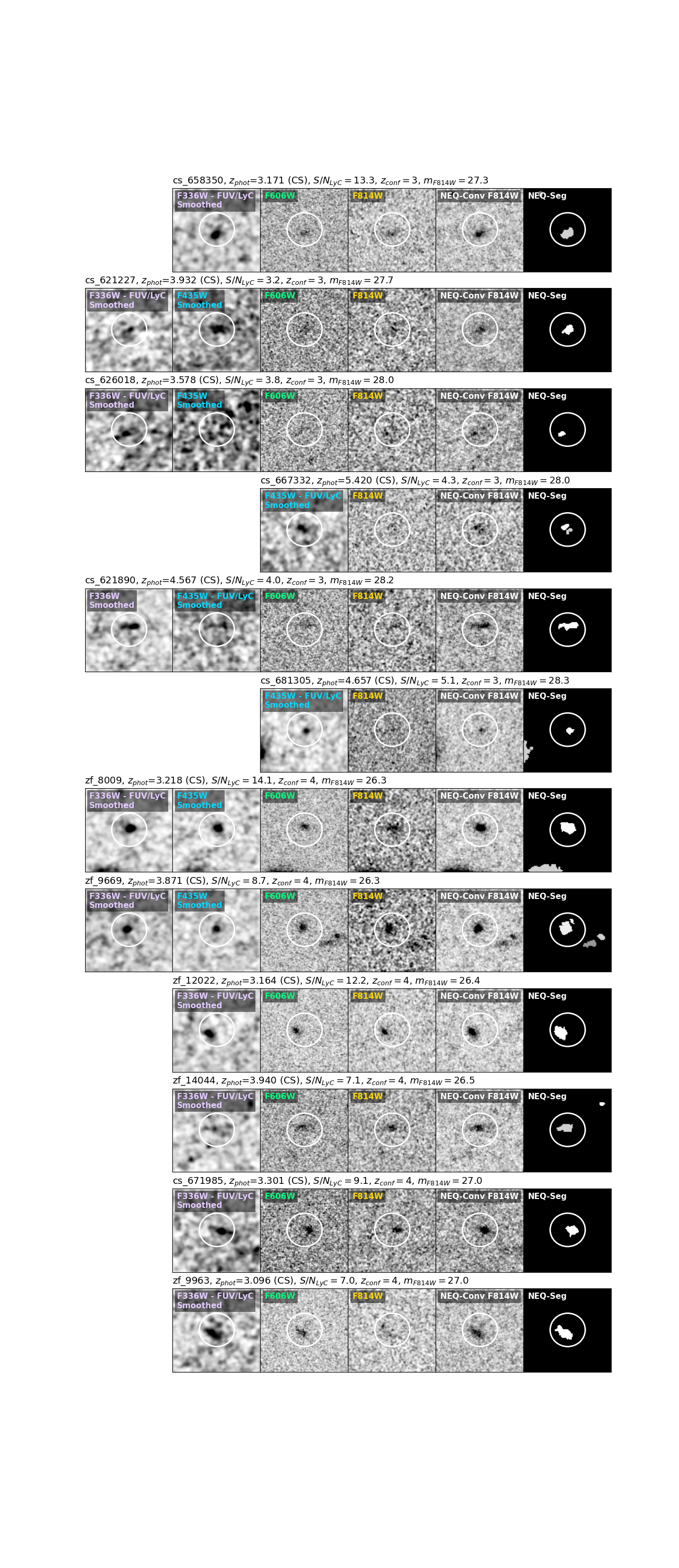}
    \caption{Photometric LCG candidates in Group C (continues on next page) with the same labeling convention as Figure \ref{fig:top3}. These are photometrically detected candidates for which we do not have reliable redshifts ($z_{\rm conf}=3$ or $4$). The candidates are first ordered by $z_{\rm conf}$ then m$_{\rm F814W}$. See section \ref{sec:props} for more details on candidate classifications.\small}
\end{center}
\end{figure*}

\begin{figure*}
\ContinuedFloat
\begin{center}
    \includegraphics[height=0.9\textheight, trim=115 220 95 200, clip]{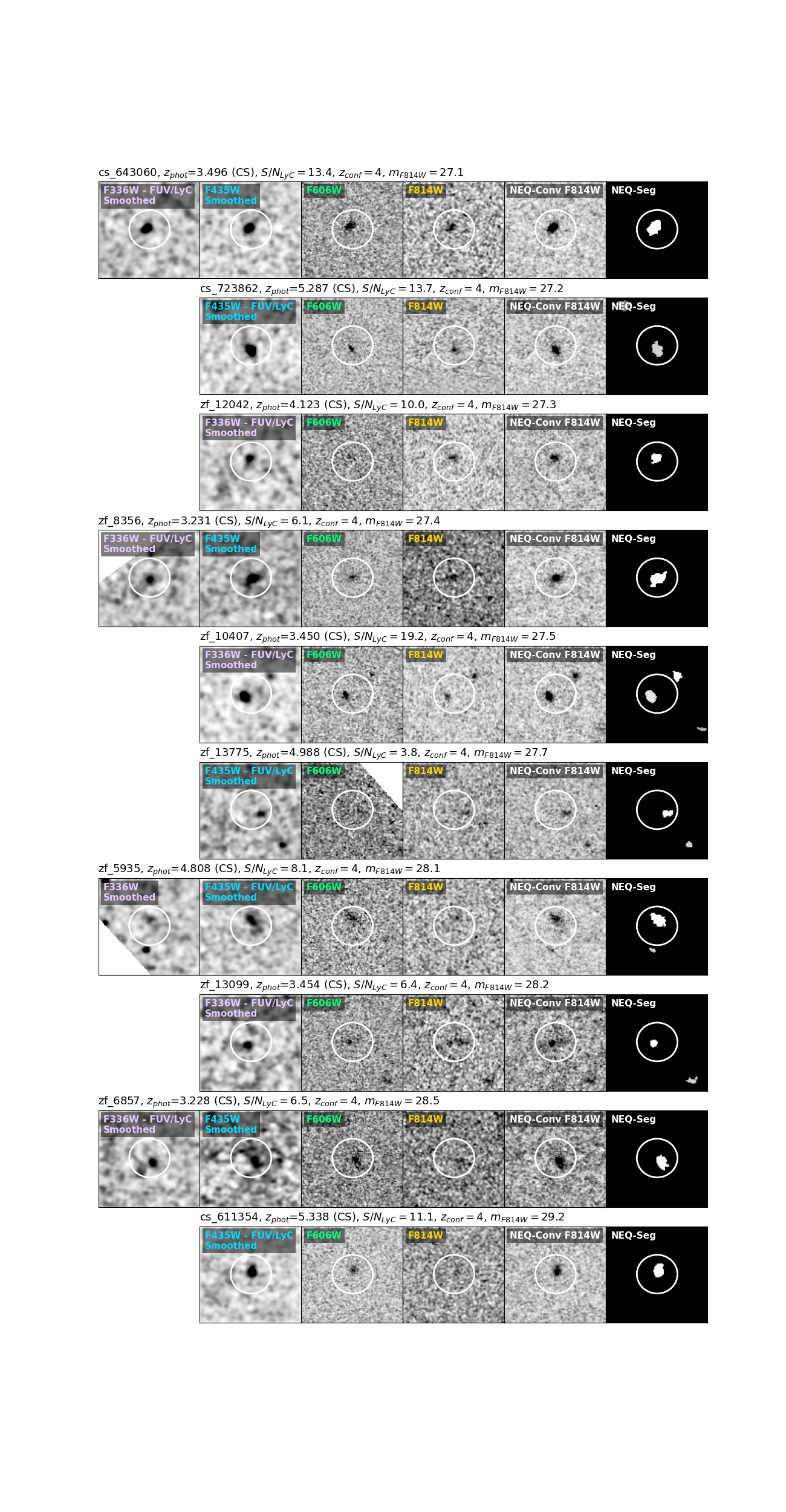}
    \caption{Continued.\small}
    \label{fig:3sigC}
\end{center}
\end{figure*}

\section{Analysis} \label{sec:an}

\subsection{Categorizing the LCG candidates} 
\label{sec:props}

We refine our LCG candidates and categorize them based on our confidence in their detection and redshift. We select only LCG candidates with $3\sigma$ flux detections in their LyC probing band (determined from the available redshifts). We then remove candidates with close neighboring sources (within $1.2^{\prime\prime}$ apertures), and those contaminated by bright saturated nearby stars. To help rule out the possibility that our candidates are active galactic nuclei we check the spectra (where available) for AGN FUV emission features. We then cross-match the candidate list against the public XMM-Newton X-ray \citep{Cappelluti2009} and Chandra COSMOS legacy \citep{Marchesi2016} catalogs within 0.5$^{\prime\prime}$ radii and find no matches. 

This refinement results in a final sample of 35 sources with 26 sources in the F336W pointings and 9 in F435W. With cross-matching and overlap, our final LCG candidate sample has redshifts from ZFOURGE (25), COSMOS (32), CANDELS (22), the spectroscopic catalog (3), and our new LRIS spectroscopic redshifts (3). This is within just our four untargeted HST fields ($\sim40$ arcmin$^2$).

The three LCG candidates for which we have Keck/LRIS spectra for, zf\_9775, zf\_11754, and zf\_14000, are the spectroscopic candidates that we categorize as Group A. We then categorize the remaining sample of 32 sources as ``photometric'' LCG candidates. We define a redshift confidence parameter ($z_{\rm conf}$) to categorize the LCG candidates, with $z_{\rm conf}=1$ being most confident and $z_{\rm conf}=4$ being least confident. 
\begin{itemize}[leftmargin=*]
\itemsep0em
    \item $\boldsymbol{z_{\rm conf}=1}$\textbf{ (most confident)}: Has at least one high-quality spectroscopic redshift and at least two redshifts total, both above the redshift for F336W or F435W to probe LyC flux ($z_{\rm lim}$). This only applies to the three Group A candidates with LRIS spectra.
    \item $\boldsymbol{z_{\rm conf}=2}$: Has more than two photometric redshifts $>z_{\rm lim}$ OR has one or more redshifts from a reliable source $>z_{\rm lim}$. We consider a reliable source a tentative public spectroscopic redshift that we have inspected (i.e., zf\_11423 see Section \ref{sec:specz}, and Figure \ref{fig:deimos}), photometric redshifts from ZFOURGE or the CANDELS average minus one standard deviation. This is Group B, which contains 10 LCG candidates.
    \item $\boldsymbol{z_{\rm conf}=3}$: Has one photometric redshift $>z_{\rm lim}$ from a less reliable source (i.e., COSMOS2015, that we found to deviate more from the other available redshifts) and no other redshifts (i.e., not enough information to rule the redshift reliable or not). We place these candidates in Group C.
    \item $\boldsymbol{z_{\rm conf}=4}$\textbf{ (least confident)}: Has one photometric redshift $>z_{\rm lim}$ from a less reliable source (COSMOS2015) and other conflicting redshifts that place the source at $<z_{\rm lim}$. We place these candidates in Group C along with $z_{\rm conf}=3$ sources, with a total of 22 LCG candidates.
\end{itemize}

In Figure \ref{fig:top3} we show 3$^{\prime\prime}$ postage stamps of the three Group A spectroscopic LCG candidates (with $z_{\rm conf}=1$), ordered by magnitude in F814W (m$_{\rm F814W}$) from brightest to faintest. In Figure \ref{fig:3sigB} we show the 10 Group B photometric LCG candidates ordered by m$_{\rm F814W}$ (with $z_{\rm conf}=2$; that includes the tentative spectroscopic redshift of zf\_11423, see Section \ref{sec:specz} and Figure \ref{fig:deimos}). In Figure \ref{fig:3sigC} we show our 22 Group C photometric LCG candidates (with $z_{\rm conf}=3$ and 4) also ordered by m$_{\rm F814W}$. 

For each of the LCG candidate figures we show the available HST bands for the targets (new F336W and F435W, CANDELS-F606W and COSMOS-F814W), the NEQ combined image, and the source-detection segmentation map of the NEQ image (NEQ-Seg) used for isophotal fluxes (right stamp). We apply minor smoothing to the blue bands by convolving the images with a 2D Gaussian kernel (1.5-pixel standard deviation) to highlight the often-faint flux. We indicate which of the blue bands contain clean LyC flux if their $>z_{\rm lim}$ redshifts are accurate. Above the row of stamps for each candidate, we show their IDs, any $>z_{\rm lim}$ redshifts, $S/N$ of their LyC flux, $z_{\rm conf}$, and m$_{\rm F814W}$ values. We over plot a $1.2^{\prime\prime}$ aperture for scale. See Figure \ref{fig:space} for the spatial distribution of the candidates split by Groups A (with IDs shown), B, and C. See Table \ref{tab:lcgs} for a summary of all the LCG candidate details and redshift information, and Table \ref{tab:lcgsphot} for their photometry.

\subsection{Sample numbers} 
\label{sec:stats}

We summarize the numbers of LCG candidates, both by LyC photometric band ($>3\sigma$ in F336W and F435W) and assigned category (Groups A, B, and C). We also indicate the candidate IDs that do not have F606W coverage which is relevant for the illustrative color-color plots in Section \ref{sec:col}. 
\begin{itemize}[leftmargin=*]
\itemsep0em
    \item LCG candidates in F336W: 26
        \begin{itemize}[leftmargin=*]
        \itemsep0em
            \item Group A: 2
            \item Group B: 9 (zf\_13676 not covered by F606W)
            \item Group C: 15
        \end{itemize}
     \item LCG candidates in F435W: 9
        \begin{itemize}[leftmargin=*]
        \itemsep0em
            \item Group A: 1 (zf\_11754 not covered by F606W)
            \item Group B: 1
            \item Group C: 7 (cs\_667332 \& cs\_681305 not covered by F606W)
        \end{itemize}
\end{itemize}

We find the following detection rates for our $>3\sigma$ LCG candidates relative to the \textit{Parent} and \textit{All-$z$} samples within their respective LyC-band footprints:
\begin{itemize}[leftmargin=*]
\itemsep0em
    \item F336W candidates/\textit{Parent}: $26/380 = 6.84\%$
    \item F435W candidates/\textit{Parent}: $9/242 = 3.72\%$
    \item F336W candidates/\textit{All-$z$}: $26/3065 = 0.85\%$
    \item F435W candidates/\textit{All-$z$}: $9/6430 = 0.14\%$
\end{itemize}

We find the following number densities for the $>3\sigma$ LCGs in different pointings:

\begin{figure*}
\begin{center}
    \includegraphics[width=0.48\textwidth, trim=15 10 10 10, clip]{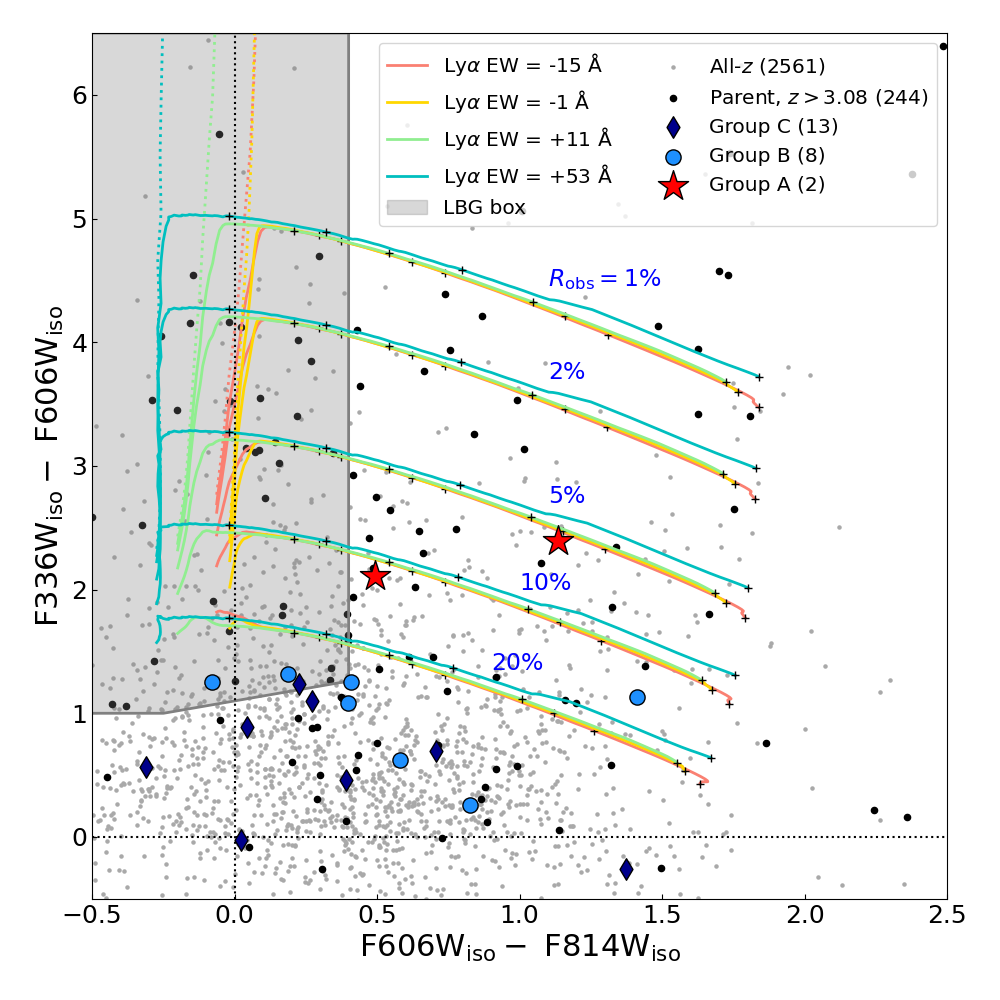}
    \includegraphics[width=0.48\textwidth, trim=15 10 10 10, clip]{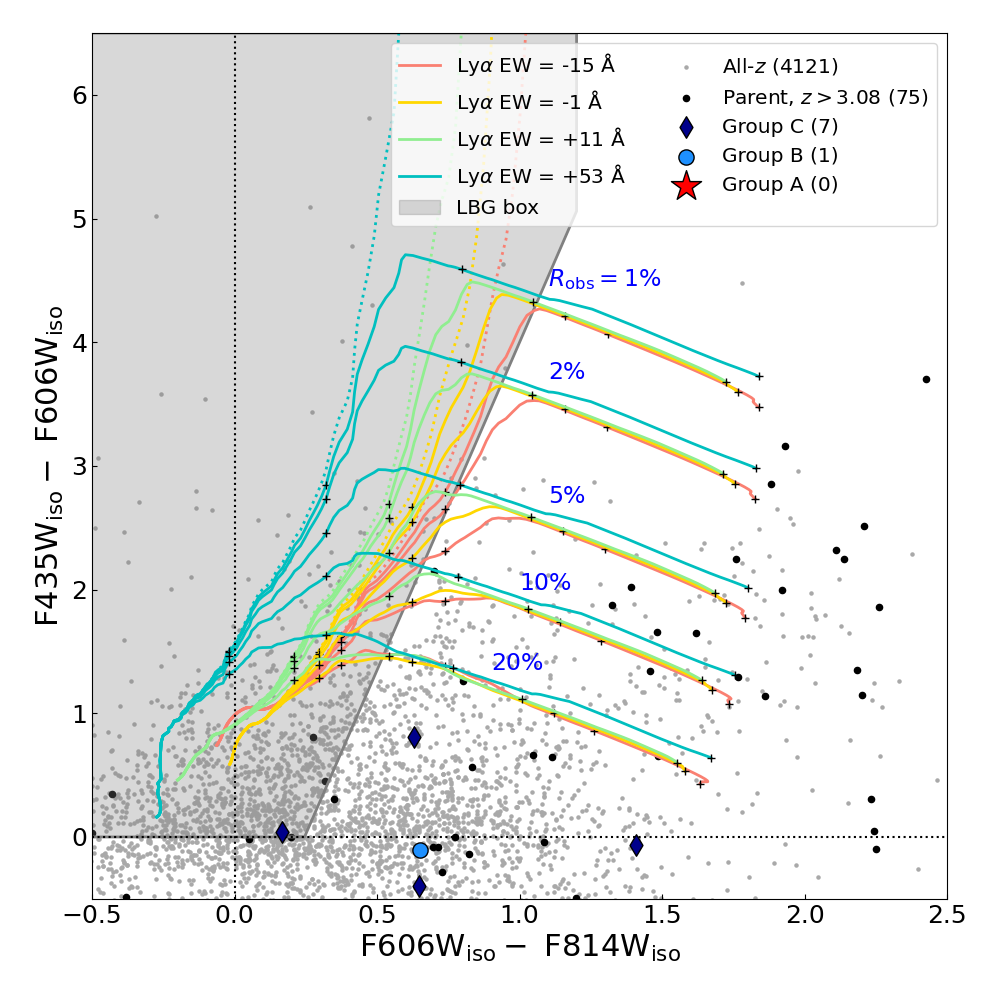}
    \includegraphics[width=0.48\textwidth, trim=15 10 10 10, clip]{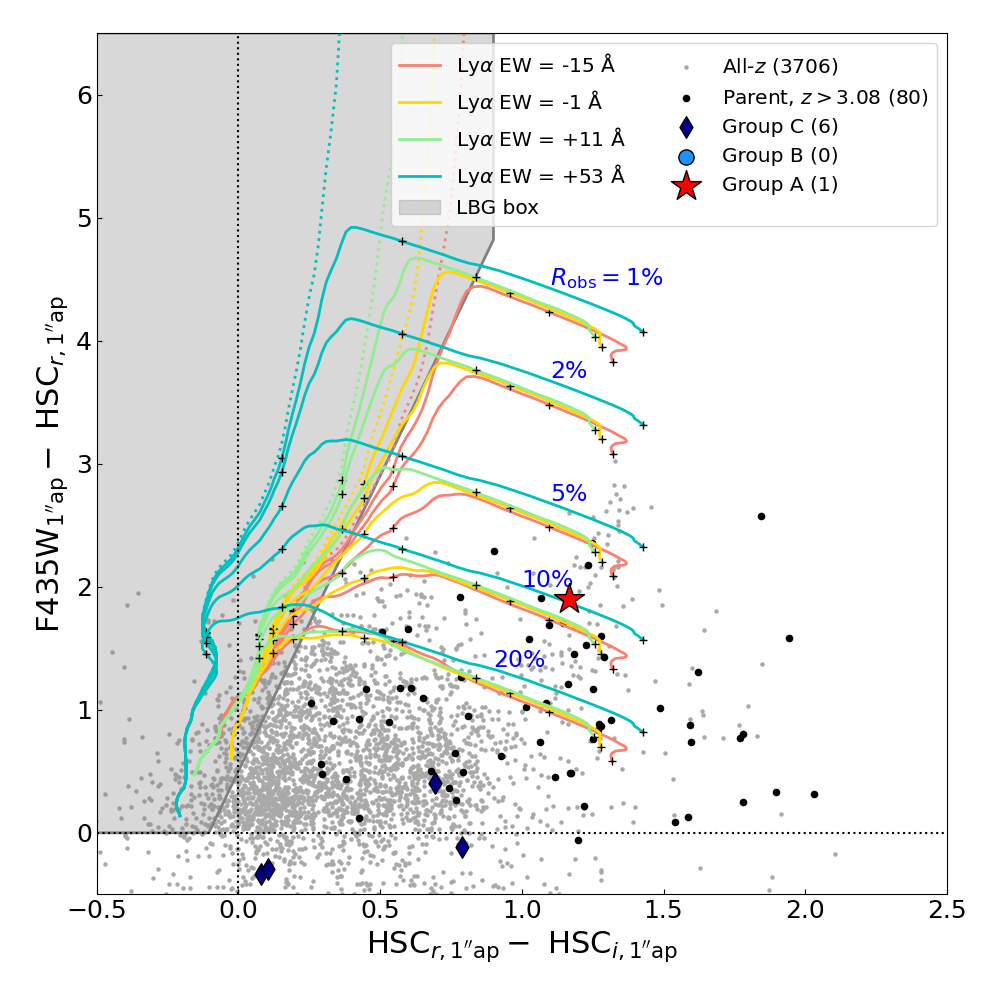}
    \caption{Color-color plots of the LCG candidates. The top row shows the isophotal fluxes for all sources that have the respective HST bands: F336W$-$F606W vs. F606W$-$F814W (top left) and F435W$-$F606W vs. F606W$-$F814W (top right). Not all the targets have F606W available (see Figure \ref{fig:space}), so we also use ground-based HSC bands (see Figure \ref{fig:lbg_bands} for a comparison and Section \ref{sec:hsc}). We show the Group A spectroscopic candidates (red stars): top left plot has zf\_9775 (upper point) and zf\_14000 (lower point), the bottom plot has zf\_11754. We also show the photometric candidates in Group B (blue circles), and Group C (dark blue diamonds). For reference, we show the \textit{Parent Sample} at a common redshift to the candidates (black points) and \textit{All-$z$ Sample} within the same HST footprints (gray points). Overlaid are evolutionary tracks of four LBG composites from \cite{Shapley2003} with increasing Ly$\alpha$ strength (red, yellow, green, turquoise) redshifted from $z=2.7$ to $5$. We show LBGs with no observed LyC flux (dotted curves) and increasing observed LyC to UV continuum flux ratio ($R_{\rm obs}=$1\%, 2\%, 5\%, 10\%, and 20\%) that peel the tracks off at different levels from the top respectively. The tracks are marked at increasing redshift (left to right) with `$+$'s at $z=3.5$, 4 and 5. Selection boxes for LBGs at $z=2\text{--}4$ galaxies (higher-redshift LBGs are off the plot) with these filter combinations are shown in gray. Most of the LCG candidates, including all the top Group A targets, sit outside of the illustrative LBG boxes. See Section \ref{sec:col}.\small}
    \label{fig:colcol}
\end{center}
\end{figure*}

\begin{itemize}[leftmargin=*]
\itemsep0em
    \item F336W candidates ($z>3.087$, two 15-orbit images, m$_{\rm AB}=29.77, 30.11$ at $5\sigma$ depth): $\sim1.8$ arcmin$^{-2}$,
    \item F435W candidates ($z>4.391$):
    \begin{itemize}[leftmargin=*]
        \itemsep0em
        \item F435W\_1a (5 orbits, m$_{\rm AB}=28.52$ at $5\sigma$):\\ $\sim0.1$~arcmin$^{-2}$,
        \item F435W\_1b (10 orbits, m$_{\rm AB}=29.11$ at $5\sigma$):\\ $\sim0.4$~arcmin$^{-2}$,
        \item F435W\_2 (15 orbits, m$_{\rm AB}=29.78$ at $5\sigma$):\\ $\sim0.4$~arcmin$^{-2}$.
    \end{itemize}
\end{itemize}

\subsection{Color-color plots} 
\label{sec:col}

Color selections are often used to identify high-redshift galaxies and LBGs \citep[e.g.][]{Steidel1996b}. We therefore plot our LCG candidates on color-color plots with the available bands to show their colors. However, these filter combinations are not the normal bands used for high-redshift LBG selection, nor are they optimal for it. Given that we do not select any galaxies based on colors, we use these purely for illustrative purposes. 

Figure \ref{fig:colcol} shows the LCG candidates split by category: Group A (red stars), Group B (blue circles), and Group C (dark blue diamonds). We show the \textit{Parent Sample} of all galaxies $>z_{lim}$ (black points) and the \textit{All-$z$ Sample} within our HST footprints (gray points). The evolutionary tracks for four \cite{Shapley2003} LBG composites with increasing Ly$\alpha$ strength (red, yellow, green, turquoise) redshifted from $z=2.7$ to $5$ are shown. We show LBGs with no LyC flux (dotted curves) and increasing LyC flux as observed from Earth (rest $<912~\text{\AA}$) as compared to observed rest $\sim1500~\text{\AA}$ flux \citep[i.e., $R_{\rm obs}=$1\%, 2\%, 5\%, 10\%, 20\%; see][]{Cooke2014} that peel the tracks off at different colors from the top. Note that these tracks are not at derived $f_{\rm esc}$ values but observed flux ratios (see Section \ref{sec:fesc} for more details). The tracks are marked at increasing redshift (left to right) with `$+$'s at $z=3.5$, 4 and 5. 

We show the color-color plots for the HST bands (Figure \ref{fig:colcol} top row) with colors calculated from our isophotal magnitudes. The isophotal magnitudes are extracted within the same footprints on each band for each respective source (see Figures \ref{fig:top3} to \ref{fig:3sigC}). The F336W$_{\rm iso}-$F606W$_{\rm iso}$ vs. F606W$_{\rm iso}-$F814W$_{\rm iso}$ plot (top left) shows all but one of our LCG candidates selected within the F336W band. The missing candidate is zf\_13676 from Group B (a lower-priority candidate) as it is not covered by F606W. The F435W$_{\rm iso}-$F606W$_{\rm iso}$ vs. F606W$_{\rm iso}-$F814W$_{\rm iso}$ plot (top right) is missing one of the top Group A candidates (zf\_11754) as it does not have F606W coverage. We therefore use our derived correction (see Section \ref{sec:hsc}, Equation \ref{eq:hsc}) to directly compare our HSC PSF- and aperture-matched F435W fluxes with ground-based HSC bands (bottom plot). The F435W$_{\rm 1^{\prime\prime}ap}-$HSC$_{r\_ \rm  1^{\prime\prime}ap}$ vs. HSC$_{r\_ \rm  1^{\prime\prime}ap}-$HSC$_{i\_ \rm  1^{\prime\prime}ap}$ shows zf\_11754 (red star). Some LCG candidates (mostly in Group C) are not shown on the axes ranges we present. This may indicate contamination of the lower confidence LCG candidates if they are that far away from the \textit{All-$z$} cloud (gray points).

LBG selection boxes for $z=2\text{--}4$ galaxies (higher-redshift LBGs are off the plot) with these filter combinations are shown in gray in Figure \ref{fig:colcol}. The boxes are comprehensive to span possible LBG regions. However, they are purely illustrative and should not be used for the clean selection of LBGs. We start with the boxes for a similar filter combination presented in \cite{Alavi2016} (F336W$-$F435W vs. F435W$-$F814W). The box is defined by the location of color-color evolution tracks of star-forming galaxies for these filters (not shown here, see \citealt{Alavi2016}). The star-forming galaxy tracks are model predictions from \cite{Bruzual2003} synthetic stellar populations with constant star formation, $0.2\ Z_{\odot}$, 100 Myr ages and color excesses E(B$-$V)=[0,0.1,0.2,0.3]. We apply the \cite{Madau1995} cosmic opacity prescription and dust extinction law from \cite{Calzetti2000}. We translate these model-based LBG boxes in color space to our filter combinations. We then expand them to encompass the \cite{Shapley2003} tracks (dotted lines shown here) as needed. Most of the LCG candidates, including all the top Group A targets, sit outside of the comprehensive illustrative LBG boxes.

\subsection{Escape fraction estimates} 
\label{sec:fesc}

In determining $f_{\rm esc}$, the escape fraction of ionizing UV flux relative to the non-ionizing continuum, many assumptions are required that have significant degeneracies. These include the SED of a galaxy, its star-formation history (SFH), metallicity, and assumed line-of-sight IGM distribution through which it is observed. There are often a variety of escape-fraction related parameters quoted in the literature which can further complicate the picture. 

The simplest observational parameter for measuring LyC emission is the flux ratio of the observed LyC to observed UV continuum \citep[$R_{\rm obs}(\lambda)$; see][]{Cooke2014}. We measure $R_{\rm obs}=4\%, 3\%$ and $9\%$ for zf\_9775, zf\_11754 and zf\_14000 respectively (see Table \ref{tab:fesc} for a summary). These values assume the LyC flux to be in F336W (for zf\_9775 and zf\_14000) or F435W (zf\_11754), and the UV continuum to be F814W for the redshifts ($z\sim3.7\text{--}4.4$) of our sources (see Figure \ref{fig:lbg_bands}). The $R_{\rm obs}$ values vary slightly from the color-color plot approximations (where they overlap with the tracks) given the different filter combinations and galaxy properties.

\begin{figure*} 
\includegraphics[width=0.7\textwidth, trim=80 10 60 50, clip]{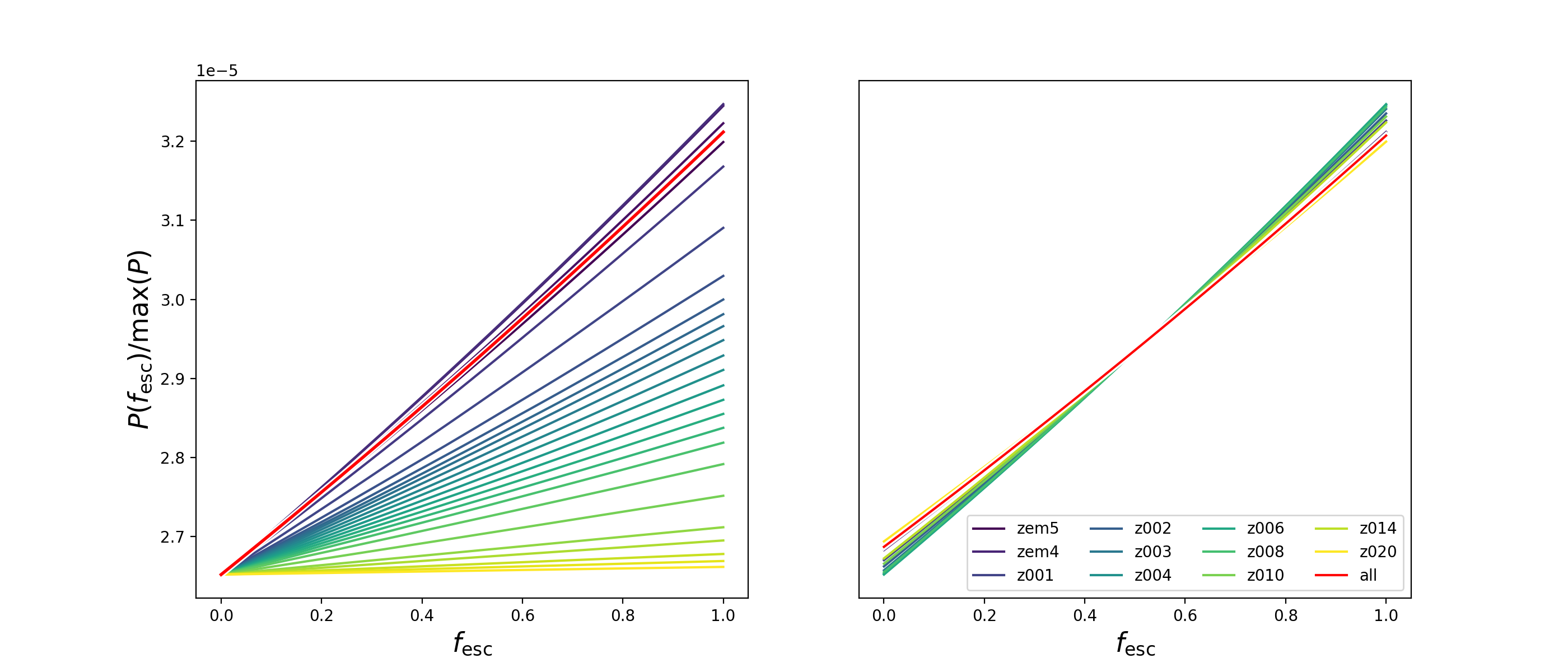}
\includegraphics[width=0.3\textwidth]{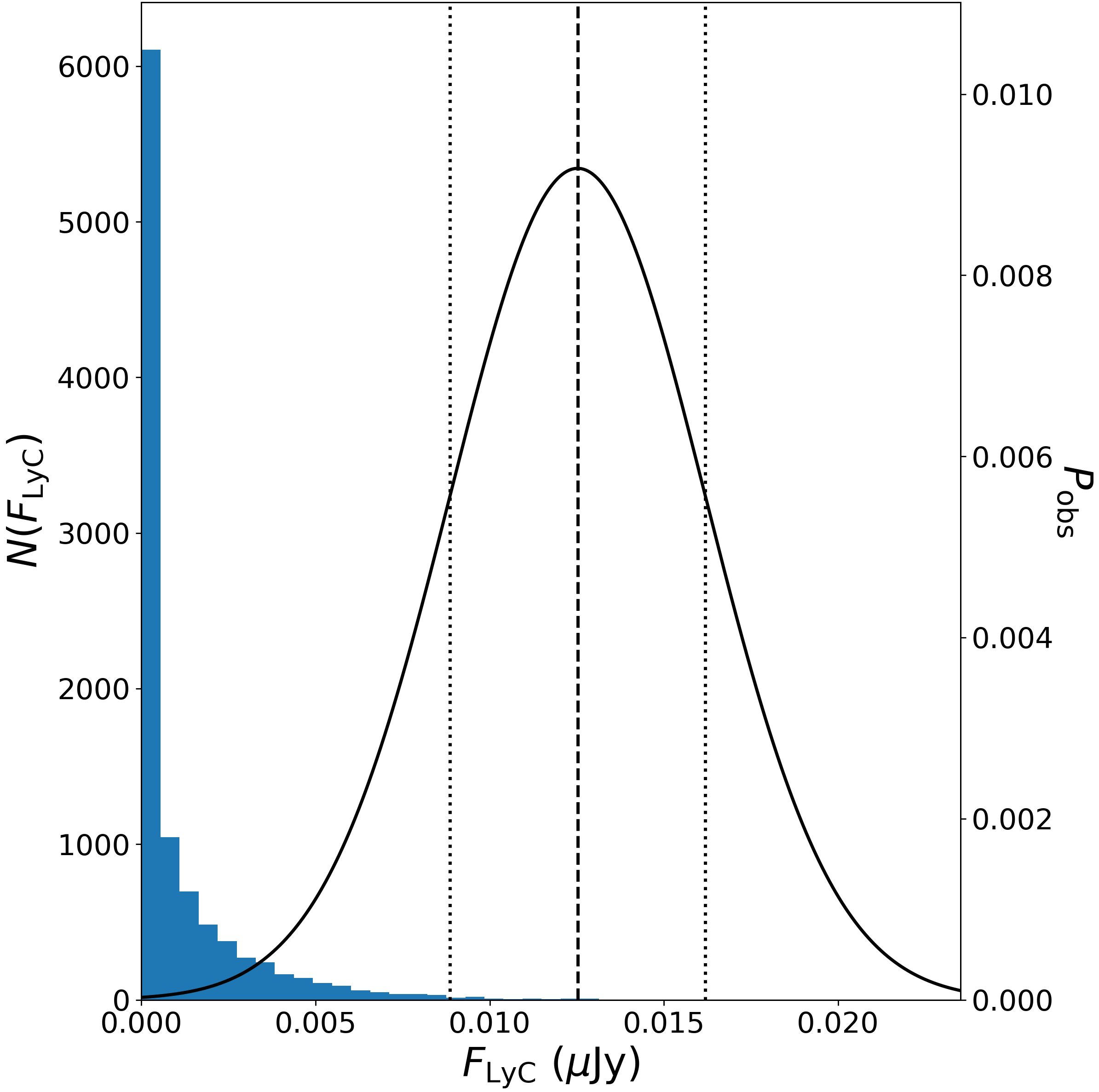}
\caption{Escape fraction ($f_{\rm esc}$) PDFs for zf\_11754. 10,000 IGM transmission curves were generated for $z=4.470$ (see Figure \ref{fig:fesctau}) and applied to every BPASS model along with 10,000 values of $f_{\rm esc}$ between 0 and 1. The resulting F435W flux was then calculated. The probability that each individual mock observation is consistent with the observed flux is determined by taking the value at the model flux for a Gaussian distribution (dashed line: observed flux; dotted lines: FWHM flux error). \textit{Left:} PDFs (scaled by the maximum probability) at each 25 ages for the closest matching BPASS model metallicity (zem5 $= Z_{*}=1e^{-5}$, $\log(\rm age) = 6.0$), and the weighted average at this fixed metallicity (red line). \textit{Middle:} Weighted average for each metallicity. The red line shows the weighted average considering all 275 metallicities and ages. \textit{Right:} Histogram of output F435W fluxes (blue) across all 10,000 sight lines and 10,000 $f_{\rm esc}$ values ($10^8$ trials). The observational data for zf\_11754 is represented by a Gaussian (black curve) with mean of its observed LyC flux in F435W ($\mu$, dashed line) and width of the flux error ($\sigma$, dotted lines). The panels suggest the most probable $f_{\rm esc}$ for zf\_11754 is $>100\%$ as there is no peak reached in the PDFs within the constrained 0--1 range. See Section \ref{sec:fesc} and Appendix \ref{sec:appfesc}. \small} 
\label{fig:fesclines}
\end{figure*}

The parameter required to constrain cosmic reionization is the absolute $f_{\rm esc}$, or $f_{\rm esc}^{\rm abs}$. This is the fraction of escaping LyC photons, produced by O and B stars in star-forming galaxies, that reach the IGM without being absorbed by the interstellar or circumgalactic medium \citep[CGM; e.g.,][]{Wyithe2007, Wise2009}. However, direct measurements of this value are not possible due to the severe attenuation of ionizing photons by the IGM. The quantity used to constrain $f_{\rm esc}^{\rm abs}$ from observations is the relative $f_{\rm esc}$ \citep[$f_{\rm esc}^{\rm rel}$;][]{Steidel2001},
\begin{equation} 
f_{\rm esc}^{\rm rel} = \frac{(F_{\rm LyC}/F_{\rm 1500})_{\rm obs}}{(L_{\rm LyC}/L_{\rm 1500})_{\rm int}} \rm exp(\tau^{\rm LyC}_{\rm IGM}).
\label{eq:fescrel}
\end{equation}
Here $(F_{\rm LyC}/F_{\rm 1500})_{\rm obs}$ is the observed rest-frame LyC to UV flux ratio, $(L_{\rm LyC}/L_{\rm 1500})_{\rm int}$ is the ratio of the intrinsic ionizing to non-ionizing luminosity density, and $\tau^{\rm LyC}_{\rm IGM}$ is the redshift-dependent IGM attenuation along the line of sight. Both $(L_{\rm LyC}/L_{1500})_{\rm int}$ and $\tau^{\rm LyC}_{\rm IGM}$ are highly model and assumption dependent. This $f_{\rm esc}^{\rm rel}$ parameter can be converted to an $f_{\rm esc}^{\rm abs}$ estimate \citep[proposed in][]{Inoue2005, Siana2007} using 
\begin{equation}
f_{\rm esc}^{\rm abs} = f_{\rm esc}^{\rm rel} \times 10^{-0.4(k_{\rm 1500}E(B-V))},
\label{eq:fescabs}
\end{equation}
where $k_{\lambda}$ is the reddening law \citep{Calzetti2000} and $E(B-V)$ is the total dust attenuation \citep[see e.g.,][for a summary]{Mestric2020}.

We determine $f_{\rm esc}^{\rm abs}$ for the three Group A candidates using the equations above as is traditionally done in the literature. We first determine average IGM transmission of the LyC probing filter using the commonly adopted models from \cite{Inoue2014} ($T^{\rm LyC}_{\rm IGM,I14}$). These models consider only the IGM contributions along the line of sight and do not include attenuation by the CGM. We find average $T^{\rm LyC}_{\rm IGM,I14}= 0.0118$, 0.0261, and 0.0429 for zf\_9775, zf\_11754, and zf\_14000 respectively. The intrinsic ratio between the luminosity in the LyC band over the non-ionizing band is measured for the best fitting Binary Population and Spectral Synthesis \citep[BPASSv2.1;][]{Eldridge2017} model for each candidate. We determine $f_{\rm esc}^{\rm rel}$ using Equation \ref{eq:fescrel} and since our best fitting $E(B-V) = 0$, $f_{\rm esc}^{\rm abs} = f_{\rm esc}^{\rm rel}$, so no further correction for dust is applied. With the attenuation values from \cite{Inoue2014}, we find escape fractions of $f_{\rm esc,I14}^{\rm abs}= 8.75, 3.63$, and 5.87 for zf\_9775, zf\_11754, and zf\_14000 respectively.

We then calculate a new average IGM transmission of the LyC probing filter ($T^{\rm LyC}_{\rm IGM}$) by generating 10,000 sight lines that include contributions from both the IGM and CGM (see Appendix \ref{sec:appfesc} for more details). We find average $T^{\rm LyC}_{\rm IGM}= 0.0015$, 0.0141, and 0.0125 for zf\_9775, zf\_11754, and zf\_14000 respectively. Using the same calculation as before with these new attenuation estimates, we find $f_{\rm esc}^{\rm abs}= 111.3, 9.8$, and 20.23 for zf\_9775, zf\_11754, and zf\_14000 respectively. Note that these are not percentages but fractions so are well over 100\% (i.e., 980\% for zf\_11754) and are not physical given these assumptions.

\begin{deluxetable*}{|c|c|c|c|c|c|c|c|c|c|c|}
\tabletypesize{\footnotesize}
\tablewidth{0pt}
\tablecaption{LyC and UV continuum fluxes and escape fraction parameters for the Group A candidates.}
\label{tab:fesc}
\tablehead{ID & LyC$^{1}$ & UV Cont.$^{1}$ & $F_{\rm LyC}$ $^{2}$ & $F_{\rm 1500}$ $^{2}$ & $R_{\rm obs}$ $^{3}$ & $\langle T^{\rm LyC}_{\rm IGM,I14} \rangle$ $^{4}$ & $f_{\rm esc,I14}^{\rm abs}$ $^5$ & $\langle T^{\rm LyC}_{\rm IGM} \rangle$ $^{6}$ & $f_{\rm esc}^{\rm abs}$ $^7$ & $f_{\rm esc}^{\rm PDF}$ $^8$ \\ & Filter & Filter & ($\mu$Jy) & ($\mu$Jy) & $(F_{\rm obs}^{\rm LyC}/F_{\rm obs}^{\rm UV})$ & & & (this work) & &}
\startdata
zf\_9775 & F336W & F814W & $0.031\pm0.003$ & $0.79\pm0.04$ & 0.04 & 0.0118  & 8.75 & 0.0015 & 111.3 & 1$^{9}$ \\
zf\_11754 & F435W & F814W & $0.013\pm0.004$ & $0.39\pm0.04$ & 0.03 & 0.0261 & 3.63 & 0.0141 & 9.8 & 1$^{9}$ \\
zf\_14000 & F336W & F814W & $0.035\pm0.002$ & $0.39\pm0.02$ & 0.09 & 0.0429 & 5.87 & 0.0125 & 20.23 & 1$^{9}$ \\
\enddata
\tablenotetext{}{1. See filters in Figure \ref{fig:lbg_bands}. 2. Fluxes from the corresponding LyC and UV continuum filters. 3. Observed LyC to UV continuum flux ratio \citep[see][]{Cooke2014}. 4. Average IGM attenuation for the LyC band determined using models from \cite{Inoue2014} that do not include CGM. 5. Absolute escape fraction that uses \cite{Inoue2014} $\langle T^{\rm LyC}_{\rm IGM,I14} \rangle$ (from Equations \ref{eq:fescrel} \& \ref{eq:fescabs}), not percentages. 6. Average IGM transmission for the LyC band of our 10,000 generated sight lines. 7. Absolute escape fraction using our $\langle T^{\rm LyC}_{\rm IGM} \rangle$ (from Equations \ref{eq:fescrel} \& \ref{eq:fescabs}), not percentages. 8. Probabilistic escape fraction (see Section \ref{sec:fesc}, Appendix \ref{sec:appfesc}). 9. Value constrained 0 to 1, most likely $f_{\rm esc}^{\rm PDF} >1$ (see Section \ref{sec:fesc}).}
\end{deluxetable*}

To further investigate $f_{\rm esc}$ of the Group A candidates, we also adopt a statistical approach to generate realistic $f_{\rm esc}$ probability distribution functions (PDFs), $f_{\rm esc}^{\rm PDF}$, that account for some of the sources of uncertainty. We use the probabilistic method described in \cite{Bassett2022} that we outline in more detail in Appendix \ref{sec:appfesc}. Probabilistic approaches to estimating $f_{\rm esc}$ have also been adopted in previous studies \citep{Vasei2016, Vanzella2016}. We show the $f_{\rm esc}$ PDFs for one LCG candidate, zf\_11754, in Figure \ref{fig:fesclines} (left panel). The PDFs are scaled by the maximum probability (max(P)) assuming the model flux is equal to the observed flux. The PDFs are shown at each of the 25 BPASS model ages and for the best fitting model (zem5~=~$Z_{*}=1e^{-5}$, $\log(\rm age) = 6.0$) and the weighted average at this fixed metallicity (red line). We also show the probability-weighted average $f_{\rm esc}$ PDF at each metallicity (Figure \ref{fig:fesclines}, middle panel). The red line shows the weighted average for all 275 BPASS metallicities and ages. See Figure \ref{fig:fescmetal} for grids of weights for each BPASS model for zf\_11754.

The $f_{\rm esc}$ estimates are constrained from 0 to 1, and the maximum is at 1. These plots show that the most consistent $f_{\rm esc}^{\rm PDF}$ for zf\_11754 is 100\%. The right panel of Figure \ref{fig:fesclines} shows a histogram of fluxes (blue) across all 10,000 sight lines for the redshift of zf\_11754 ($z=4.470$) and 10,000 $f_{\rm esc}$ values between 0 and 1 ($10^8$ trials). The observational data for zf\_11754 are represented by a Gaussian (black curve) with mean of its observed LyC flux in F435W ($\mu$) and width of the flux error ($\sigma$). Where the blue histogram tails off on the right is $f_{\rm esc}^{\rm PDF} = 100\%$ and high transmission. The peak for zf\_11754, i.e., its highest probability $f_{\rm esc}^{\rm PDF}$, sits even further right of that tapered tail. The right panel therefore shows that the actual probability of this estimate being realistic is low and that it is likely $>100\%$. See Table \ref{tab:fesc} for summary of all $f_{\rm esc}$ values and related parameters.

From the full range of $f_{\rm esc}$ and IGM transmission combinations tested, $\sim1/10,000$ sight lines would produce a high enough transmission to obtain the observed LyC flux provided $f_{\rm esc}=1$. Given the very low likelihood of this happening, we deem that the more likely explanation is that the $f_{\rm esc}>1$. If adopting a similar approach to \cite{Inoue2014} where the CGM is not included in the sight lines, the likelihood of this high-transmission and high-escape combination only slightly increases and the models become less physical. We are unable to provide absolute probabilities of these values being possible since they would depend on the intrinsic PDF of $f_{\rm esc}$ for a population of star-forming galaxies at the redshifts of the Group A candidates ($z=3.7\text{--}4.4$), which is unknown. All of the probabilities we present are relative, so we therefore quote only the $f_{\rm esc}$ value that has the highest relative probability within the framework of our analysis.

Quantifying the errors is not possible as the PDF analysis indicates the most likely true value of $f_{\rm esc}^{\rm PDF}$ is $>100\%$, and therefore not constrained. The main limitation here is that even in the 99$^{\rm th}$ percentile, IGM transmission is $\sim0.2$ in the F435W band at the redshift of zf\_11754 (see Appendix \ref{sec:appfesc}, Figure \ref{fig:fesctau}). Even for the most transparent sight lines with the largest intrinsic LyC fluxes and $f_{\rm esc} = 1$ as inputs, the resulting flux of the model is $>3$ times the measured uncertainty below the measured flux (i.e., they are too faint). This is also the case for the other two spectroscopic candidates zf\_9775 and zf\_14000, so we only present the plots for zf\_11754 to illustrate the method. An additional caveat for zf\_14000 is that from the full array of SEDs, there are no models that fit the data well, whereas the best-fit models to zf\_11754 and zf\_9775 are realistic. We therefore have additional doubts about the validity of the $f_{\rm esc}^{\rm PDF}$ measurement, and therefore LyC detection, of zf\_14000.

%------------------
%DISC & CONC
%------------------
\section{Discussion} \label{sec:disc}

\subsection{Escape fraction comparison} \label{sec:discfesc}

The most likely value of $f_{\rm esc}^{\rm abs}$ and $f_{\rm esc}^{\rm PDF}$ for all three Group A spectroscopic candidates is $>100\%$, and therefore not physical with these models and assumptions. It is possible that some assumptions that go into the $f_{\rm esc}^{\rm PDF}$ calculations may not be appropriate for our candidates. Many of the assumptions are those commonly adopted by others for the more traditional $f_{\rm esc}$ estimates (e.g., $f_{\rm esc}^{\rm abs}$ calculated with Equation \ref{eq:fescabs}). The models \citep[BPASSv2.1;][]{Eldridge2017}, metallicity ranges \citep{Steidel2018, Fletcher2019, Bassett2021a}, and IGM sight line calculations \citep{Inoue2014, Steidel2018, Bassett2021a} are all based on established methods but do not include strong emission lines in the SEDs, only exponentially declining SFHs. Although as the best fitting model for all three Group A candidates is ``zero-age" for BPASS (equivalent to a single, instantaneous burst with an age of $10^6$ years), the choice of SFH will not affect the fact that $f_{\rm esc}^{\rm PDF}$ is unconstrained at $>100\%$ (see Appendix \ref{sec:appfesc}),

The statistical $f_{\rm esc}^{\rm PDF}$ estimate approach presented here has been tested on a spectroscopically confirmed $z = 3.64$
galaxy \citep[VUDS 511227001;][]{LeFevre2015} in \cite{Bassett2022}. CLAUDS \citep{Sawicki2019} $U$-band shows a LyC detection for this galaxy. The resulting PDF has a peak within the 0 to 1 range, with $f_{\rm esc}^{\rm PDF}=0.45^{+0.39}_{-0.28}$. If we derive many more possible sight lines ($10^7$ rather than $10^4$), it is possible that at least one would be transparent enough to match the observed flux for our candidates. However, all three would be unlikely to be within our four untargeted HST pointings. 

$f_{\rm esc}$ is a valuable parameter that is highly unconstrained with estimates in the literature ranging from $0\text{ to}>100\%$. The clean sample of 13 KLCS LyC emitters \citep{Steidel2018, Pahl2021} have $f_{\rm esc}^{\rm abs}=0.21\pm0.03$, with a full sample estimate (including 107 non-LyC detected $z\sim3$ galaxies) of $f_{\rm esc}^{\rm abs}=0.06\pm0.01$. In \cite{Mestric2020}, they present two promising LCG candidates with spectroscopic redshifts, high-resolution optical HST imaging (to identify contaminants), and CLAUDS $U$-band LyC detections. \cite{Mestric2020} estimate $f_{\rm esc}^{\rm abs}$ for these two candidates in the range $\sim5\% \text{--}73\%$ and $\sim30\%\text{--}93\%$. The spectroscopically confirmed LyC emitter, \textit{Ion2} at $z=3.2$ \citep{Vanzella2015, deBarros2016, Vanzella2016}, has an $f_{\rm esc}^{\rm abs}$ range of $\sim50\%\text{--}100\%$ depending on IGM attenuation. For the confirmed LyC emitter Q1549-C25 at $z=3.15$, \cite{Shapley2016} find $f_{\rm esc}^{\rm abs} \geq 0.51$ at 95\% confidence, but that $f_{\rm esc}^{\rm abs} \leq 1$ is at a 45\% confidence. \textit{Ion3} at $z=4.0$ has a quoted $f_{\rm esc}^{\rm rel}\sim60\%$ \citep{Vanzella2018}. An additional five $z\sim3\text{--}3.6$ galaxies from LACES have average LyC $f_{\rm esc}^{\rm abs}=0.35\pm0.14$ \citep{Nakajima2020}. While a lensed $z=2.5$ dwarf galaxy was found to have $f_{\rm esc}^{\rm abs} \simeq28\text{--}57\%$ from imaging \citep{Bian2017}. The lensed ``Sunburst Arc'' at $z=2.37$ \citep{Rivera-Thorsen2017} has line-of-sight $f_{\rm esc}^{\rm abs} = 32^{+2}_{-4}\%$, although the global escape fraction is expected to be much lower \citep{Rivera-Thorsen2019}.

This large uncertainty can exist even for individual sources as measured in simulations \citep[e.g.,][]{Paardekooper2015, Trebitsch2017, Rosdahl2018}. The true form of $f_{\rm esc}^{\rm PDF}$ could also vary by several factors, including galaxy mass, which further complicates its accurate estimation \citep[e.g.,][]{Finkelstein2019, Naidu2020, Bassett2021a}. The broad range of $f_{\rm esc}$ estimates in the literature, and even for each individual source, makes it hard to infer meaningful interpretations for the ionizing flux contributions from high-redshift galaxies. There only exists a relatively small sample of individually detected LyC candidates and they make up a diverse population of chance detections. 

A key conclusion from \cite{Mestric2020} was that only the cleanest lines of sight could work for observing LCGs. Therefore, these detections tell us as much about the random line of sight of the galaxies as it does about the galaxies themselves. It is likely that a population of high-redshift ($z\sim3\text{--}5$) galaxies may hold some of the most valuable clues about the sources that reionized the universe due to their proximity in time, and likely their physical properties. However, given the severe attenuation of LyC photons by the intervening IGM, this population remains incredibly elusive. 

Only in large statistical numbers can we average over possible IGM sight lines. Even then, many of those that we observe in the LyC may exist along the cleanest lines of sight. Non-detections of LyC may be the best path forward for constraining the global ionizing photon budget. The KLCS clean non-detection ($<3\sigma$) sample of 107 galaxies has a $f_{\rm esc}^{\rm abs}=0.05\pm0.01$ \citep{Steidel2018, Pahl2021}.  \cite{Bian2020} stack 54 Ly$\alpha$ emitters at $z\simeq3.1$ and find no visible LyC but 3$\sigma$ upper limits of $f_{\rm esc}^{\rm abs}<14\text{--}32\%$. LBG stacks from \cite{Ji2020} give upper limits of $f_{\rm esc}^{\rm abs}\lesssim0.63\%$. The spectroscopic assessment in \cite{Cooke2014} gives $f_{\rm esc}^{\rm abs}$ upper limits of 0.086 for $z\sim3$ galaxies. Using a simplified stacking method of $f_{\rm esc}^{\rm abs}$ for galaxies without AGN at $z\sim2.3\text{--}4.3$, \cite{Smith2018, Smith2020} find limits of 0.08, 0.06, 0.57 for F225W, F275W and F336W respectively from their stacks. \cite{Mestric2021} analyze quoted upper limits in the literature and find global $f_{\rm esc}^{\rm abs}$ values of 0.086 at $2\lesssim z\leq3$, 0.105 at $3\leq z\leq4$ and 0.084 at $2\lesssim z<6$.

\subsection{Do the LCG candidates have line-of-sight low-redshift interlopers?} \label{sec:disclowz}

Even with deep HST LyC imaging and secure LRIS redshifts for zf\_9775 and zf\_11754, the LCGs presented in this paper remain candidates. This is in part due to the quality of the spectra given consistently poor weather conditions for our observations (see Table \ref{tab:spec}) and slit positioning on our targets. Nevertheless, from the spectra and imaging alone, these appear to be promising candidates. However, the $f_{\rm esc}^{\rm abs}$ values and statistical $f_{\rm esc}^{\rm PDF}$ estimates (that explore a broader range of possible assumptions than are usually adopted), we find that the most likely $f_{\rm esc}$ is $>100\%$ and therefore not physical given these models and assumptions. 

The candidates shown on the color-color plots (Figure \ref{fig:colcol}) mostly sit outside the comprehensive LBG region. This was hypothesized by \cite{Cooke2014} and found with a small number of LCG candidates by \cite{Mestric2020}. The Group A candidates sit on the LBG tracks with observed LyC flux artificially added ($R_{\rm obs}\sim5\%$ for zf\_9775, $R_{\rm obs}\sim10\%$ for zf\_11754 and zf\_14000, see Table \ref{tab:fesc} for calculated values). This could imply that these LCGs are similar to LBGs that exhibit LyC flux \cite[as shown in KLCS;][]{Steidel2018, Pahl2021} or they have flux contamination.

Are the Group A spectroscopic LCG candidates contaminated by low-redshift interlopers? Spatially resolving the LyC in high-resolution imaging is one of the best ways to identify low-redshift contaminants \citep[e.g.,][]{Vanzella2012}. We find that the chance of contamination is higher than the LCG candidate detection rates of sources in the field. We derive $1.3\%$ to $5.0\%$ probabilities of $3\sigma$ overlapping sources in our deep HST F336W and F435W images probing the LyC (see Section \ref{sec:cont}, Table \ref{tab:psf}). The redshift-independent contamination rates are higher than the detection rates relative to the \textit{All-$z$ Sample} from the combined redshift catalog ($0.85\%$ in F336W, $0.14\%$ in F435W). Therefore, it is possible that our candidates are high-redshift sources contaminated by low-redshift interlopers and our selection method is biased towards finding these chance alignments.

Our blue HST images ($\sim29\text{--}30$ mag depths at 5$\sigma$) are deeper than other searches for LyC emitters. Comparable surveys from \cite{Siana2015} reach $\sim29.27$ mag at 5$\sigma$ and \cite{Smith2018, Smith2020} reach $\sim29.11$ mag at 5$\sigma$ in F336W \citep[with data from][]{Oesch2018}. Deeper bluer filter images could help to resolve the question of possible contamination. However, with deeper images, picking up faint foreground sources is more likely \citep[e.g.,][]{Alavi2014}. At the high redshift of our candidates ($z\sim3.7\text{--}4.4$) the sight line is long and therefore the chance of coincident sources is higher, compared with $z\sim1$ \citep[e.g.,][]{Siana2007, Siana2010}. 

Higher signal-to-noise spectra with maximal spatial separation of components along the slit of zf\_9775 might help us identify the true nature of the candidate. If detected, emission lines in the LyC part of the spectrum could confirm this double source as a low-redshift interloper rather than a physical merger. The LyC emission for zf\_11754 is also slightly offset from the optical emission and could also be contaminated. zf\_14000 is very centered but this is the Group A candidate with the largest uncertainties with respect to its nature, redshift, and $f_{\rm esc}$ estimates. However, it may be that LyC flux does not trace non-ionizing flux, for example mergers can disrupt gas in galaxies, inhomogeneously exposing star-forming regions.

At higher redshifts, the IGM opacity increases significantly. This further reduces the chance of identifying strong LyC detection, particularly at the redshifts of our Group A spectroscopic candidates ($z\sim3.7\text{--}4.4$). For zf\_9775 and zf\_14000, the band cleanly sampling their LyC flux is F336W (zf\_9775 also has F435W which contains some forest lines), which would be around $\sim560\text{--}695~\text{\AA}$. The expected LyC flux is low in F336W, e.g., intrinsic: $F_{\rm int}=0.034$ $\mu$Jy, attenuated: $F_{\rm att}=5.2e^{-5}\pm 3.2e^{-4}$ $\mu$Jy, $\approx34.6\pm6.6$ mag. Although, \textit{Ion3} does show faint LyC down to $750~\text{\AA}$ \citep{Vanzella2018}. The results presented in this work are a cautionary tale that even with a blue spectroscopic redshift, detection in a ``LyC'' probing band, and HST imaging to remove contamination, the true nature of the sources may still not be clear.

\subsection{Future prospects \& need for spectroscopic follow-up} \label{sec:discspec}

The Group A spectroscopic candidates are the three brightest galaxies of all our LCG candidates, with isophotal m$_{\rm F814W}<25$ mag. Given the poor to moderate observing conditions (with poorer-than-optimal seeing), obtaining spectra for galaxies fainter than m$_{\rm F814W}\sim25$ proved challenging. Of the three top candidates, additional improved spectra are required to confirm them as genuine LCGs. A slit orientation to maximally spatially resolve the two components of zf\_9775 is needed. A slit for zf\_11754 placed to avoid its bright neighboring sources is also required. We have already attempted to target these three Group A candidates with these optimized slit positions (2020B\_N168, PI Prichard, weathered out). We were recently re-awarded this time for January 2022 (2021B\_N87, PI Prichard).

The remaining Group B and Group C photometric candidates are relatively faint with isophotal optical magnitudes ($\gtrsim26$ mag, down to 28.3 mag). Note that the isophotal magnitudes are a conservative clean flux estimate (best for accurate colors) that may not include all the flux of a source. In good conditions, with seeing $<1^{\prime\prime}$, we expect to get redshifts in 4--6 hours down to m$_{\rm F814W}\sim26$. If a candidate has a strong line, assumed to be Ly$\alpha$ in emission, a redshift can be determined for a source fainter than m$_{\rm F814W}\sim26$.

High $S/N$ spectroscopy is essential for identifying genuine LCGs. In a recent complimentary LCG study with MOSFIRE, \cite{Bassett2022} found that the ZFOURGE photometric redshifts for all seven LCG candidates were overestimated by $\delta z\sim 0.2$, most likely due to template mismatch. These candidates were therefore ruled out as LCGs as their true spectroscopic redshift placed them at a redshift such that the $U$-band was not cleanly sampling the LyC. Although the ZFOURGE stated accuracy is $\sim2\%$ errors \citep{Straatman2016}, \cite{Bassett2022} concluded that the search for LCGs is biased towards finding overestimated redshifts (with $94\%$ confidence). The results of this paper further emphasize that HST data are required for removing low-redshift interlopers, and demonstrate the requirement of blue spectra to aid with the removal of contaminants.

Redder wavelengths can be used to confirm redshifts, but may miss blue features to aid with contaminant removal. We have had supporting programs at redder wavelengths using Keck/MOSFIRE (2018B\_W151, PI Bassett) that focused on the [OIII]/[OII] relation with LyC \citep{Bassett2019, Bassett2022}. The combination of MOS observations spanning the bluest wavelengths does provide the best opportunity for getting redshifts and removing low redshift contaminants efficiently, along with the opportunity to spectroscopically detect the LyC in optimal observing conditions if bright enough. Keck/LRIS is not the only blue MOS option we have tried, we were awarded telescope time with the Large Binocular Telescope/MODS (LBT) that was also weathered out (see Table \ref{tab:spec} for a summary). Integral field units (IFUs) could be valuable for exploring spectroscopic properties of individual LyC emission regions and for removing contaminants. However, this requires high pixel resolution ($<1^{\prime\prime}$) for these small sources and is less efficient for building up statistical samples.

Additional information in the blue part of the spectrum can help reveal the nature of the detection and if it is contaminated. This expensive spectroscopic campaign will be handled with superior efficiency for galaxies such as our candidates with the new class of blue MOS instruments on 30 m telescopes. Additionally, the proposed Keck Wide-Field Imager \citep[KWFI;][]{Gillingham2020} would provide photometry down to $\sim29$ mag in $u$ in just $\sim5$ hours and $\sim3$ hours in $g$ over an area $\sim150\times$ that of HST to rapidly identify photometric candidates for spectroscopic follow-up. This combination could efficiently detect and confirm statistical samples of LCGs that could help to constrain valuable cosmic reionization parameters.

\section{Summary \& Conclusions} \label{sec:conc}

We present new deep WFC3/F336W and ACS/F435W HST images (with depths $\sim29\text{--}30$ mag at 5$\sigma$), that we use to identify a population of high-redshift LCG candidates. We process the WFC3/UVIS images with our new custom pipeline that reduces noise, gradients, hot pixels and offsets in the images to produce the cleanest HST UV data possible. Some of our custom reduction steps for WFC3/UVIS have now been adopted by the WFC3 team and are available for download directly from MAST for all existing data. We apply additional corrections to the F336W images that are not included on MAST but are easily applied to data with a new public reduction notebook\footnote{\url{https://github.com/lprichard/HST_FLC_corrections}}.

We collate a comprehensive \textit{Parent Sample} of possible candidates without relying on the Lyman-break technique. We use photometric redshifts from the ZFOURGE survey, COSMOS, CANDELS, and publicly available spectroscopic redshifts from 16 surveys within our new HST footprints. Although no catalog will be free from selection bias, this is one of the most comprehensive and agnostic LCG searches possible with the available data. There are 3065 objects at all redshifts in the two F336W footprints and 6430 in the three F435W footprints (our \textit{All-$z$ Sample}). 380 sources are at $z>3.087$ in F336W and 242 at $z>4.391$ in F435W, the redshift limits ($z_{\rm lim}$) for the bands to cleanly probe LyC (\textit{Parent Sample}). We visually inspect all sources in the \textit{Parent Sample} multiple times to obtain a long list of possible candidates (77) for spectroscopic confirmation.

The bulk of awarded spectroscopic observing time was weathered out or had less-than-optimal conditions. We present the spectra obtained for some of our sources with Keck/LRIS ($\sim 1.5$ nights). We acquire spectroscopic redshifts for the three brightest LCG candidates (i.e., Group A spectroscopic candidates): zf\_9775 at $z=4.365$, zf\_11754 at $z=4.470$, and zf\_14000 at $z=3.727$. The list of candidates is cleaned and we select only those with $>3\sigma$ LyC detections. The remaining 32 sources are photometric LCG candidates that we split by redshift confidence. Group B has 10 LCG candidates with relatively confident photometric (and one low-quality public spectroscopic) redshift. Group C has 22 candidates with more unreliable redshifts. We summarize the main results of this work below.

\begin{itemize}[leftmargin=*]
\itemsep0em 
    \item We find a $6.84\%$ detection rate of our $>3\sigma$ LCG candidates in the F336W footprints and $3.72\%$ in the F435W footprints relative to the \textit{Parent Sample}. We find number densities of $\sim1.8$ arcmin$^{-2}$ at $z>3.087$ in the F336W pointings and $\sim0.1\text{--}0.4$ arcmin$^{-2}$ at $z>4.391$ in the F435W pointings (at varying depths).
    \item We determine conservative redshift-independent LyC contamination probabilities of $1.8\%\text{--}3.3\%$ depending on the field. These are higher than the LCG candidate to \textit{All-z Sample} detection rates ($0.85\%$ for F336W, $0.14\%$ for F435W). Most of the candidates are in Group C, whose redshifts we are least confident in and that are likely to be dominated by low-redshift interlopers. 
    \item The HST images and Keck/LRIS spectroscopy for the three Group A spectroscopic candidates imply that these detections could be genuine. We determine $f_{\rm esc}$ using the traditional ($f_{\rm esc}^{\rm abs}$) and a statistical ($f_{\rm esc}^{\rm PDF}$) approach. The results imply that the most likely escape fraction estimates for all three spectroscopic candidates for both methods is $>100\%$, and therefore not physical given the models and assumptions used.
    \item From the purely illustrative color-color plots, most candidates sit outside of the LBG box. They lie on tracks with observed LyC to UV continuum flux ratios ($R_{\rm obs}$). Calculated directly from their LyC (F336W or F435W) to UV continuum (F814W) bands are $R_{\rm obs}=4\%, 3\%$ and $9\%$ for zf\_9775, zf\_11754 and zf\_14000 respectively. This could imply that the LCG candidates are similar to LBG galaxies with LyC emission, or that they have flux contamination.
    \item The three Group A spectroscopic LCG candidates and 32 Group B and C photometric LCG candidates require spectroscopic follow up to confirm their true nature. Even with spectroscopic redshifts, UV-HST imaging and multi-band photometry, more information is required to confirm these as genuine LCGs. The future KWFI instrument will be valuable for efficiently identifying new faint LCG candidates. The new class of 30 m telescopes will be able to spectroscopically confirm the faintest candidates presented here so that larger statistical samples can be built up for constraining cosmic reionization parameters. 
\end{itemize}

\begin{deluxetable*}{|c|c|c|c|c|c|c|c|c|c|c|c|c|}
\tabletypesize{\footnotesize}
\tablewidth{0pt}
\tablecaption{LCG candidate IDs, coordinates, assigned groups, LyC band and limit, and redshifts.}
\label{tab:lcgs}
\tablehead{ID$^{1}$ & Group$^{2}$ & RA & Dec & LyC$^{3}$ & $z_{\rm lim}$ $^{4}$ & $z_{\rm phot}$ $^{5}$ & $z_{\rm phot}$ $^{6}$ & $\overline{z_{\rm phot}}$ $^{7}$ & $\Delta z_{\rm phot}$ $^{8}$ & $z_{\rm spec}$ $^{9}$ & $z_{\rm spec}$ $^{10}$ & $z_{\rm conf}$ $^{11}$ \\ & & & & Pointing & & (ZF) & (CS) & (CA) & (CA) & (lit.) & (LRIS) &}
\startdata
zf\_9775 & A & 150.102478 & 2.281554 & F336W\_2 & 3.087 & 4.266 & 4.246 & 4.231 & 0.074 & 0.446$^{12}$ & 4.365 & 1 \\
zf\_11754 & A & 150.047912 & 2.303804 & F435W\_2 & 4.391 & 4.453 & 0.705 & - & - & - & 4.470 & 1 \\
zf\_14000 & A & 150.208221 & 2.328479 & F336W\_1 & 3.087 & 3.953 & 0.301 & - & - & 3.779$^{13}$ & 3.727 & 1 \\
zf\_11423 & B & 150.201065 & 2.300642 & F336W\_1 & 3.087 & 0.796 & 0.716 & 0.813 & 0.094 & 3.655$^{13}$ & - & 2 \\
zf\_12283 & B & 150.192413 & 2.309807 & F336W\_1 & 3.087 & 3.520 & 0.372 & 3.592 & 0.126 & - & - & 2 \\
zf\_8333 & B & 150.114257 & 2.266746 & F336W\_2 & 3.087 & 3.223 & 3.117 & 0.285 & 0.056 & - & - & 2 \\
zf\_8461 & B & 150.118164 & 2.268226 & F336W\_2 & 3.087 & 3.197 & 0.443 & 0.350 & 0.016 & - & - & 2 \\
zf\_13836 & B & 150.196579 & 2.326127 & F336W\_1 & 3.087 & 3.622 & - & 0.338 & 0.324 & - & - & 2 \\
zf\_14198 & B & 150.196670 & 2.330944 & F336W\_1 & 3.087 & 3.826 & 2.180 & 0.619 & 0.020 & - & - & 2 \\
zf\_13312 & B & 150.201431 & 2.320581 & F336W\_1 & 3.087 & 3.159 & 3.165 & 2.748 & 1.283 & - & - & 2 \\
zf\_6745 & B & 150.115646 & 2.249375 & F435W\_1a & 4.391 & 5.461 & - & 1.213 & 0.772 & - & - & 2 \\
zf\_14060 & B & 150.211822 & 2.329048 & F336W\_1 & 3.087 & 3.158 & - & - & - & - & - & 2 \\
zf\_13676 & B & 150.220871 & 2.324596 & F336W\_1 & 3.087 & 4.779 & 4.557 & - & - & - & - & 2 \\
cs\_658350 & C & 150.207800 & 2.300505 & F336W\_1 & 3.087 & - & 3.171 & - & - & - & - & 3 \\
cs\_621227 & C & 150.106355 & 2.244453 & F336W\_2 & 3.087 & - & 3.931 & - & - & - & - & 3 \\
cs\_626018 & C & 150.111087 & 2.251412 & F336W\_2 & 3.087 & - & 3.578 & - & - & - & - & 3 \\
cs\_667332 & C & 150.035636 & 2.314277 & F435W\_2 & 4.391 & - & 5.419 & - & - & - & - & 3 \\
cs\_621890 & C & 150.121013 & 2.245430 & F435W\_1a & 4.391 & - & 4.567 & - & - & - & - & 3 \\
cs\_681305 & C & 150.011024 & 2.334490 & F435W\_2 & 4.391 & - & 4.656 & - & - & - & - & 3 \\
zf\_8009 & C & 150.114654 & 2.263270 & F336W\_2 & 3.087 & 0.833 & 3.218 & 0.656 & 0.328 & - & - & 4 \\
zf\_9669 & C & 150.098083 & 2.280273 & F336W\_2 & 3.087 & 0.415 & 3.871 & 0.377 & 0.029 & - & - & 4 \\
zf\_12022 & C & 150.193847 & 2.306703 & F336W\_1 & 3.087 & 0.652 & 3.164 & 0.723 & 0.038 & - & - & 4 \\
zf\_14044 & C & 150.199768 & 2.32866 & F336W\_1 & 3.087 & 0.545 & 3.939 & 0.483 & 0.017 & - & - & 4 \\
cs\_671985 & C & 150.192170 & 2.321238 & F336W\_1 & 3.087 & - & 3.301 & 2.025 & 0.565 & - & - & 4 \\
zf\_9963 & C & 150.10466 & 2.284251 & F336W\_2 & 3.087 & 0.332 & 3.095 & 0.436 & 0.052 & - & - & 4 \\
cs\_643060 & C & 150.104563 & 2.277120 & F336W\_2 & 3.087 & - & 3.496 & 0.579 & 0.194 & - & - & 4 \\
cs\_723862 & C & 150.109444 & 2.398033 & F435W\_1b & 4.391 & - & 5.286 & 2.249 & 0.154 & - & - & 4 \\
zf\_12042 & C & 150.222595 & 2.307420 & F336W\_1 & 3.087 & 1.080 & 4.122 & - & - & - & - & 4 \\
zf\_8356 & C & 150.13797 & 2.267169 & F336W\_2 & 3.087 & 2.009 & 3.231 & 1.303 & 0.639 & - & - & 4 \\
zf\_10407 & C & 150.205856 & 2.289908 & F336W\_1 & 3.087 & 1.061 & 3.45 & - & - & - & - & 4 \\
zf\_13775 & C & 150.044174 & 2.326125 & F435W\_2 & 4.391 & 0.632 & 4.988 & - & - & - & - & 4 \\
zf\_5935 & C & 150.118850 & 2.241288 & F435W\_1a & 4.391 & 2.460 & 4.807 & 2.024 & 0.549 & - & - & 4 \\
zf\_13099 & C & 150.195251 & 2.318458 & F336W\_1 & 3.087 & 0.312 & 3.453 & 1.869 & 1.692 & - & - & 4 \\
zf\_6857 & C & 150.125915 & 2.250819 & F336W\_2 & 3.087 & 2.447 & 3.227 & 0.371 & 0.457 & - & - & 4 \\
cs\_611354 & C & 150.107760 & 2.229506 & F435W\_1a & 4.391 & - & 5.338 & 1.687 & 0.369 & - & - & 4 \\
\enddata
\tablenotetext{}{1. Prefixes show ID and coordinate survey origin (zf\_* is ZFOURGE, cs\_* is COSMOS). 2. LCG group assigned in this work, described in Section \ref{sec:props}. 3. HST filter probing LyC flux and the pointing name. 4. Redshift limit above which the LyC filter will cleanly sample LyC flux. 5. ZFOURGE photometric redshift \citep{Straatman2016}. 6. COSMOS2015 photometric redshift \citep{Laigle2016}. 7. CANDELS photometric redshift average \citep{Nayyeri2017}. 8. CANDELS photometric redshift standard deviation \citep{Nayyeri2017}. 9. Spectroscopic redshifts from the literature (see Section \ref{sec:zcats}). 10. New Keck/LRIS spectroscopic redshift presented in this paper. 11. Redshift confidence parameter defined in Section \ref{sec:props}. 12. Low-confidence spectroscopic redshift from 3D-HST \citep{Brammer2012, Momcheva2016}, see Section \ref{sec:specz} and Appendix \ref{sec:appspec}. 13. Low-confidence spectroscopic redshift from DEIMOS 10K \citep{Hasinger2018}, see Section \ref{sec:specz} and Appendix \ref{sec:appspec}.}
\end{deluxetable*}

\begin{deluxetable*}{|c|c|c|c|c|c|c|c|c|c|c|}
\tabletypesize{\footnotesize}
\tablewidth{0pt}
\tablecaption{LCG candidate isophotal area, isophotal magnitude, and signal-to-noise ($S/N$) values.}
\label{tab:lcgsphot}
\tablehead{ID & Isophotal$^{1}$ & m$_{\rm F336W}$ & $S/N$ & m$_{\rm F435W}$ & $S/N$ & m$_{\rm F435W}$ $^{2}$ & m$_{\rm F606W}$ & $S/N$ & m$_{\rm F814W}$ & $S/N$ \\ & Area {\scriptsize(pix)} & & {\scriptsize(F336W)} & & {\scriptsize(F435W)} & {\scriptsize(HSC match)} & & {\scriptsize(F606W)} & & {\scriptsize(F814W)}}
\startdata
zf\_9775 & 391 & $27.68\pm0.11$ & 9.1 & $27.25\pm0.18$ & 5.7 & $27.39\pm0.18$ & $25.29\pm0.04$ & 23.4 & $24.15\pm0.05$ & 18.3 \\
zf\_11754 & 323 & - & - & $28.65\pm0.31$ & 3.4 & $28.14\pm0.31$ & - & - & $24.92\pm0.09$ & 11.0 \\
zf\_14000 & 126 & $27.52\pm0.06$ & 16.6 & - & - & - & $25.41\pm0.04$ & 26.1 & $24.92\pm0.05$ & 21.1 \\
zf\_11423 & 196 & $28.29\pm0.15$ & 6.9 & - & - & - & $27.17\pm0.25$ & 4.2 & $25.75\pm0.11$ & 9.1 \\
zf\_12283 & 204 & $27.95\pm0.12$ & 8.9 & - & - & - & $26.70\pm0.12$ & 8.8 & $26.29\pm0.22$ & 4.7 \\
zf\_8333 & 193 & $29.07\pm0.31$ & 3.4 & $27.74\pm0.21$ & 5.1 & $27.70\pm0.21$ & $27.09\pm0.20$ & 5.2 & $26.60\pm0.41$ & 2.5 \\
zf\_8461 & 156 & $27.90\pm0.09$ & 11.2 & $27.26\pm0.12$ & 8.7 & $27.40\pm0.12$ & $26.65\pm0.11$ & 9.7 & $26.73\pm0.37$ & 2.8 \\
zf\_13836 & 106 & $28.23\pm0.11$ & 9.4 & - & - & - & $27.15\pm0.17$ & 6.2 & $26.75\pm0.20$ & 5.2 \\
zf\_14198 & 128 & $28.10\pm0.10$ & 9.9 & - & - & - & $27.48\pm0.26$ & 4.1 & $26.90\pm0.29$ & 3.7 \\
zf\_13312 & 112 & $28.92\pm0.22$ & 4.9 & - & - & - & $27.60\pm0.27$ & 3.9 & $27.41\pm0.41$ & 2.5 \\
zf\_6745 & 80 & $28.36\pm0.10$ & 10.3 & $28.06\pm0.20$ & 5.3 & $27.88\pm0.20$ & $28.16\pm0.39$ & 2.7 & $27.51\pm0.60$ & 1.8 \\
zf\_14060 & 74 & $28.61\pm0.13$ & 7.8 & - & - & - & $28.35\pm0.48$ & 2.2 & $27.52\pm0.38$ & 2.8 \\
zf\_13676 & 96 & $28.50\pm0.17$ & 6.1 & - & - & - & - & - & $27.57\pm0.31$ & 3.4 \\
cs\_658350 & 127 & $27.76\pm0.08$ & 13.2 & - & - & - & $27.51\pm0.39$ & 2.7 & $27.29\pm0.39$ & 2.7 \\
cs\_621227 & 111 & $29.53\pm0.34$ & 3.1 & $27.97\pm0.23$ & 4.6 & $27.83\pm0.23$ & $28.37\pm0.44$ & 2.4 & $27.72\pm0.86$ & 1.2 \\
cs\_626018 & 57 & $29.62\pm0.28$ & 3.7 & $29.33\pm0.56$ & 1.9 & $28.34\pm0.56$ & $29.39\pm0.82$ & 1.3 & $27.99\pm0.72$ & 1.4 \\
cs\_667332 & 61 & - & - & $29.23\pm0.25$ & 4.2 & $28.32\pm0.25$ & - & - & $28.01\pm0.73$ & 1.4 \\
cs\_621890 & 119 & $28.20\pm0.11$ & 9.1 & $28.15\pm0.27$ & 3.9 & $27.92\pm0.27$ & $28.23\pm0.51$ & 2.1 & $28.21\pm1.36$ & 0.7 \\
cs\_681305 & 44 & - & - & $29.20\pm0.21$ & 5.1 & $28.31\pm0.21$ & - & - & $28.32\pm0.82$ & 1.3 \\
zf\_8009 & 191 & $27.52\pm0.07$ & 14.0 & $27.16\pm0.12$ & 8.5 & $27.33\pm0.12$ & $27.11\pm0.18$ & 5.8 & $26.26\pm0.28$ & 3.8 \\
zf\_9669 & 170 & $28.19\pm0.12$ & 8.7 & $27.71\pm0.18$ & 5.7 & $27.69\pm0.18$ & $26.90\pm0.13$ & 8.0 & $26.27\pm0.26$ & 4.1 \\
zf\_12022 & 159 & $27.76\pm0.08$ & 12.2 & - & - & - & $27.07\pm0.15$ & 7.0 & $26.36\pm0.19$ & 5.6 \\
zf\_14044 & 154 & $28.39\pm0.15$ & 7.1 & - & - & - & $27.24\pm0.23$ & 4.6 & $26.49\pm0.21$ & 4.9 \\
cs\_671985 & 108 & $28.32\pm0.11$ & 9.0 & - & - & - & $27.22\pm0.19$ & 5.5 & $26.95\pm0.29$ & 3.6 \\
zf\_9963 & 159 & $28.44\pm0.15$ & 6.9 & - & - & - & $27.21\pm0.18$ & 5.7 & $26.98\pm0.51$ & 2.0 \\
cs\_643060 & 143 & $27.76\pm0.08$ & 13.4 & $27.28\pm0.12$ & 8.9 & $27.42\pm0.12$ & $27.24\pm0.15$ & 7.0 & $27.07\pm0.40$ & 2.6 \\
cs\_723862 & 106 & - & - & $27.39\pm0.07$ & 13.6 & $27.49\pm0.07$ & $27.39\pm0.25$ & 4.2 & $27.19\pm0.32$ & 3.3 \\
zf\_12042 & 72 & $28.39\pm0.10$ & 9.9 & - & - & - & $28.65\pm0.80$ & 1.3 & $27.27\pm0.27$ & 3.8 \\
zf\_8356 & 152 & $28.28\pm0.17$ & 6.1 & $27.13\pm0.10$ & 10.3 & $27.31\pm0.10$ & $27.39\pm0.25$ & 4.2 & $27.35\pm0.72$ & 1.5 \\
zf\_10407 & 100 & $27.47\pm0.05$ & 19.2 & - & - & - & $27.13\pm0.16$ & 6.4 & $27.54\pm0.43$ & 2.5 \\
zf\_13775 & 54 & - & - & $29.50\pm0.28$ & 3.7 & $28.38\pm0.28$ & $28.85\pm1.27$ & 0.8 & $27.74\pm0.53$ & 2.0 \\
zf\_5935 & 145 & $28.32\pm0.22$ & 4.7 & $27.27\pm0.13$ & 8.1 & $27.41\pm0.13$ & $27.75\pm0.32$ & 3.3 & $28.06\pm1.25$ & 0.8 \\
zf\_13099 & 40 & $29.01\pm0.17$ & 6.3 & - & - & - & $28.55\pm0.40$ & 2.7 & $28.16\pm0.55$ & 1.9 \\
zf\_6857 & 123 & $28.44\pm0.16$ & 6.4 & $27.66\pm0.18$ & 5.7 & $27.66\pm0.18$ & $27.31\pm0.21$ & 4.9 & $28.50\pm2.03$ & 0.5 \\
cs\_611354 & 109 & - & - & $27.15\pm0.09$ & 11.0 & $27.32\pm0.09$ & $27.39\pm0.21$ & 5.0 & $29.20\pm3.14$ & 0.3 \\
\enddata
\tablenotetext{}{1. Measured from the source-detection segmentation map with 0.03$^{\prime\prime}$ pixel scale (see Section \ref{sec:phot}). 2. F435W isophotal fluxes converted to HSC PSF-matched 1$^{\prime\prime}$ aperture fluxes (see Section \ref{sec:hsc}).}
\end{deluxetable*}

%------------------
%ACKNOWLEDGEMENTS
%------------------
\acknowledgments
\noindent We thank the anonymous reviewer for a helpful report. This research is based on observations made with the NASA/ESA HST obtained from the STScI, which is operated by the Association of Universities for Research in Astronomy, Inc., under NASA contract NAS 5–26555. These observations are associated with program ID 15100 (PI Cooke) and we acknowledge financial support for this work from HST. We also acknowledge some additional support from HST program ID 15647 (PI Teplitz) for work related to the final data calibrations presented here. This work is also based on observations taken by the CANDELS Multi-Cycle Treasury Program with the NASA/ESA HST.

This work was supported by three NASA Keck PI Data Awards (2018B\_N188 \& 2019B\_N010, PI Rafelski \& 2020B\_N168, PI Prichard), administered by the NASA Exoplanet Science Institute. This work was also made possible with funding from STScI's Director's Discretionary Research Funding (grant ID: D0001.82500).

Part of this research was funded by the Australian Research Council Centre of Excellence for All-sky Astrophysics in 3 Dimensions (ASTRO-3D), CE170100013, the Australian Research Council Centre of Excellence for All-sky Astrophysics (CAASTRO), CE110001020, and the Australian Research Council Centre of Excellence for Gravitational Wave Discovery (OzGrav), CE170100004.

Data presented herein were obtained at the W. M. Keck Observatory from telescope time allocated to NASA through the agency's scientific partnership with the California Institute of Technology and the University of California. The Observatory was made possible by the generous financial support of the W. M. Keck Foundation. The authors wish to recognize and acknowledge the very significant cultural role and reverence that the summit of Maunakea has always had within the indigenous Hawaiian community. We are most fortunate to have the opportunity to conduct observations from this mountain.

This research made use of \textsc{Montage}. It is funded by the National Science Foundation under Grant Number ACI-1440620, and was previously funded by the NASA's Earth Science Technology Office, Computation Technologies Project, under Cooperative Agreement Number NCC5-626 between NASA and the California Institute of Technology.

This paper is based in part on data collected at the Subaru Telescope and retrieved from the HSC data archive system, which is operated by Subaru Telescope and Astronomy Data Center (ADC) at National Astronomical Observatory of Japan. Data analysis was in part carried out with the cooperation of Center for Computational Astrophysics (CfCA), National Astronomical Observatory of Japan.

% %%%%%%%%%%%%%%%%%%%%%%%%%%%%%%%%%%%%%%%%%
%------------------
%BIBLIOGRAPHY
%------------------
\vspace{5mm}
\facilities{HST(WFC3, ACS), Keck:I (LRIS), Subaru (HSC)}

\software{\textsc{astropy} \citep{Astropy2013}, \textsc{autoslit}, \textsc{calwf3 v3.6.0} \citep{Anderson2020, Anderson2021}, \textsc{DrizzlePac} \citep{Gonzaga2012, Hoffmann2021}, IRAF, \textsc{Montage}, \textsc{photutils}, and codes from R. Bassett (on \href{https://github.com/robbassett}{GitHub}), L. Prichard (on \href{https://github.com/lprichard}{GitHub}), M. Revalski (on \href{https://github.com/mrevalski}{GitHub}), and B. Sunnquist (on \href{https://github.com/bsunnquist/uvis-skydarks}{GitHub}).}

%------------------
%APPENDIX
%------------------
\appendix
\restartappendixnumbering
\renewcommand{\thefigure}{A\arabic{figure}}
\renewcommand{\theHfigure}{A\arabic{figure}}
\setcounter{figure}{0}

\section{WFC3/UVIS darks pipeline improvements}
\label{sec:appdarks}

\begin{figure}
\begin{center}
\includegraphics[width=0.485\textwidth, trim=15 10 10 10, clip]{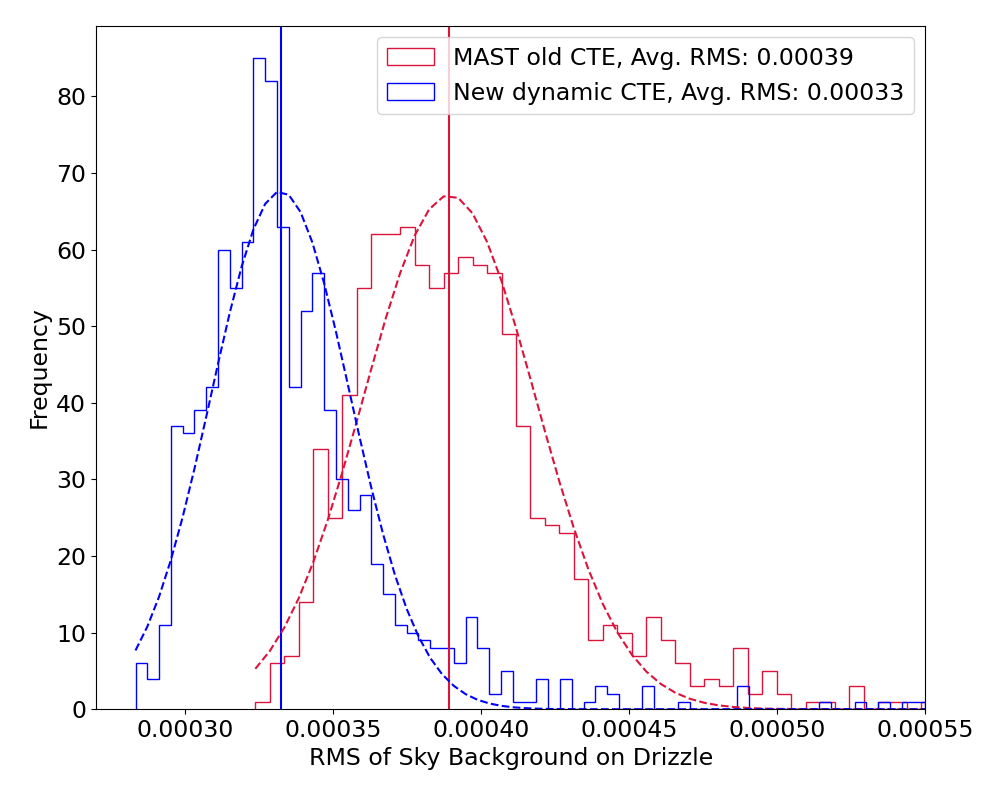}
\includegraphics[width=0.5\textwidth, trim=15 10 10 40, clip]{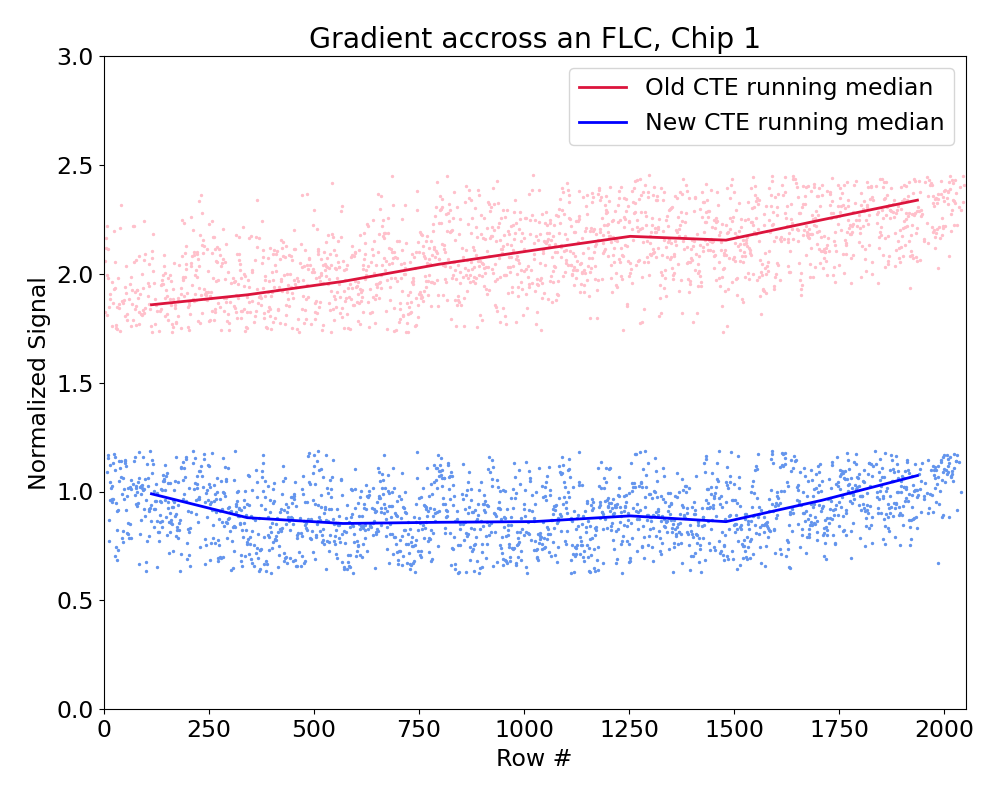}
\includegraphics[width=0.49\textwidth, trim=15 10 10 10, clip]{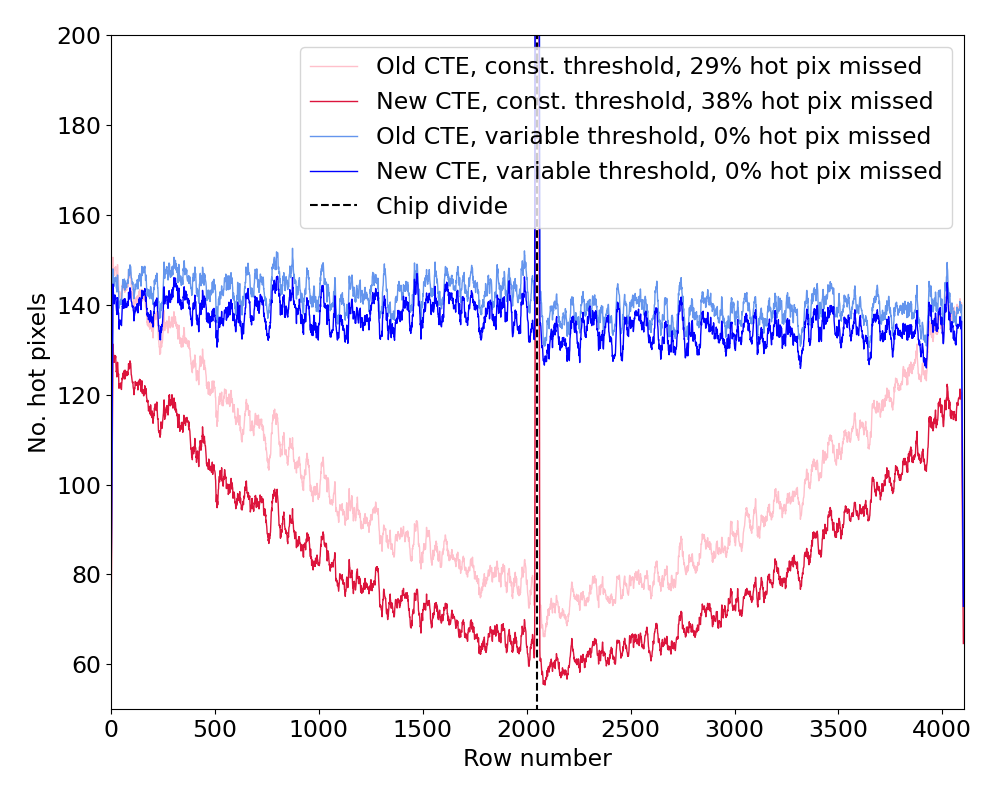}
\caption{\textit{Top left:} Comparison of RMS measured from images processed with the old CTE code (red) vs. the new CTE code (blue). The RMS of the background is measured from 1000 random apertures on sky (avoiding sources). The new CTE code significantly reduces the background with an average of $3.3\times10^{-4}$ e/s from $3.9\times10^{-4}$ e/s, valuable for the detection of faint sources. \textit{Top right:} Signal gradients across one chip of an individual FLC. The median of each row (points) and the running medians (solid lines) are shown for the old (red) and new (blue) CTE codes. The new CTE reduces the gradients over the chips. \textit{Bottom:} Number of hot pixels identified as a function of row number across chip 2 (left) and chip 1 (right). The previous method using a constant threshold (red lines) misses $\sim$30\% of hot pixels. Using a new variable threshold (blue lines) identifies the correct number of hot pixels across each entire chip. Bold and faint lines show the new and old CTE corrections respectively. \small}
\label{fig:cte}
\end{center}
\end{figure}

We make several improvements to the WFC3/UVIS reduction pipeline \citep{Bourque2016} for which we provide details here. The WFC3/UVIS channel has a charge coupled device (CCD) detector. Photons landing on the array are converted to charge which is measured by being transferred across the array to the ``readout'' amplifiers. Due to degradation of the detector over time from radiation, charge traps develop, reducing the charge transfer efficiency (CTE). This effect is corrected after the fact \citep[e.g.,][]{Anderson2020, Anderson2021} to recover as much of the original flux as possible. 

The detector also has imperfections such as hot pixels. The WFC3/UVIS CCD undergoes anneals (brief heating of the detector every $\sim$month) to reduce the number of hot pixels. In each anneal cycle (period between anneals), the detector has a different ``fingerprint''. Darks and data are typically processed as soon as possible after being observed. This speed means that darks from the previous anneal cycle (using a different detector fingerprint than the observations) are used to calibrate the data which diminishes the quality of the previously available data products.

The new CTE code \citep[\textsc{calwf3 v3.6.0};][]{Anderson2020, Anderson2021} optimizes parameters for the data to mitigate read noise, which reduces background noise and gradients across the images. Figure \ref{fig:cte} (top left panel) shows the RMS measured in 1000 randomly placed apertures on sky (avoiding sources) of a drizzled image processed with the old CTE code (red) and the new code (blue). The background noise is significantly reduced with the new code. The CTE corrections are applied to both the science data and the darks in the reduction. Figure \ref{fig:cte} (top right panel) shows the signal gradients across one chip of a single FLC. We show the median signal of each row (dots) and the running median of the row values (lines) for the old (red) and new (blue) CTE code. The new CTE correction results in much flatter signal across the chips. The new CTE code along with the use of concurrent darks (from the same anneal cycle as the data are observed) are now included as standard by WFC3, and are available from MAST (as of May 2021).

We developed a novel flagging routine to identify hot pixels more effectively. The method used by MAST adopts a constant threshold, and due to the imperfect CTE correction, increasingly more hot pixels are missed further from the readout. To correct for this loss of hot pixels caused by CTE degradation, we derive a variable threshold for every chip to identify roughly a constant number of hot pixels across the chip as expected. The most accurate number of hot pixels is closest to the readout with the shortest distance to travel across the array. We therefore take this value to be the true number of hot pixels, and find the threshold that selects just below this value for each set of 50 rows. We then fit a function to those values to derive a variable threshold conversion for each chip.

Figure \ref{fig:cte} (bottom panel) shows the number of hot pixels as a function of row number with the readout locations at the edges of the plot. We show the hot pixels identified using a constant threshold (red) and the new variable threshold method (blue) for the old (faint lines) and new CTE code (bold lines). The new CTE code increases the number of hot pixels missed as the flux pedestal subtracted is larger, so more pixels are below the previous constant threshold. The constant threshold misses around $\sim30\%$ of the expected hot pixels. For a detailed step-by-step run through of all the steps we use for the WFC3/UVIS reduction, see our pipeline and supplementary codes on GitHub\footnote{\url{https://github.com/lprichard/hst_wfc3_uvis_reduction}}.

We flag negative divots adjacent to the readout cosmic rays (ROCRs) in the FLC data quality (DQ) arrays to prevent them from propagating through to the drizzled images. As these cosmic rays appear during readout, they fall closer to the amplifier than they appear in the FLCs, which causes over-corrections by the CTE code. Due to their nature, these flags increase with distance from the amplifier while the array is being readout, and affect a total of $\sim0.2\%$ detector pixels. We flag divots within five pixels of a CR hit (away from the readout direction) that are $3\sigma$ below the image mean as bad detector pixels (value of 4 in the DQ arrays).

We equalize the overall signal levels in the four amplifier quadrants of each FLC. Typically, small bias offsets exceeding one electron can exist between these regions, causing discontinuities in the background. To remove these differences, we calculate the offsets between the median level in each individual quadrant and the average of all four quadrants, multiply the difference by the flat field, and subtract these offsets from the original data. As the zeropoints are determined as an average over all four quadrants, the equalization of the amps does not affect our photometry as we have a roughly even distribution of sources in the images.

\begin{figure*}
\begin{center}
\includegraphics[width=0.7\textwidth]{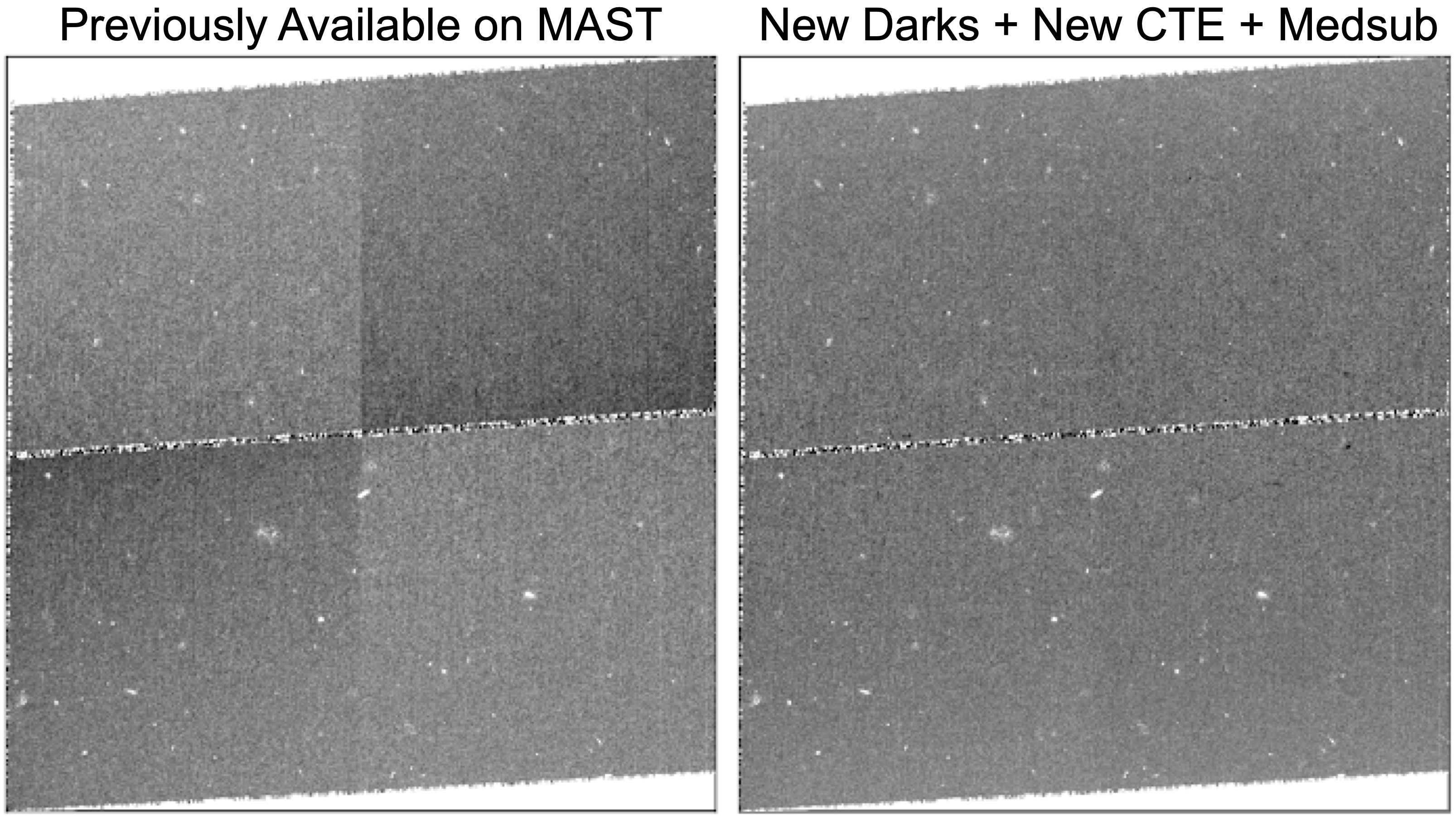}
\caption{Comparison of a reduced CTE-corrected single-visit exposure (FLC) using the old and new reduction methods. An FLC previously available on MAST is shown on the left. On the right is an FLC reduced with the new darks (including the new hot pixel method) and new CTE (that reduces noise and gradients). An additional correction to equalize the bias offsets between the four quadrants is also applied (medsub). \small}
\label{fig:flcs}
\end{center}
\end{figure*}

Figure \ref{fig:flcs} shows a comparison of an FLC previously available on MAST (left) and one processed using the new concurrent darks, the new CTE, and the amplifier bias offsets equalized (\texttt{medsub}). Gradients, noise, offsets, and blotchy patterns are greatly reduced using the new reduction methods. The new CTE and concurrent darks are now applied as standard by STScI and available from MAST. The additional corrections can be easily applied to reduced FLCs from MAST using a new publicly available notebook\footnote{\url{https://github.com/lprichard/HST_FLC_corrections}}.

\renewcommand{\thefigure}{B\arabic{figure}}
\renewcommand{\theHfigure}{B\arabic{figure}}
\setcounter{figure}{0}

\section{Ancillary spectra}
\label{sec:appspec}

\begin{figure*} 
\begin{center}
\includegraphics[width=\textwidth]{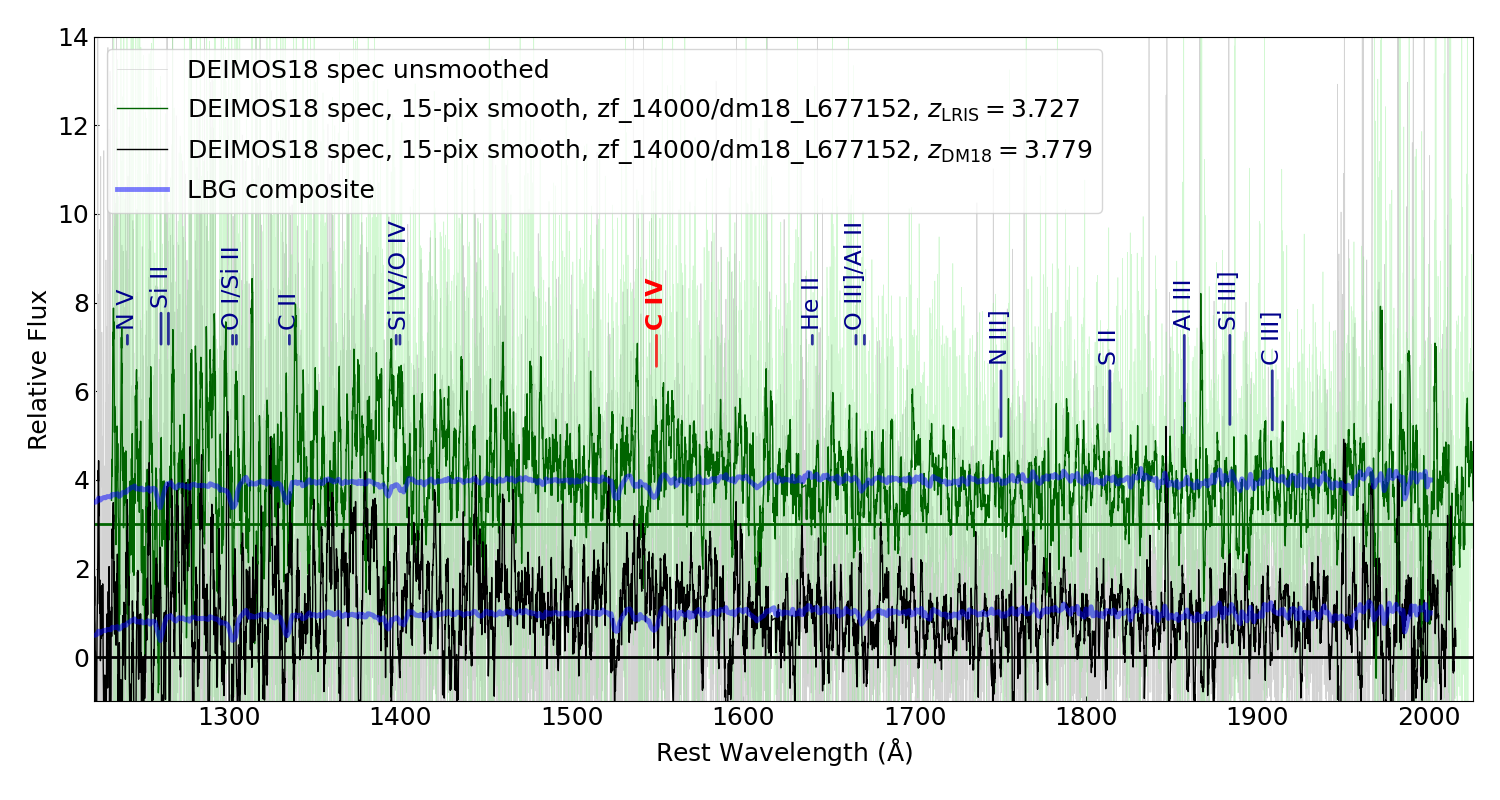}\\
\includegraphics[width=\textwidth]{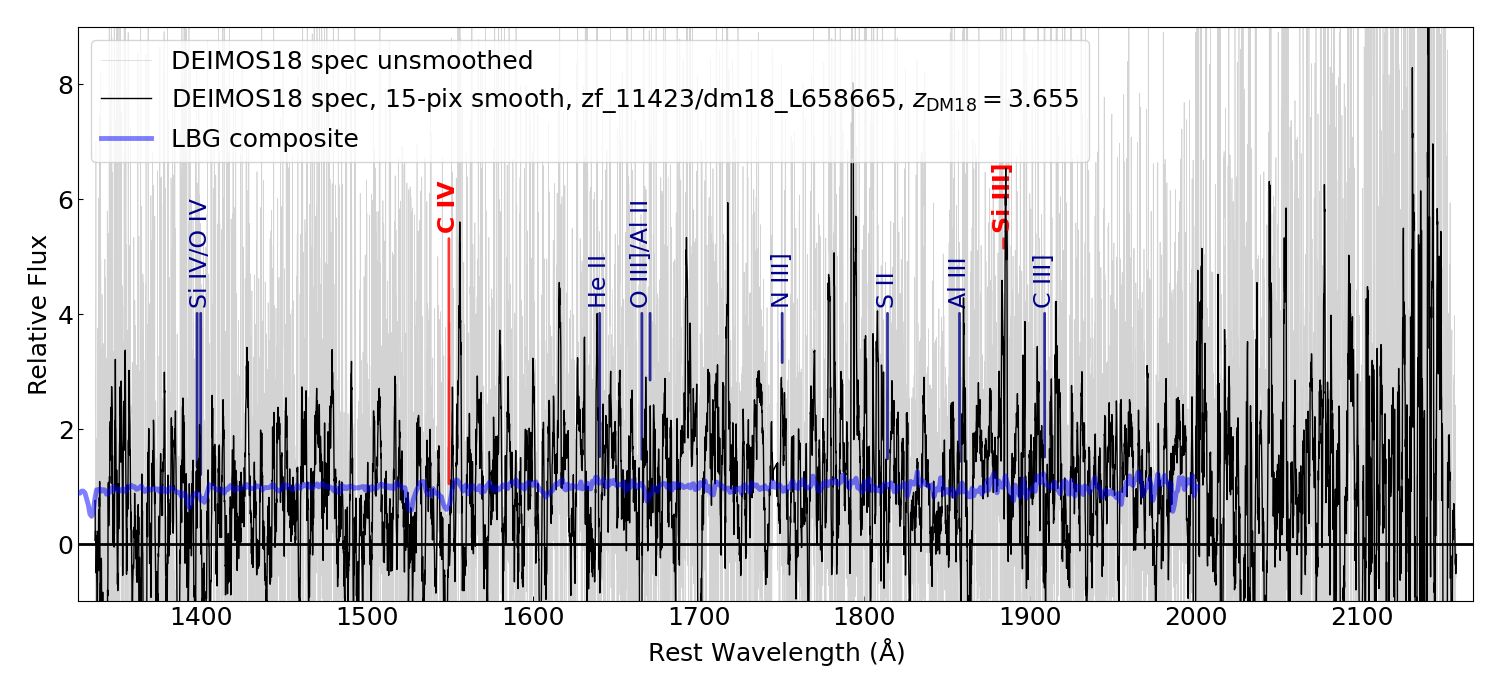}
\caption{DEIMOS 10K spectra \citep[DM18;][]{Hasinger2018} of two LCG candidates. We show the unsmoothed spectra (pale background spectra), 15-pixel boxcar-smoothed spectra (dark green/black lines), LBG composite (blue) from \cite{Shapley2003}, and spectral features (vertical lines). \textit{Top:} DM18 spectrum of a Group A spectroscopic candidate, zf\_14000 (DM18 ID: L677152). The redshift we determine from the LRIS spectrum (Figure \ref{fig:lris}) is $z=3.727$. We show the DM18 spectrum shifted to the redshift from LRIS (green) and the spectrum at the DM18 redshift of $z_{\rm DM18}=3.779$ (black) for comparison. We highlight a potentially real feature which adds weight to the LRIS redshift (red). \textit{Bottom:} DM18 spectrum of a Group B candidate, zf\_11423 (DM18 ID: L658665, $z_{\rm DM18}=3.655$). We highlight two potentially real spectral features (red) that might add weight to the DM18 redshift but with low confidence.\small}
\label{fig:deimos}
\end{center}
\end{figure*}

\begin{figure*}
\begin{center}
\includegraphics[width=\textwidth, trim=20 24 10 20, clip]{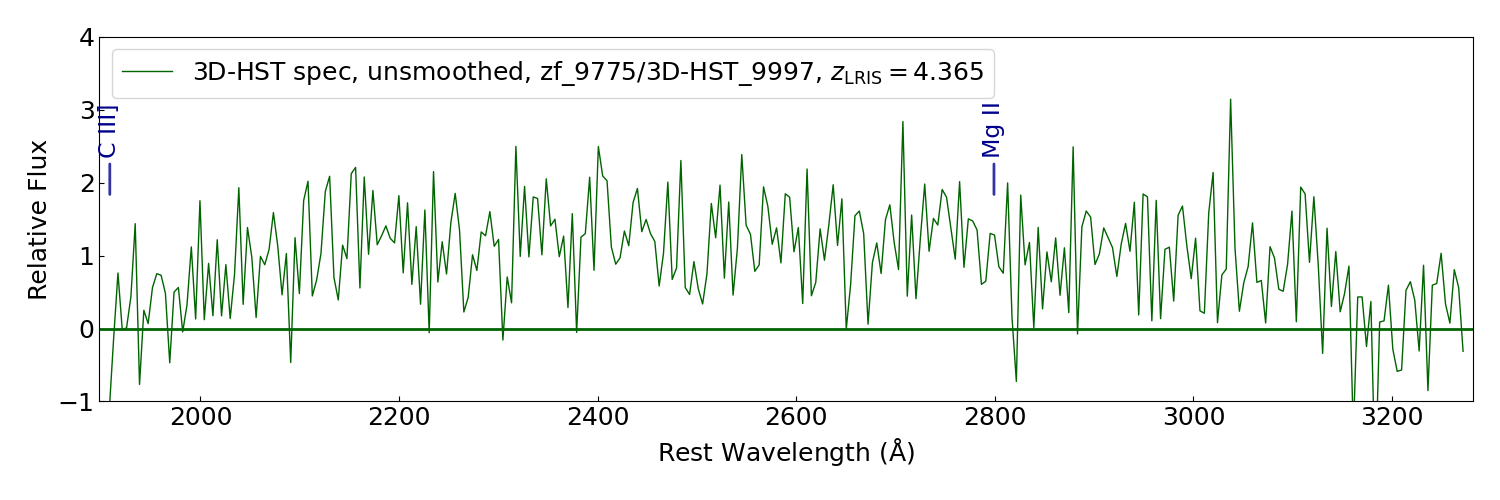}\\
\includegraphics[width=\textwidth, trim=20 24 10 20, clip]{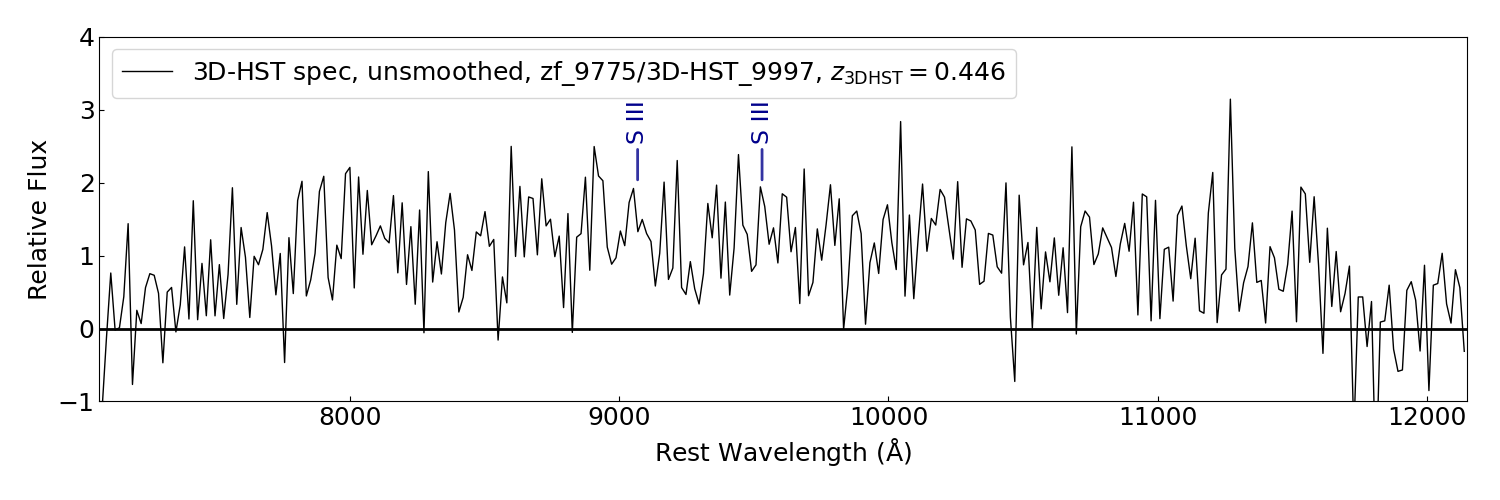}
\caption{3D-HST spectrum \citep{Brammer2012,Momcheva2016} for zf\_9775 (3D-HST ID: 9997) redshifted to the LRIS redshift ($z_{\rm LRIS}=4.365$, green, top panel), and that quoted in 3D-HST ($z_{\rm 3DHST}=0.446$, black, bottom panel). The 3D-HST spectrum does not provide sufficient evidence for either redshift, we therefore deem the 3D-HST redshift invalid due to the confident redshift we derive from the LRIS spectrum (Figure \ref{fig:lris}). \small}
\label{fig:3dhst}
\end{center}
\end{figure*}

We show the DEIMOS 10K \citep[DM18;][]{Hasinger2018} spectra for two of the LCG candidates in Figure \ref{fig:deimos}. Both the unsmoothed spectra (pale background spectra) and 15-pixel boxcar-smoothed spectra (dark green/black) are shown overlaid. We show the DM18 spectrum of a Group A spectroscopic candidate zf\_14000 (DM18 ID: L677152, $z_{\rm DM18}=3.779$; top plot). We shift the spectrum to our LRIS redshift (green) and the DM18 redshift (black) to compare. We highlight one potentially real spectral feature in red (C IV) that may add weight to the LRIS redshift, although with low confidence.

The lower panel of Figure \ref{fig:deimos} shows the DM18 spectrum of one of the Group B photometric candidates zf\_11423 (DM18 ID: L658665, $z_{\rm DM18}=3.655$). We highlight two potentially real spectral features in red. However, as the spectrum is noisy and is shown heavily smoothed here, we are not confident of the DM18 redshift. This galaxy therefore remains in Group B with the same confidence as our photometric candidates. Both spectra have assigned quality flags from DM18 of 1 (lowest quality), and a quality flag in the combined spectroscopic catalog of 2 (from 1 uncertain to 4 best) based on a comparison of available redshifts in each of the 16 catalogs. 

We show the unsmoothed 3D-HST G141 grism spectrum \citep{Brammer2012, Momcheva2016} of the Group A spectroscopic candidate  zf\_9775 in Figure \ref{fig:3dhst}. The same spectrum is shown at rest wavelengths assuming the redshift derived from our LRIS spectrum ($z_{\rm LRIS}=4.365$, green, top panel) and that quoted in 3D-HST ($z_{\rm 3DHST}=0.446$, black, bottom panel). There are no redshift quality flags provided by the 3D-HST survey. However, as the zf\_9775 3D-HST redshift is derived from a single $<1\sigma$ S III line, it is low confidence. The 3D-HST grism spectra are not visually inspected. A recent paper performed visual inspection of a large number of 3D-HST spectra and found that a quarter of the published redshifts are not correct \citep{Henry2021}. Additionally, redshifts for zf\_9775 from ZFOURGE, COSMOS, average CANDELS, and LRIS are all in close agreement with a stated redshift of $z\sim4.2\text{--}4.4$. The 3D-HST redshift ($z=0.446$) deviates from these values significantly.

\renewcommand{\thefigure}{C\arabic{figure}}
\renewcommand{\theHfigure}{C\arabic{figure}}
\setcounter{figure}{0}

\section{Escape fraction simulations}
\label{sec:appfesc}

\begin{figure}
\begin{center}
\includegraphics[width=0.5\textwidth]{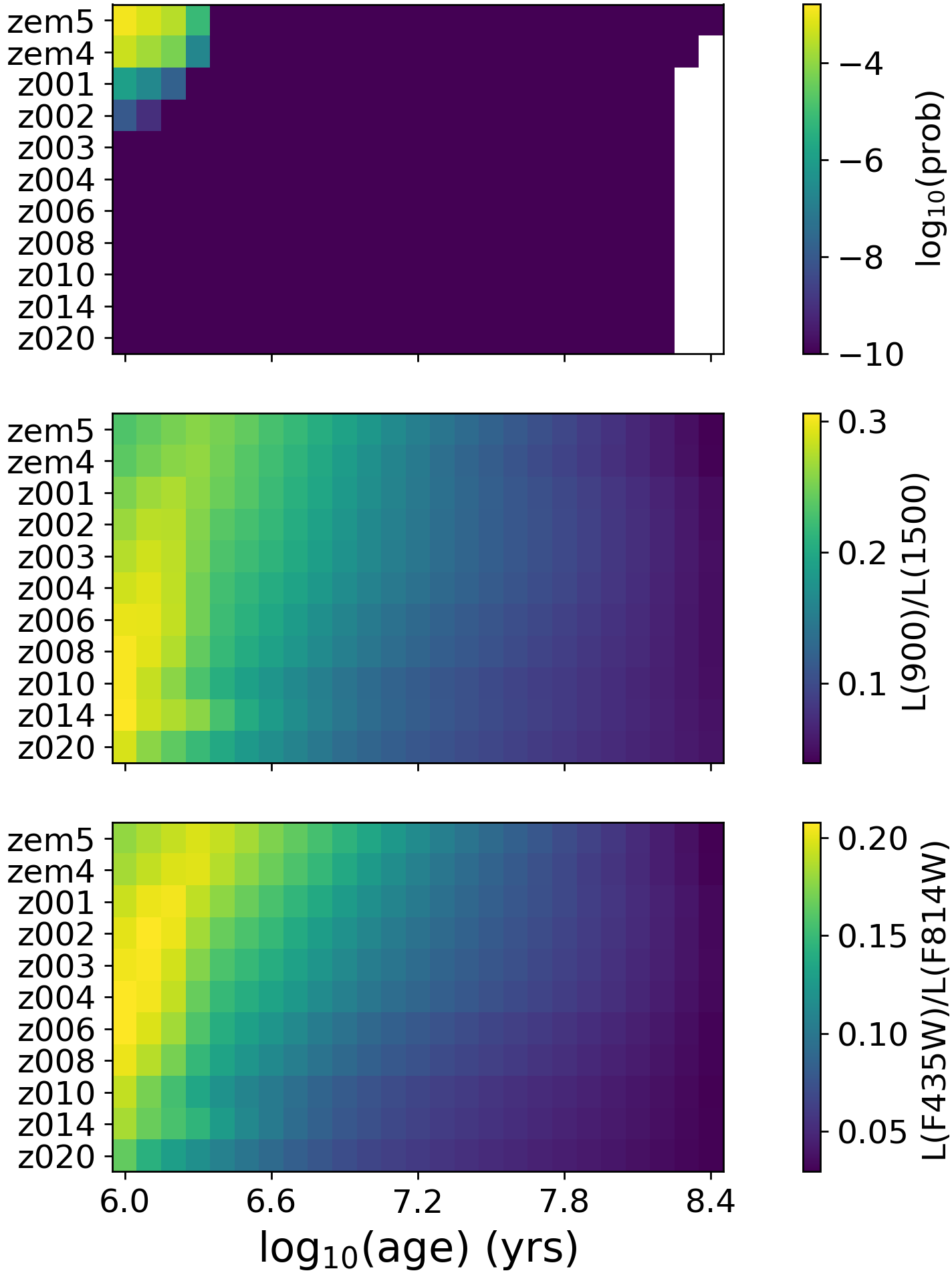} 
\caption{Log(probability) for zf\_11754 at each age and metallicity of the BPASS models (top panel). The LCG candidate strongly prefers a young, metal-poor template. These grids are used to determine the best E(B-V) to the ZFOURGE photometry for exponentially declining star-formation rate (SFR) BPASS models at a range of metallicities and ages. For both zf\_11754, zf\_9775 and zf\_14000 the best E(B-V) is always 0. The lower two panels show the intrinsic L(900)/L(1500) and L(F435W)/L(F814W) ratios for every single model age and metallicity. \small} 
\label{fig:fescmetal}    
\end{center}
\end{figure}

\begin{figure*}
\includegraphics[width=\textwidth]{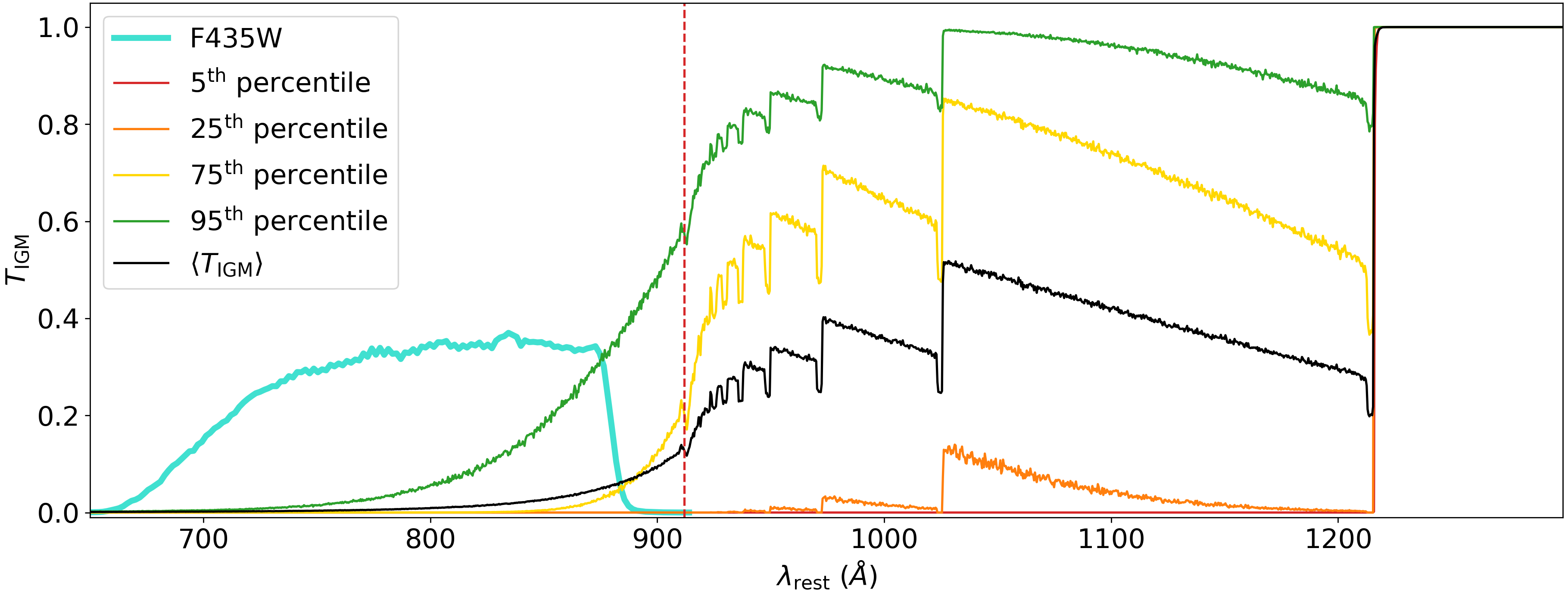}
\caption{IGM transmission curves for zf\_11754 ($z=4.470$), the highest redshift spectroscopic LCG candidate in Group A. Shown is the median (black line) of the 10,000 IGM curves generated. The 5$^{\rm th}$ (red), 25$^{\rm th}$ (orange), 75$^{\rm th}$ (yellow), and 95$^{\rm th}$ (green) percentiles are also shown. The Lyman break ($\sim912~\text{\AA}$) is shown (red dashed line). Overlaid is the LyC probing filter for this candidate F435W (turquoise). The curves generated for all three of the Group A spectroscopic candidates are very similar due to their close redshifts. The median and percentile IGM curves do not represent individual lines of sight. \small} 
\label{fig:fesctau}
\end{figure*}

We summarize the key steps for estimating $f_{\rm esc}$ PDFs for the three Group A spectroscopic LCG candidates. This method is detailed in full in \cite{Bassett2022}. See Section \ref{sec:fesc} for a summary of the results.
\begin{enumerate}[leftmargin=*]
\itemsep0em
    \item We use a range of possible SED templates from BPASS \citep[v2.1;][]{Eldridge2017}. We combine these with a single, exponentially declining SFH with $e$-folding time 0.1 Gyr. We also use BPASS models with metallicity $Z_{*}=0.001$--0.020 \citep[as in][]{Steidel2018, Fletcher2019, Bassett2021a}, initial mass function slope $\alpha=-2.35$, and stellar-mass limit $300~\rm M_{\odot}$. We consider only SED models with ages $\lesssim2.5\times10^8$ years ($(L_{900}/L_{1500})_{\rm int} > 0.05$) as fainter LyC is highly unlikely to be detected in galaxies at $z\simeq3.7\text{--}4.4$.
    \item We estimate E(B$-$V) at each BPASS model age. We apply each E(B$-$V) value to a \cite{Reddy2016} attenuation curve, determine photometric fluxes of attenuated models using the ZFOURGE filters and minimize $\chi^2$. With the optimal E(B$-$V) for a given age, we calculate the likelihood that the model represents our LCG candidates. This produces a best matched model (age, metallicity, and E(B$-$V)) for our photometry which we use for calculating the $f_{\rm esc}$ PDFs. \label{step:bestf}
    \item We measure the $\log(\rm probability)$ of each LCG candidate for every age and metallicity of the BPASS models in our range. We also measure the intrinsic LyC to non-ionizing UV continuum luminosity ratio $(L_{\rm 900}/L_{\rm 1500})_{\rm int}$ and the measured $(L_{\rm LyC}/L_{\rm F814W})$ where $L_{\rm LyC}$ is either F336W or F435W depending on the LCG candidate. The resulting weights for zf\_11754 are shown in Figure \ref{fig:fescmetal}. The best fitting models for all three Group A candidates is ``zero-age'' for BPASS which is equivalent to a single, instantaneous burst with an age of $10^6$ years (i.e., only one template is used). For these zero-age best-fit models, the choice of SFH will not change the fact that the result is unconstrained (i.e., still $>100\%$, see Section \ref{sec:fesc}). An investigation into the difference of model SFH for the probabilistic $f_{\rm esc}^{\rm PDF}$ method is therefore not appropriate for these candidates and is beyond the scope of this paper. 
    \item To generate an $f_{\rm esc}$ PDF, for each SED model we need to sample the PDF of IGM transmission $T_{\rm IGM}$. 10,000 IGM sight lines are produced using the \textsc{taoist-mc} code\footnote{\url{https://github.com/robbassett/TAOIST_MC}} \citep[described in full in][]{Bassett2021a}, based on methods from \cite{Inoue2014, Steidel2018}. These are produced for each redshift of the Group A spectroscopic candidates. Given their relatively close redshifts, these have very similar distributions and we show just one distribution for zf\_11754 in Figure \ref{fig:fesctau}. The median and percentile IGM curves do not represent individual sight lines.
    \item We then determine the probability that the photometric LyC flux and error is consistent with $f_{\rm esc}$ considering the SED models and IGM transmission curves. We apply our ensemble of 10,000 IGM curves and 10,000 $f_{\rm esc}$ values (between 0 and 1) to every BPASS model. We measure the probability ($P_{\rm obs}$) that a given mock observation is consistent with the observations. We take the value at a model flux for a Gaussian distribution with center of the observed flux and FWHM of the error. 
    \item The probability of any $f_{\rm esc}$ value is taken as the median across all 10,000 sight lines. The PDFs of $f_{\rm esc}$ are for an individual SED age, metallicity, and E(B$-$V). We therefore use our best fitting models from Step \ref{step:bestf} to generate the most appropriate $f_{\rm esc}$ PDF (and corresponding $f_{\rm esc}^{\rm PDF}$ estimate) for each of our sources. 
\end{enumerate}

\bibliography{lcg_cosmos.bib}{}
\bibliographystyle{aasjournal}

\end{document}